\documentclass[sigconf, nonacm=true]{acmart}
\makeatletter
\def\@ACM@checkaffil{% Only warnings
    \if@ACM@instpresent\else
    \ClassWarningNoLine{\@classname}{No institution present for an affiliation}%
    \fi
    \if@ACM@citypresent\else
    \ClassWarningNoLine{\@classname}{No city present for an affiliation}%
    \fi
    \if@ACM@countrypresent\else
        \ClassWarningNoLine{\@classname}{No country present for an affiliation}%
    \fi
}
\makeatother
\usepackage{thm-restate}
\usepackage{booktabs} % For formal tables
\usepackage{color,soul}
\usepackage{fancyvrb}
\usepackage{amsmath}
\usepackage{mathtools}
\usepackage{float}
\usepackage{bbm}
\usepackage{tcolorbox}
\usepackage{graphicx}
\usepackage{lipsum}
\usepackage{xfrac}
\usepackage{colortbl}
\usepackage{multirow}
\usepackage{balance}
\usepackage{paralist}
\usepackage{tabularx}
\usepackage{etoolbox}
\usepackage{caption}
\usepackage{subcaption}
\usepackage{algorithm}
\usepackage{algpseudocode}

\usepackage{enumitem}

\usepackage{tikz}
\usepackage{array}
\usepackage{xcolor}
\usepackage{multicol}

\usepackage{enumitem}
\usepackage{makecell}
\usepackage{tikz}
\definecolor{bgred}{RGB}{255,210,205}
\definecolor{bgblue}{RGB}{210,220,255}
\definecolor{bgyellow}{RGB}{255,255,109}
\definecolor{bggrey}{RGB}{223,223,225}
\definecolor{bgpink}{RGB}{255,202,239}
\definecolor{bgsky}{RGB}{182,219, 255}
\definecolor{purple}{RGB}{180,0,180}

\definecolor{hvygreen}{RGB}{0,128,0}
\definecolor{hvypink}{RGB}{102,0,51}
\definecolor{forstgreen}{RGB}{10,150,10}

\AtBeginDocument{%
  }

\definecolor{hvygreen}{RGB}{0,128,0}
\definecolor{hvypink}{RGB}{102,0,51}

\renewcommand{\epsilon}{\varepsilon}
\renewcommand{\phi}{\varphi}

\newcommand{\AC}{{\mathrm{\mathcal{AC}}}}

\newcommand{\diversityTabEE}{{\mathrm{Div}}}
\newcommand{\intTabEE}{{\mathrm{Int}}}
\newcommand{\sufTabEE}{{\mathrm{Suf}}}

\newcommand{\interest}{{\mathrm{Int}_p}}
\newcommand{\suf}{{\mathrm{Suf}_p}}
\newcommand{\diversity}{{\mathrm{Div}_p}}

\ifdef{\removecomments}
{
    \newcommand{\tova}[1]{}
    \newcommand{\kathy}[1]{}
    \newcommand{\amir}[1]{}
    \newcommand{\ron}[1]{}

}
{
    \newcommand{\annotation}[1]{[[[#1]]]}
    
    \newcommand{\kathy}[1]{\textbf{\small\textcolor{hvypink}{\annotation{(Kathy)~#1}}}{\typeout{#1}}}
    \newcommand{\tova}[1]{\textbf{\small\textcolor{purple}{\annotation{(Tova)~#1}}}{\typeout{#1}}}
    \newcommand{\amir}[1]{\textbf{\small\textcolor{hvygreen}{\annotation{(Amir)~#1}}}{\typeout{#1}}}
    \newcommand{\ron}[1]{\textcolor{blue}{\annotation{(Ron)~#1}}}{\typeout{#1}}

}
\newcommand{\eat}[1]{}

\usepackage[nameinlink, noabbrev, capitalize]{cleveref}

\usepackage{algorithm}
\usepackage{algpseudocode}

\algnewcommand{\LineComment}[1]{\State \textit{\textcolor{blue}{// #1}}}

\newtheorem{theorem}{Theorem}[section]
\newtheorem{problem}{Problem}
\newtheorem*{problem*}{Problem}

\def\OPT{\ensuremath{\mathrm{OPT}}}

\def\R{\ensuremath{\mathbb{R}}}

\def\cA{\ensuremath{\mathcal{A}}}
\def\cD{\ensuremath{\mathcal{D}}}
\def\cM{\ensuremath{\mathcal{M}}}

\def\cR{\ensuremath{\mathcal{R}}}

\def\cY{\ensuremath{\mathcal{Y}}}
\def\cZ{\ensuremath{\mathcal{Z}}}

\renewcommand{\Pr}{\mathop{\bf Pr\/}}

\newcommand{\set}[1]{\ensuremath{\{ #1 \}}}

\newcommand{\eps}{\varepsilon}

\newcommand{\norm}[1]{\left\|#1\right\|}
\newcommand{\abs}[1]{\left\vert#1\right\vert}

\newcommand{\one}{\mathbf{1}}

\newcommand{\suchthat}{{\;|\;}}

\newcommand{\proj}{\ensuremath{\pi}}

\newcommand{\Gumb}[1]{\mathrm{Gumbel}\left(#1\right)}

\newcommand{\dom}{\mathrm{dom}}

\newcommand{\score}{\mathrm{Score}}
\newcommand{\globscore}{\mathrm{GlScore}}

\newcommand{\hist}{{\ensuremath{h}}}
\newcommand{\cnt}{\mathit{cnt}}

\newcommand{\Explainer}{\mathsf{Exp}}

\newcommand{\nbr}{{\sim}}

\newcommand{\numclust}{{\abs{C}}}

\def\ourframework{$\mathsf{DPClustX}$}
\def\dpnaive{DP-Naive}

\def\dptabee{DP-TabEE}

\def\tvd{\mathrm{TVD}}

\newcommand{\paratitle}[1]{\vspace{1mm}\noindent\textbf{{#1}.}}

\newcommand\plabel[1]{\phantomsection\label{#1} } % This helps refernce lines in presence of multiple algorithms

\newcolumntype{C}{>{\centering\let\newline\\\arraybackslash\hspace{0pt}}m{3.5cm}}

\newif\ifpaper
\paperfalse

\begin{document}

\title{Differentially Private Explanations for Clusters}

\settopmatter{authorsperrow=4}
\author{Amir Gilad}
\affiliation{%
  \institution{Hebrew University}
  % \city{Jerusalem}
  % \country{Israel}
  }
% \email{amirg@cs.huji.ac.il}

\author{Tova Milo}
\affiliation{%
  \institution{Tel Aviv University}
  }
% \email{milo@post.tau.ac.il}

\author{Kathy Razmadze}
\affiliation{%
  \institution{Tel Aviv University}
}
% \email{kathyr@mail.tau.ac.il}

\author{Ron Zadicario}
\affiliation{%
  \institution{Tel Aviv University}
  }  
% \email{ronzadicario@mail.tau.ac.il}
%% The "title" command has an optional parameter,
%% allowing the author to define a "short title" to be used in page headers.
% \title{Differentially Private Histogram-Based Clustering Explanations}

\begin{abstract}

The dire need to protect sensitive data has led to various flavors of privacy definitions. Among these, Differential privacy (DP) is considered one of the most rigorous and secure notions of privacy, enabling data analysis while preserving the privacy of data 
contributors. 
One of the fundamental tasks of data analysis is clustering , which is meant to unravel hidden patterns within complex datasets. However, interpreting clustering results poses significant challenges, and often necessitates an extensive analytical process. 
Interpreting clustering results under DP is even more challenging, as analysts are provided with noisy responses to queries, and longer, manual exploration sessions require additional noise to meet privacy constraints.
While increasing attention has been given to clustering explanation frameworks that aim at assisting analysts by automatically uncovering the characteristics of each cluster, such frameworks may also disclose sensitive information within the dataset, leading to a breach in privacy.

To address these challenges, we present \ourframework, a framework that provides explanations for black-box clustering results while satisfying DP. \ourframework\ takes as input the sensitive dataset alongside privately computed clustering labels, and outputs a global explanation, emphasizing prominent characteristics of each cluster while guaranteeing DP.
We perform an extensive experimental analysis of \ourframework\ on real data, showing that it provides insightful and accurate explanations even under tight privacy constraints.

\end{abstract}

%%
%% The code below is generated by the tool at http://dl.acm.org/ccs.cfm.
%% Please copy and paste the code instead of the example below.
%%
% \begin{CCSXML}
% <ccs2012>
%  <concept>
%   <concept_id>00000000.0000000.0000000</concept_id>
%   <concept_desc>Do Not Use This Code, Generate the Correct Terms for Your Paper</concept_desc>
%   <concept_significance>500</concept_significance>
%  </concept>
%  <concept>
%   <concept_id>00000000.00000000.00000000</concept_id>
%   <concept_desc>Do Not Use This Code, Generate the Correct Terms for Your Paper</concept_desc>
%   <concept_significance>300</concept_significance>
%  </concept>
%  <concept>
%   <concept_id>00000000.00000000.00000000</concept_id>
%   <concept_desc>Do Not Use This Code, Generate the Correct Terms for Your Paper</concept_desc>
%   <concept_significance>100</concept_significance>
%  </concept>
%  <concept>
%   <concept_id>00000000.00000000.00000000</concept_id>
%   <concept_desc>Do Not Use This Code, Generate the Correct Terms for Your Paper</concept_desc>
%   <concept_significance>100</concept_significance>
%  </concept>
% </ccs2012>
% \end{CCSXML}

%%%%%%% Ron: I added this:%%%%%%%%%
\settopmatter{printfolios=true}
%%%%%%%%%%%%%%%%%%%%%%%%%%%%%%%%%%%
% \input{sections/revision-plan}

\maketitle
% \pagenumbering{arabic} % Set numbering style (arabic, roman, etc.)
% \setcounter{page}{1}
% \input{sections/intro}
\section{Introduction}\label{sec: intro}
% \ag{Ron - please add missing citations and change the template to ACM SIGMOD.}\ron{Changed to sigmod. TODO: citations}
Sensitive data collection has never been more prevalent, from fitness trackers~\cite{GabrieleC20} and financial institutions~\cite{oyewole2024data}, to healthcare and insurance facilities~\cite{AbouelmehdiHK18}. 
Such data require special treatment to allow for its safe and secure usage without exposing individuals to harm due to the presence of their information in the data. 
Differential Privacy (DP)~\cite{dwork2006differential,dwork2014algorithmic} has emerged as the gold standard for handling such issues. The crux of DP is to allow sensitive data to be used while bounding the privacy leakage and to offer utility bounds on the obtained results. Indeed, DP has been adopted by multiple companies~\cite{Abowd18,dwork2019differential} and governmental organizations~\cite{erlingsson2014rappor,ding2017collecting,tang2017privacy}. 

Among the many operations in data analysis, {\em clustering} plays a pivotal role in uncovering hidden patterns and providing actionable insights from data. To ensure privacy, researchers have worked extensively to adapt clustering techniques to comply with DP~\cite{gupta2010differentially, stemmer2018differentially,  su2016differentially,  ghazi2020differentially}. Under DP, the true clusters must be obfuscated and slightly distorted to prevent individual information leakage, often at the cost of accuracy. 

Clustering algorithms often operate as black boxes,
offering little insight into the reasoning behind their results. Hence, it is challenging for end users to comprehend this reasoning, or draw meaningful conclusions from the results based on domain knowledge \cite{hu2024interpretable}. The additional requirement to adhere to strict DP standards further amplifies this complexity. 
To account for this, previous work has focused on {\em explaining non-private clustering algorithms}~\cite{Cluster-Explorer,moshkovitz2020explainable,  esfandiari2022almost, makarychev2022explainable, gamlath2021nearly} and aimed to provide succinct and interpretable summaries of the properties of each cluster by showing how it varies from the other clusters. 
When considering privacy, it is likely that lack of access to the data is accompanied by lack of access to the clustering algorithm, requiring an approach that is suitable for black-box clustering algorithms.

Works that focus on providing explanations for black-box clustering results are often histogram-based approaches~\cite{TabEE,FEDEX}, which is a popular form of explanation in other settings as well, including visualization recommendations 
\cite{wongsuphasawat2016voyager,luo2018deepeye,lee2021lux}
and automated insight extraction from data \cite{tang2017extracting, amer2021exploring}.
With these approaches, users get a {\em histogram for each cluster} that focuses on a specific attribute and graphically shows how the data associated with the cluster differs from the rest of the data. Yet, the approaches that generate such explanations cannot be directly used in the DP setting. 

First, these approaches choose histograms based on quality functions, such as interestingness~\cite{TabEE, hilderman2013knowledge, sarawagi1998discovery, FEDEX}, sufficiency~\cite{TabEE, dasgupta2022framework}, and diversity~\cite{TabEE, youngmann2022guided},
which require significant distortion under DP. That is, the required noise scale is large relative to their range, making it impractical to obtain a reliable explanation, as the ranking of explanations by quality is unlikely to be preserved after adding noise.
Second, existing methods generate all explanation options before choosing the ones with the highest scores. However, in the DP setting, this strategy quickly becomes infeasible because it requires producing private histograms for every attribute and cluster, necessitating excessive distortion to ensure DP compliance.
Third, evaluating the quality of the explanation based on noisy histograms introduces excessive noise, as each bin is injected with independent noise, which accumulates and leads to an inaccurate quality assessment. 

In light of this, {\em we propose \ourframework, a novel framework for generating histogram-based explanations of black-box clustering results under DP.} 
\ourframework\ is inspired by previous work~\cite{TabEE,FEDEX} and addresses the above limitations as follows. 
(1) We first prove that previous approaches cannot be applied directly, as existing quality functions for histogram-based explanations would have to be significantly distorted to satisfy DP.
%, making the attainment of a reliable explanation impractical.} 
Then, we develop DP tailored variants that enable the generation of high-quality and privacy-preserving explanations. 
(2) To minimize privacy costs in the explanation selection process, we evaluate the \emph{attribute} quality directly, privately selecting high-scoring explanation attributes based on the sensitive dataset with our novel quality function. 
We then generate noisy histograms exclusively for the selected attributes, leveraging previous work on DP histograms \cite{dong2020optimal, dwork2006calibrating}.
However, each explanation corresponds to an assignment of histograms to clusters, and its quality is evaluated globally across all clusters. Therefore, the search space for the best clustering explanation, which encompasses all possible assignments, is considerably large.

\begin{figure}
        \centering
        \includegraphics[width=1\linewidth]{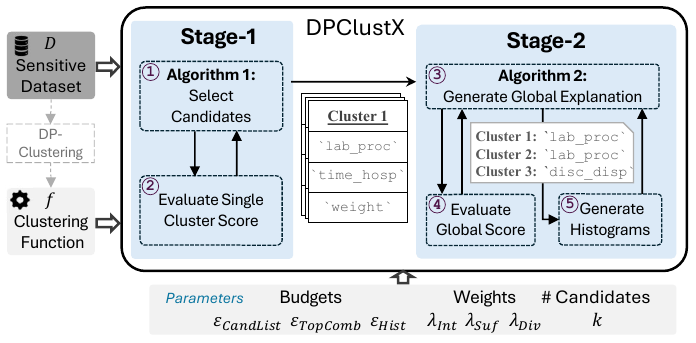}
        \caption{System architecture of \ourframework. Algorithm 1 selects candidate attributes for each cluster (1) using the single-cluster score function (2). Algorithm 2  selects a high-quality attribute combination from these candidates (3) using the global score function (4), and generates noisy histograms only for the selected attributes (5). 
        }
        \label{fig:system_architecture}
        % \vspace{-2pt}
    \end{figure}

Hence, (3) we adapt techniques from prior work~\cite{TabEE} and develop a DP mechanism that constructs a smaller {\em candidate  set} for each cluster, from which the clustering explanation is privately derived.
To this end, we adapt the idea of a single-cluster score function which is used to rank the attributes for each cluster by their explanation quality for that cluster~\cite{TabEE} and tailor it to the DP setting. 
This function is adapted to DP so that the noise added to it will still render the results useful in filtering attributes. 
However, to naively select the top explanation attributes 
% \ag{Explaining attributes or Explanation attributes?}\ron{changed}
for each cluster, one would need to apply a DP mechanism for privately selecting a single high quality attribute
% EM (\Cref{def:exp-mech} \ag{Can you explain this intuitively without referring to the technical def. here?})
multiple times, with each iteration requiring a re-calculation of noisy scores for all remaining candidates. Instead, we utilize the One-shot Top-$k$ mechanism~\cite{durfee2019practical}, which computes the noisy scores {\bf once} and then releases the top $k$ candidates. This further reduces execution times, thereby enhancing the interactive user experience.
An illustration of the \ourframework\ framework, summarizing these steps, is given in \Cref{fig:system_architecture}. 
Our experimental study confirms that \ourframework\ generates insightful and accurate explanations even under tight privacy constraints, demonstrating robustness to attribute correlations, variations in the number of clusters, and maintaining reasonable execution times

\begin{figure}
\centering
\footnotesize 
        \begin{tabular}{|l|l|l|l|l||l|} 
        \hline
        \rowcolor{gray!30} 
        % \textbf{\texttt{age}} &  \textbf{\texttt{gender}} & \makecell{ \textbf{\texttt{num\_lab\_}} \\ \textbf{\texttt{procedures}}} & \textbf{\texttt{diag\_1}} & \textbf{\dots} & \textbf{\texttt{cluster}} \\
         \textbf{\texttt{age}} &  \textbf{\texttt{gender}} & \textbf{\texttt{lab\_proc}} & \textbf{\texttt{diag\_1}} & \textbf{\dots} & \textbf{\texttt{cluster}} \\
        \hline
            $\mathtt{[60,70)}$ & \texttt{Female} & $\mathtt{[40, 50)}$ & \texttt{Circulatory} & \dots & $\mathtt{1}$ \\
            $\mathtt{[60,70)}$ & \texttt{Female} & $\mathtt{[0,10)}$ & \texttt{Diabetes} & \dots & $\mathtt{2}$ \\
             $\mathtt{[70,80)}$ & \texttt{Male} & $\mathtt{[40,50)}$ & \texttt{Injury} & \dots & $\mathtt{2}$ \\
            \dots & \dots & \dots & \dots & \dots & \dots \\
            \hline
        \end{tabular}
    
\caption{Example of the Diabetes dataset.}
\label{fig:example_diabetes_data}
\end{figure}

\begin{example}\label{example:top_exp_diabates}
Consider an analyst working with the Diabetes dataset \cite{diabetes} (a subset of the columns is illustrated in \Cref{fig:example_diabetes_data}), 
aiming to identify groups of patients with similar medical records using the DP-$k$-means algorithm \cite{su2016differentially}. While DP-$k$-means privately provides cluster centers (see the last column in \Cref{fig:example_diabetes_data} associating each tuple with a cluster), it does not offer additional insights about the clusters. Instead of exhausting the privacy budget through a manual EDA session, the analyst employs \ourframework\ to generate histogram-based explanations. These explanations reveal significant patterns, such as the fact that Cluster 1 consists primarily of individuals who underwent a higher number of lab procedures, as shown in \Cref{fig:hist_clust_example_diabetes}. Comparing the cluster distribution of \texttt{lab\_proc} with the remaining data, we see that values outside Cluster 1 are concentrated in the middle and lower ranges, peaking at $[40,50)$. In contrast, Cluster 1 values are concentrated in the upper range, peaking at $[60,70)$. This suggests that the clustering algorithm groups individuals with a higher number of lab procedures in Cluster 1. 
For simplicity, we have attached an LLM generated textual description of the histogram in \Cref{fig:textual_example1}. 

    \begin{figure}
        \begin{subfigure}{1\linewidth}
            \centering
            \includegraphics[scale=0.48]{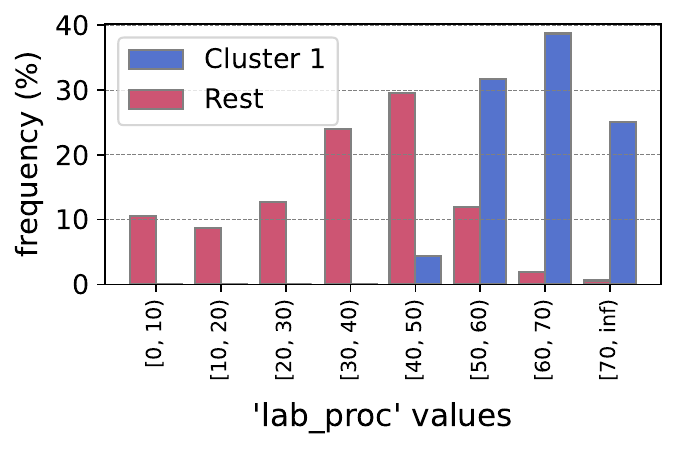} 
             \caption{Part of a histogram-based explanation for the Diabetes dataset. The selected attribute, \texttt{lab\_proc}, specifies the number of lab procedures an individual underwent.
            }
         \label{fig:hist_clust_example_diabetes}
        \end{subfigure}
        
        \begin{subfigure}{1\linewidth}
        \vspace{3mm}
            \begin{tcolorbox}[colback=white,left=1.5pt,right=1.5pt,top=0pt,bottom=0pt]
        \textbf{Textual description:}  
        The `\texttt{lab\_proc}` column values
        differ significantly. Values outside Cluster 1 are concentrated in the lower
        and mid-range ($\mathbf{85\%}$ below $\mathbf{50}$), while Cluster 1 contains mainly higher
        values ($\mathbf{95\%}$ above $\mathbf{50}$).
        \vspace{-2pt}
\end{tcolorbox}
     \caption{Textual description of the histogram explanation in \Cref{fig:hist_clust_example_diabetes}.}
      \label{fig:textual_example1}
        \end{subfigure}
        \caption{Cluster explanation with its textual description.}
    \end{figure}

\end{example}

% \begin{figure*}[t]
%     \begin{subfigure}[t]{0.52\linewidth}
%         \centering
%         \includegraphics[width=1\linewidth]{img/sys_arch.pdf}
%         \caption{\ourframework' system architecture}
%         \label{fig:system_architecture}
%         \vspace{0pt}
%     \end{subfigure}
%     \hfill
%     % Subfigure 3
%     \begin{subfigure}[t]{0.45\linewidth}
%            \centering
%             \includegraphics[width=1\linewidth]{img/3_0_diabetes2_clust_explanation.pdf} 
%            \caption{Part of a histogram-based explanation for the Diabetes dataset. The selected attribute, \texttt{lab\_proc}, specifies the number of lab procedures an individual underwent. \ag{Change the size of the intervals to be smaller}\ron{changed}}
%          \label{fig:hist_clust_example_diabetes}
%     \end{subfigure}

%      \caption{The \ourframework\; framework: system architecture and output example.}
% \end{figure*}

\paragraph{Our contributions}\label{par: contributions} Our contributions are summarized as follows:
\begin{itemize}[noitemsep,topsep=0pt,leftmargin=*]
    % \item We formally define privacy-preserving histogram-based explanation mechanisms for black-box clustering methods (\Cref{sec:prob_setting}) \ag{Need to see if this is not going to upset reviewers if listed as novel.}\ron{Let's see if we can phrase it as a contribution (from meeting of 29.12)}.
    \item We formulate the problem of finding a high scoring privacy-preserving histogram-based explanation for black-box clustering methods, with the main challenge being the private selection of high-quality attributes (\Cref{sec:prob_setting}).
    % \ag{Your auxiliary text is overlapping with the left text column. Can you place it to the right?}
    \item We design quality functions for histogram-based explanations inspired by the notions of interestingness, sufficiency, and diversity that are adapted for the DP setting (\Cref{sec:metrics}). 
    % \item An open-source implementation of the framework as a Python library. \cite{dpclustex_github}
    \item We develop \ourframework, the first framework designed to generate histogram-based explanations for clustering results under DP (\Cref{sec: alg}), equipped with formal guarantees. 
    \item We provide a comprehensive experimental study demonstrating the effectiveness of \ourframework, showing that its explanations align closely with non-private explanations even under a strict privacy budget (\Cref{sec: exp}). 
\end{itemize}

\section{Preliminaries}\label{sec: prelim}
In this section, we introduce our notations, and review existing concepts from DP used in this work.

\paragraph{Data} 
% \ag{Do not make titles bold.} 
Given a single table schema $R(A_1,\dots,A_d)$, $R$ is a relation name and $\cA=\set{A_1,\dots,A_d}$
denotes the set of attributes. Each attribute $A_i$ has discrete, finite, and data-independent domain $\dom(A_i)$. The full domain of $R$ is $\dom(R)=\dom(A_1)\times \dots \times \dom(A_d)$. An instance (dataset) $D$ of a relation $R$ is a bag (multiset) whose elements are tuples in $\dom(R)$. We let $\cD$ denote the set of all datasets of the relation $R$, i.e., $\cD = \{D \mid D \subseteq \dom(A_1)\times \dots \times \dom(A_d), |D| < \infty\}$.
The number of tuples in $D$ is denoted as $\abs{D}$. $\proj_A(D)$ is the projection of $D$ onto the attribute $A$.
For a dataset $D$ and an attribute $A \in \cA$, we let $\dom_{D}(A)$ denote the {\em active domain} of $A$ with respect to $D$, i.e, the subset of $\dom(A)$ of values appearing in $\proj_A(D)$ at least once.
 we denote by $\cnt_{A=a}(D)$ the number of occurrences of value $a$ in $\proj_A(D)$, and by $\hist_A(D)$
 %$\in(\N\cup\set{0})^{\dom(A)}$ 
 the histogram of the dataset $D$ over attribute $A$. That is, $\hist_A(D)[a] = \cnt_{A=a}(D)$ for any $a\in \dom(A)$. For visualizations, we use normalized histograms, where each count is replaced by its proportion. 

\begin{example}\label{example:notations}
    Consider the Diabetes dataset, illustrated in \Cref{fig:example_diabetes_data}, which contains $47$ attributes, including \texttt{age}, \texttt{gender}, \texttt{lab\_proc}, and \texttt{diag\_1}. The domain $\dom(\mathtt{lab\_proc})$ comprises 8 values, each representing a range of lab procedure counts. 
    \Cref{fig:hist_clust_example_diabetes} illustrates two histograms derived from the Diabetes dataset.  For instance, the blue bars represent the histogram $h_{\mathtt{lab\_proc}}(D_1)$, where $D_1$ is the cluster explained.
     Here, $\dom_{D_1}(\mathtt{lab\_proc})$ comprises 4 values, as no tuple in $D_1$  has $\mathtt{lab\_proc} < 40$.
\end{example}

\paratitle{Histogram-based explanation (HBE)} A clustering of $D$ is a partition into disjoint subsets $\set{D_c\subseteq D \mid c\in C}$, each assigned with a cluster label $c\in C$. That is, $D_{c} \cap D_{c'} = \emptyset$ for $c\neq c'$ and $\bigcup_{c\in C} D_c = D$,
% \ag{
% %Add that $D_{c_i} \cap D_{c_j} = \emptyset$ and their union is the entire $D$.
% Why not just write $C_1, ..., C_k$? why use $D_c$? This would simplify notation}
An single-cluster HBE candidate consists of two histograms on a specified attribute: the histogram of the cluster values, and the histogram of the values outside the cluster (illustrated in \Cref{fig:hist_clust_example_diabetes}). A global HBE candidate is a set of single-cluster HBE candidates, with a candidate for each cluster. Formally,
\begin{definition}[Single cluster HBE candidate \cite{TabEE, FEDEX}]\label{def:single-candidate}
For a dataset $D$ and a cluster label $c\in C$, a single-cluster HBE candidate $e_c$ is a tuple $(c, A,\hist_{A}(D \setminus D_c), \hist_{A}(D_c))$
where $A\in \cA$.
\end{definition}

% \ag{Example}
\begin{example}
    \Cref{fig:hist_clust_example_diabetes} illustrates a single cluster HBE candidate for Cluster 1 in the Diabetes dataset \cite{diabetes}, using the \texttt{lab\_proc} attribute. The (normalized) histogram of the cluster values is shown in blue, and for the remaining data in red.
\end{example}

\begin{definition}[Global HBE candidate \cite{TabEE}]\label{def:global-candidate} Given $D$, clustered into $\set{D_c \mid c\in C}$, a global HBE candidate is a set $\set{e_c \mid c\in C}$, where $e_c$ is a single-cluster HBE candidate for $D_c$.
\end{definition}

An HBE should capture unique patterns that characterize each cluster, and distinguish it from other clusters.
Building upon \cite{FEDEX, dasgupta2022framework, youngmann2022guided, hilderman2013knowledge}, the work of \cite{TabEE} proposed the aspects of \emph{interestingness}, \emph{sufficiency}  and \emph{diversity} for evaluating the quality of an HBE. Briefly, interestingness is quantified as the distributional shift between the inner cluster distribution and the full dataset in a given attribute. Sufficiency represents the extent to which the HBE captures the patterns of the underlying clustering, and diversity measures the overall distinctiveness among explanations. We discuss  these measures further in \Cref{sec:metrics}.

\subsection{Differential Privacy}
\label{subsec:prelim-dp}
We next review preliminary background from the DP literature used in this work.  
DP ensures that the distribution of outputs does not significantly change when the algorithm is applied to any two {\em neighboring databases}. 
% DP enables querying a private database without compromising privacy by ensuring that the distribution of outputs does not significantly change when the algorithm is applied to any two distinct but {\em neighboring databases}. 
% (\Cref{def:nbr-datasets}). The parameter $\eps$ is referred to as the privacy budget, and a lower $\eps$ value implies a lower privacy loss. In our privacy analysis we use the composition results stated in \Cref{prop:composition}.

\begin{definition}[Neighboring Datasets~\cite{dwork2006differential}]\label{def:nbr-datasets}
Two datasets $D$ and $D'$ are called \emph{neighboring} (denoted $D\nbr D'$) if $D'$ can be obtained from $D$ by removing or adding one tuple.
% , i.e,  $\abs{(D\setminus D') \cup (D'\setminus D)} = 1$.
\end{definition}

\begin{definition}[$\eps$-Differential Privacy (DP) \cite{dwork2006differential, dwork2014algorithmic}]\label{def:DP}
    A randomized mechanism $\cM$ is said to satisfy $\eps$-DP if for any neighboring datasets $D\nbr D'$  and any set of possible outputs $S\subseteq Range(\cM)$, 
    \[
       \Pr[\cM(D)\in S] \le e^\eps \cdot \Pr[\cM(D')\in S],
    \]
    where the probability is over the randomness of $\cM$.
\end{definition}
\begin{proposition}\label{prop:composition}
The following holds for DP \cite{dwork2006calibrating, dwork2014algorithmic}:
\begin{itemize}
    \item \textit{Sequential Composition:} Let $\cM_1:\cD\to\cY$ and $\cM_2:\cD\times\cY\to\cZ$. 
    Suppose $\cM_1$ satisfies $\eps_1$-DP and for every $y\in\cY$, $\cM_2(\cdot,y)$ satisfies  $\eps_2$-DP (as a function of its first input). Define $\cM_3:\cD\to\cZ$ by $\cM_3(D) = \cM_2(D,\cM_1(D))$. Then, $\cM_3$ satisfies $(\eps_1+\eps_2)$-DP 
    % \ag{Is this phrasing with two inputs to $\cM_2$ something that you took from another paper? What do you mean by `$\eps_2$-DP (as a function of its first input)'?}\ron{
    % Yes. for example Theorem B.1 in the book , or Bun and Steinke's CDP paper}.
    \item \textit{Parallel Composition:}  Suppose $\cM_1:\cD\to\cY$ satisfies $\eps_1$-DP and  $\cM_2:\cD\to\cZ$ satisfies $\eps_2$-DP. Let $\cD_1,\cD_2\subseteq \dom(R)$ be disjoint subsets of the input domain. Define $\cM':\cD\to\cY\times \cZ$ by $\cM'(D)=(\cM_1(D\cap\cD_1),\cM_2(D\cap\cD_2))$. Then, $\cM'$ satisfies $\max\set{\eps_1, \eps_2}$-DP.
    \item \textit{Post-processing:}  Suppose $\cM:\cD\to\cY$ satisfies $\eps$-DP. Then, for any function $g:\cY\to\cZ$ (deterministic or randomized), The mechanism defined by $g(\cM(D))$ satisfies $\eps$-DP.  
\end{itemize}

\end{proposition}

To quantify the noise that has to be injected in order for mechanisms to satisfy DP, we first define then notion of sensitivity, 

\begin{definition}[Sensitivity~\cite{dwork2014algorithmic}]\label{def:sensitivity}
For a set of candidates $\cR$, Let $q: \cD \times \cR \to\R$ be a quality function that given $D\in \cD$, defines a score for every $r\in \cR$. The sensitivity of $q$ is
$$
\Delta_q = \sup_{r\in\cR}\sup_{D\nbr D'} \abs{q(D)-q(D')}
$$
\end{definition}

The exponential mechanism \cite{mcsherry2007mechanism} is a DP primitive for privately releasing the top item from a set of candidates with respect to a quality function that depends on the sensitive dataset.
The mechanism injects noise to the selection process, and outputs each candidate with probability proportional to its score. 

\begin{definition}[The Exponential Mechanism (EM) \cite{mcsherry2007mechanism}]\label{def:exp-mech}
Given $D\in\cD$, a set of candidates $\cR$, a quality function  $q:\cD\times \cR \to\R$, and a privacy parameter $\eps$, the exponential mechanism  $\cM_E$ selects and  outputs $r\in \cR$ with probability proportional to 
    $\exp\left( \frac{\eps \cdot q(D,r)}{2\Delta_q} \right)$ 
\end{definition}

\begin{theorem}
[\cite{mcsherry2007mechanism}]\label{thm:exponential_mech}
The exponential mechanism satisfies $\eps$-DP.
\ifpaper
\else
Moreover,
\[
\Pr \left[ \cM_E(D,\cR,q,\eps) \le \max_{r\in \cR}q(D,r) -  \frac{2\Delta_q}{\eps} (\ln\abs{\cR} + t )\right] \le e^{-t}.
\] 
\fi

\end{theorem}

\paratitle{The One-shot Top-k Mechanism}
In \ourframework, we utilize the \textit{One-shot Top-k mechanism} \cite{durfee2019practical} to privately select top-$k$ explanation attributes for each cluster. This mechanism adds independent Gumbel noise\footnote{The CDF of the Gumbel distribution $\Gumb{\sigma}$ is $F(z) = \exp(-\exp(-z/\sigma))$.}
to each of the true scores with scale $\sigma = 2\Delta_\score \cdot k/\eps$, where $\Delta_\score$ is the sensitivity of the score function (\Cref{def:sensitivity}). Then, it reorders
all the candidates in a descending order by their noisy scores, and outputs the first $k$ candidates.
It satisfies $\eps$-DP since it is identical to iteratively applying the EM $k$ times \cite{durfee2019practical}, where each satisfies $\eps/k$-DP. 
Therefore, by sequential composition (\autoref{prop:composition}) it satisfies  overall $\eps$-DP.

\paratitle{Differentially private histograms} \label{para:private-hist}
DP mechanisms for computing private histograms are well-studied (e.g., \cite{hay2010boosting, xu2013differentially, acs2012differentially, qardaji2013understanding, lin2013information}). As \ourframework\ can be instantiated with any DP histogram generation mechanism, we denote it as $\cM_{hist}(\proj_A(D), \eps_{hist})$. It takes the column of interest $\proj_A(D)$ and a privacy budget $\eps_{hist}$, and outputs a histogram of noisy counts $\widetilde{h_A}(D)$ over $\dom(A)$, while satisfying $\eps_{hist}$-DP. Such mechanisms are accompanied by utility bounds, enabling accuracy control by translating accuracy requirements into the required privacy budget.

\paratitle{Differentially private clustering} \label{para:private-clust}
DP clustering has been extensively studied in the DP literature, with prominent approaches aiming to release  cluster centers  from a sensitive dataset $D$ while preserving DP (e.g., \cite{gupta2010differentially, stemmer2018differentially,  su2016differentially,  ghazi2020differentially}). 
In the non-private  black-box clustering explanation setting (e.g. \cite{Cluster-Explorer}), a clustering is typically modeled by a function $f: D \to C$,  assigning  each tuple in $D$ a cluster label $c\in C$. However, this modeling is inherently unsuitable for the output of DP clustering, which requires any possible output (i.e., a clustering function) to occur with similar probability for any two different but neighboring datasets. Instead, we model the output of a DP clustering algorithm as function $f: \dom(R) \to C$. This definition encompasses DP mechanisms that release centers, as the fixed centers define a cluster assignment for any tuple in $\dom(R)$, while also accommodating other approaches, such as user-defined predicates or future clustering techniques.% With this motivation, we define:

\section{Problem Statement}
\label{sec:prob_setting}
% We begin by outlining the problem addressed in this work, starting with the privacy setup which is inspired by prior work on DP black-box classifier explanations~\cite{patel2022model}. Subsequently, we present the formal problem definitions.
We begin by describing the privacy setup. Subsequently, we present the formal problem definitions.

An HBE mechanism (\Cref{def:HBE-mech}) preserves privacy if for any clustering function $f:\dom(R)\to  C$, its distribution of outputs does not change much when we add a single tuple to the dataset. Note that we assume only \emph{black-box} access to $f$, and therefore make no assumptions regarding the formation or structure of the clusters. 
\begin{definition}\label{def:HBE-mech}
An HBE mechanism $\Explainer(D,f)$ is an algorithm that takes a dataset $D$ and a clustering function $f:\dom(R)\to  C$, and outputs a global HBE (\Cref{def:global-candidate}) for the clustering 
of $D$ defined by $f$.
We say that $\Explainer$ satisfies $\eps_{exp}$-DP if for any clustering function $f$, for any pair of neighboring datasets $D \nbr D'$, and any set of explanations $S\subseteq Range(\Explainer)$, it holds $\Pr[\Explainer(D,f)\in S] \le e^{\eps_{exp}} \cdot \Pr[\Explainer(D',f)\in S]$.
\end{definition}

Note that a clustering algorithm may output different clustering functions for two neighboring datasets. However, \Cref{def:HBE-mech} is motivated by the sequential composition theorem
(\Cref{prop:composition})
as our approach is aimed at a privately computed $f$. 
To formally argue that the entire process of DP clustering and explanation satisfies DP by applying sequential composition, it suffices to show that if we fix the output of the clustering algorithm (i.e., a clustering function), the distribution of outputs of $\Explainer(\cdot,f)$ only changes slightly under two neighboring input datasets, \emph{with the fixed clustering function $f$}.
A similar setting has also been used for DP  classifier explanations~\cite{patel2022model}.
The resulting overall privacy guarantee is as follows.
 Let $M_{clust}(D, C)$ be a DP clustering mechanism, i.e, an algorithm that outputs a clustering function $f$ while satisfying $\eps_{clust}$-DP (e.g, DP $k$-means \cite{su2016differentially}, where $f$ is defined by the centers). 
% By the sequential composition theorem (\Cref{prop:composition}), 
By \Cref{prop:composition} (sequential composition), 
the sequential, adaptive composition $\Explainer(D,M_{clust}(D, C))$ 
satisfies $(\eps_{exp}+\eps_{clust})$-DP. 

One challenge with existing approaches is the high sensitivity of previously proposed HBE score functions, which cannot be used directly in the DP setting  (e.g., \Cref{claim:interest-high-sens}). Moreover, these functions are applied to pre-computed explanation candidates (\Cref{def:single-candidate}), and \emph{privately} computing all candidates incurs a significant waste of privacy budget.
Instead, one would hope to assess \emph{attribute} quality directly, and produce DP histograms only for attributes selected for the output.

An \emph{attribute combination} $\AC: C \to \cA$ maps each cluster label to an attribute. Thus, our goal is to find a high-quality attribute combination such that the histograms of the corresponding attributes form a high-quality HBE\footnote{While the global explanation uses one histogram per cluster, as in prior work \cite{TabEE}, our framework can be extended to output multiple histograms per cluster. However, this comes at the cost of increased complexity, as further discussed in 
\ifpaper
\cite{full_version}.
\else
\Cref{appendix:multexp}.
\fi
}.
This leads to our first goal: devising \emph{candidate attribute} quality functions with low sensitivity while preserving utility,  
where sensitivity is as defined in \Cref{def:sensitivity}. Then, our next  goal is to develop a privacy-preserving algorithm that leverages the low-sensitivity score to produce high-quality explanations.

We can now summarize the challenges addressed in this work: 
\begin{problem}[Low Sensitivity Quality Functions]\label{prob:metrics}
Find a low-sensitivity, global quality function $\globscore$ that maps a sensitive dataset $D$, a clustering function $f: \dom(R) \to C$, and an attribute combination $\AC$
to a real number.
\end{problem}
Once we have a low-sensitivity score function, we can rank the explanations and return the highest scoring ones. Nevertheless, we still need a private mechanism to allow us to do so with low error. 

\begin{problem}[Select Top explanation attributes]\label{prob:mult}
    Given a sensitive dataset $D$, a clustering function $f:\dom(R)\to  C$, 
     and a privacy budget $\eps$, find a high-scoring  attribute combination  $\AC$
     according to the global score function $ \globscore$ and output the corresponding global explanation while satisfying $\eps$-DP.
 \end{problem}

\begin{figure}
           \centering
            \includegraphics[width=1\linewidth]{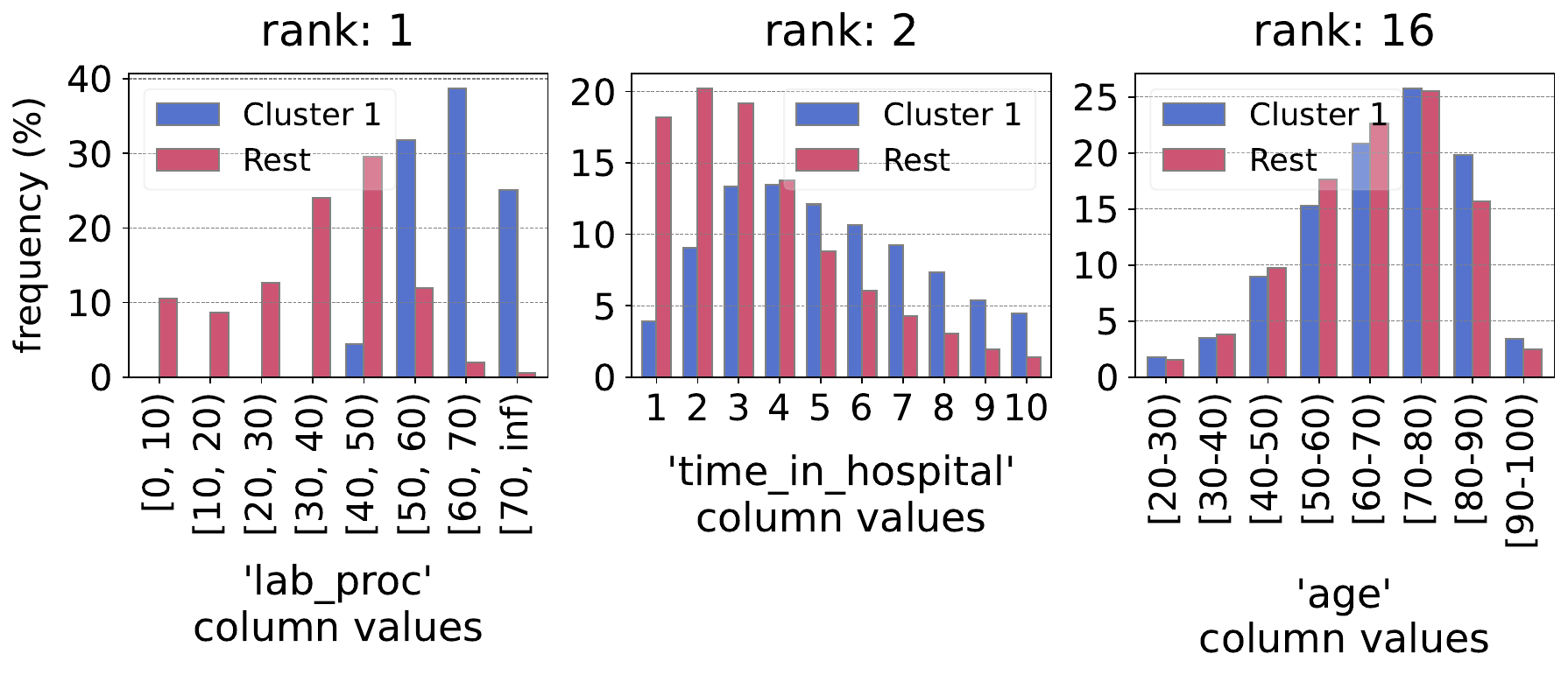} 
             \caption{Ranked explanation candidates for Cluster 1 of the Diabetes dataset. 
             % \ag{Change the ticks on the X axis to take less space}\ron{changed}
             }
         \label{fig:hist_clust_example_diabetes_ranked}
    \end{figure}

\begin{example}\label{example:hist_clust_example_diabetes_ranked}
    \Cref{fig:hist_clust_example_diabetes_ranked} gives an example of three explanation candidates for Cluster 1 of the Diabetes dataset. In this example, the top-ranked candidate is also selected to explain Cluster 1 in the final output, showcased in \Cref{example:top_exp_diabates}.
\end{example}
In the next section, we describe our solution to \Cref{prob:metrics} and then describe our approach for obtaining high-scoring explanations (\Cref{def:global-candidate}) while satisfying $\eps$-DP to solve \Cref{prob:mult}. 
\setlength{\abovedisplayskip}{1.5pt}
\setlength{\belowdisplayskip}{1.5pt}
\setlength{\abovedisplayshortskip}{0pt}
\setlength{\belowdisplayshortskip}{0pt}
\section{Low Sensitivity Quality Functions}
\label{sec:metrics}
In this section, we address \Cref{prob:metrics} by building upon prior work \cite{FEDEX, dasgupta2022framework, youngmann2022guided, hilderman2013knowledge,TabEE} and focusing on three prominent quality aspects of HBEs: Sufficiency \cite{TabEE, dasgupta2022framework}, interestingness \cite{hilderman2013knowledge, sarawagi1998discovery}, and diversity~\cite{TabEE, youngmann2022guided}. We prove that these measures are highly sensitive, making them unsuitable for a DP algorithm. These results then motivate us to devise low-sensitivity variants.

\subsection{Interestingness} 
\label{subsec: Interestingness}
The interestingness of an HBE is quantified as the distributional shift of the given attribute values between the cluster and the full-dataset \cite{TabEE, FEDEX, hilderman2013knowledge}. There are many ways to quantify distance between probability distributions. Among them, the \emph{total variation distance} (TVD) has been shown to be effective in quantifying the interestingness of HBEs \cite{TabEE}.

For a dataset $D$ and a cluster $D_c\subseteq  D$, the TVD between the distributions of values in $\proj_A(D)$ and $\proj_A(D_c)$, and thus the \emph{interestingness} of an attribute $A$ with respect to the cluster $D_c$, is defined as 
\begin{align}\label{eq:tvd}
    \tvd(\proj_A(D),\proj_A(D_c)) = \frac{1}{2}\sum_{\mathclap{a\in\dom(A)}} \abs{ \frac{\cnt_{A=a}(D)}{\abs{ D}} - \frac{\cnt_{A=a}(D_c)}{|D_c|}}
\end{align}
The global interestingness of an attribute combination $\AC$, is defined as the average of all single-cluster interestingness scores \cite{TabEE}, which we denote by $\intTabEE(D, f, \AC)$

In the context of DP, \emph{we cannot make any assumptions about the input dataset $D$ or the clustering function}, and privacy must be guaranteed for any input, regardless of the cluster size. 
 In particular, when a cluster is small, the removal of a single tuple from it can significantly change the cluster's internal distribution, which may lead to  a large change in the distribution distance, as reflected in \Cref{eq:tvd}. Given that sensitivity analysis must account for such cases, the sensitivity of this function is high. 

\begin{restatable}{proposition}{IntHighSens}
\label{claim:interest-high-sens}
The sensitivity of $\tvd$ is at least $\frac{1}{2}$ and its range is $[0,1]$.
\end{restatable}

Intuitively, since this function outputs a number between $0$ and $1$, its sensitivity is relatively high. While the proof of \Cref{claim:interest-high-sens} is provided in
\ifpaper
\cite{full_version}, 
\else
\Cref{appendix:interest}, 
\fi
we illustrate the issue in the following example.\footnote{\label{footnote:js}Previous work has also considered the Jensen-Shannon distance  \cite{lin1991divergence} as an interestingness measure. We show that it is highly sensitive as well, making it unsuitable for the privacy setting. The proposition and proof appear in
\ifpaper
\cite{full_version}.
\else
\Cref{appendix:interest}.
\fi
}

\begin{example}[The Issue with Interestingness sensitivity]
Suppose  $D$  is a dataset of size $100,000$ with a binary attribute $A$.
Suppose further that $95\%$ of individuals in the dataset we have $A=1$.
Assume that the clustering is imbalanced, with a cluster $D_c$ that contains only $1$ tuple $t$, and $t[A]=0$.
In this case, from \Cref{eq:tvd}, the interestingness score of attribute $A$ for explaining cluster $D_c$ is 
$
    \frac{1}{2}(\abs{0.95 - 0} + \abs{0.05 -1}) = 0.95.
$
however, suppose a new tuple $t'$ that satisfies $t'[A]=1$ is added to cluster $D_c$.
Now the interestingness in \Cref{eq:tvd} becomes 
$
    \frac{1}{2}(\abs{95,001 /100,001  - 0.5} + \abs{5000 /100,001  -0.5}) \approx 0.45.
$
Note that while we only added a single tuple, it led to a change of  $\approx 0.95 -0.45 = 0.5$ in the interestingness function.
\end{example}

\paratitle{Low-Sensitivity Interestingness}
We propose a new interestingness function that is inspired by the interestingness of \cite{TabEE}, but has lower sensitivity. 
% \ag{For each of these functions, you need to describe the changes that you've made to the function that you started with and why it still captures the interestingness/sufficiency/...}

\begin{definition}[Low Sensitivity Interestingness] \label{def:low-sens-interest}
For a dataset $D$ and a clustering function $f:\dom(R)\to C$, define the interestingness of an attribute $A$ for explaining the cluster $D_c = \set{ t\in D \suchthat f(t)=c}$ as 
$$
\interest( D, f, c, A) = \frac{1}{2} \sum_{a\in \dom(A)} \abs{ \cnt_{A=a}(D_c) - \frac{|D_c|}{\abs{ D}}\cdot \cnt_{A=a}(D)}.
$$
\end{definition}
To gain insight into the low sensitivity formulation, note that $\interest( D, f, c, A) = |D_c|\cdot \tvd(\proj_A(D),\proj_A(D_c))$, where $\tvd$ is the total variation distance (\Cref{eq:tvd}).
Hence, for a given cluster, the ranking of attributes by low-sensitivity interestingness (\Cref{def:low-sens-interest}) is identical to their ranking by TVD deviation, which is the sensitive interestingness \cite{TabEE}. However, intuitively, the effect of small clusters on the sensitivity is mitigated by the multiplication by $\abs{D_c}$, as now the sensitivity is $1$, but the function takes values in $[0, \abs{D_c}]$.
This new interestingness function gives us more leeway for adding the necessary noise to satisfy DP, while not distorting the attribute ranking too much when the cluster is sufficiently large.
\ifpaper
(see~\cite{full_version} for proof). 
\else
(see \Cref{appendix:interest} for proof). 
\fi
\begin{restatable}{proposition}{IntLowSens}
\label{prop:interest-sensitivity}
The sensitivity of $\interest( D, f, c, A)$ is $1$ and its range is $[0, \abs{D_c}]$.
\end{restatable}

\subsection{Sufficiency}\label{subsec:suff}

We begin by revisiting the notion of sufficiency  from previous work. 
% The motivation behind this measure is to ensure that a given explanation is relevant
% only to its associated cluster, capturing the concept of sufficiency. 
% The framework of  
Recognizing the need to ensure that a given explanation is relevant
only to its associated class, \citet{dasgupta2022framework} introduced an abstract definition of sufficiency (faithfulness) of \emph{classifier explanations}. Informally, the sufficiency of an explanation at $t\in D$ is defined as the fraction of tuples assigned the same prediction as $t$, out of all tuples for which the explanation of $t$ ``holds".
In other words, if $t$ is assigned an explanation that holds for another tuple, then that tuple should have the same classification as $t$. Then, a global sufficiency score is obtained by averaging the sufficiency value at all tuples.
However, their framework assumes the existence of a binary relation indicating whether an  explanation holds for a given tuple, and extending this idea to HBEs is not straightforward. \citet{TabEE} proposed an adaptation of the formula from \cite{dasgupta2022framework}, replacing the classifier prediction with cluster assignment, and the binary relation with a probability value derived from the HBE. Specifically, they quantify the extent to which an HBE, defined by an attribute combination $\AC$, ``holds for" a tuple $t$ that belongs to a cluster $D_c$ as
% \ag{Can you write this in words? We do not need the complex notation if you're just mentioning it.}
 the probability that uniformly random tuple sampled from $D$ conditioned on having the same  value in $\AC(c)$ as $t$, belongs to the same cluster as $t$.
% We let $ \sufTabEE(D, \AC, f)$. \ron{}
% and defer its formal definition to \cite{fullversion}.
Unfortunately, this function, which we denote $ \sufTabEE(D,f,\AC)$, is too sensitive to be of use.
\footnote{
See further discussion in
\ifpaper
\cite{full_version}.
\else
\Cref{appendix:suf}.
\fi
}
% , as follows from the following proposition:
% , which we prove in \cite{dpclustex_github}:
\begin{restatable}{proposition}{SufHighSens}
\label{claim:suf-high-sens}
The sensitivity of $\sufTabEE(D, f, \AC)$ is at least $\frac{1}{2}$ and its range is $[0,1]$.
\end{restatable}

\paratitle{Low Sensitivity Sufficiency}
% \ag{This para was located before the title. I moved it to be after it.}
Next, we introduce an alternative formulation of the sufficiency measure for the following reasons. First, the original function is too sensitive to be of use under DP. Second, the alternative formulation shows that single-attribute sufficiency can, in fact, be measured for each individual cluster, with the global sufficiency being the average of these single-cluster sufficiency functions, which simplifies the sensitivity analysis. 
% and implies that the globally most sufficient attribute combination can be obtained by independently selecting the most sufficient attribute for each cluster, a property leveraged in our algorithm.
Third, it ensures that the range of each single-cluster sufficiency matches that of the interestingness measure, $[0, \abs{D_c}]$ for a cluster $D_c$, and that the sensitivity of the modified sufficiency is also $1$, making the two directly comparable.

\begin{definition}[Low Sensitivity  Sufficiency]\label{def:local_suff}
For a dataset $D$ and a clustering function $f:\dom(R)\to C$,
define the sufficiency of an attribute $A$ for explaining the cluster $D_c$ as
    $$
        \suf( D, f, c,  A) =  \sum_{a\in\dom_{D_c}(A)} \frac{\cnt_{A=a}(D_c)^2}{\cnt_{A=a}(D)}.
    $$
\end{definition}
Note that we do not divide by zero, since the sum is only over values appearing at least once in $D_c$, and hence also in $D$.
Intuitively, from \Cref{def:local_suff}, we see that $\suf$ is maximized when values of $\dom(A)$ appearing in the cluster $D_c$ occur \emph{only} within it, reflecting maximal sufficiency of $A$, since observing a tuple's value in that attribute ``suffices" to determine its membership to $D_c$. Conversely, the function decreases when such values also appear frequently outside the cluster, indicating that $A$ is insufficient to explain $D_c$.

Our modified sufficiency is consistent with the definition in previous work, as the equality in item \textit{(1)} of \Cref{prop:suff-sensitivity} ensures that both measures induce the same ranking of attribute combinations when the dataset and clustering are fixed.
We prove:
\begin{restatable}{proposition}{SufLowSens}
\label{prop:suff-sensitivity}
For an attribute combination $\AC$, 
\begin{enumerate}
    \item  The following equality holds:
    \[\abs{D}\cdot \sufTabEE(D, f, \AC)=  \sum_{c\in C}  \suf(D, f, c, \AC(c))\]
    \item The sensitivity of $\suf( D, f, c,  A) $ is $1$ and its range is $[0, \abs{D_c}]$. 
\end{enumerate}
\end{restatable}
% Since $\suf(D, f, c, A)$ ranges within $[0, \abs{D_c}]$, its sensitivity is intuitively low.

% After defining the sufficiency score of a single attribute for a single cluster, our low-sensitivity global sufficiency is obtained by averaging the sufficiency scores of individual clusters:

% \begin{definition}[Global Sufficiency] 
% For a dataset $D$ partitioned into $D_1, \dots,  D_\numclust$ by a clustering function $f$, define the diversity score of a combination of attributes $E = (A_c)_{c\in C}$ such that $A_c$ is a candidate explanation attribute for the cluster $D_c$ as
% \begin{align*}
%     \suf(D, f, E) = \frac{1}{\numclust} \sum_{c\in C} \suf( D, f, c,  A) 
% \end{align*}
% \end{definition}

% Since $\suf$ is a convex combination of sensitivity-$1$ functions, we have the following proposition applying \Cref{lem:convex-low-sens}: 

% \begin{proposition}\label{prop:glob-suff-sensitivity}
%    $\suf$ has sensitivity bounded by $1$.
% \end{proposition}

\subsection{Diversity}
% The purpose of the diversity measure is to quantify  the overall distinctiveness among explanations.
% Due to space limitations, we provide only an intuitive introduction of the diversity measure, denoted $\diversity_{Tab}$, outline its inadequacies in the context of DP, and discuss its relation to our new measure. The reader is referred to \cite{TabEE} for the original definition, and to \cite{dpclustex_github} for our sensitivity analysis of this function. 
The diversity measure is designed to quantify the overall distinctiveness among explanations.
We provide an intuitive introduction of the diversity measure, denoted $\diversityTabEE$, 
outline its inadequacies in the context of DP, and discuss its relation to our new measure.\footnote{The reader is referred to \cite{TabEE} for the original definition, and to \Cref{appendix:proofs} for our sensitivity analysis of this function.} 

Diversity is initially defined for a pair of single-cluster explanations, and generalized to global explanations by averaging all pairwise diversities. When two single-cluster explanations utilize different attributes, the diversity for that pair attains its maximum value of $1$. If the explanations use the same attribute, the diversity is measured as the distance between the two distributions, quantifying the new knowledge gained from the additional explanation on that attribute.
However, common metrics used in previous work,  such as total-variation distance and Jensen-Shannon distance, have high sensitivity (\Cref{claim:interest-high-sens}, \Cref{footnote:js}), which intuitively means they are unsuitable for the DP setting. 

Inspired by the diversity measure of \cite{TabEE}, we introduce the following low-sensitivity pairwise diversity function: 
% \ron{TODO: motivate with an example}
\begin{definition}[Pair Diversity] \label{def:pairwise_diversity}
For a dataset $D$ and a clustering function $f$, the diversity score of a pair of attributes $A_c, A_{c'} \in \cA$, where $A_c$ (respectively $A_{c'}$) is a candidate attribute for explaining $D_c = \set{t\suchthat f(t)=c}$ (respectively $D_{c'}$), is
\begin{align*}
    &d(D, f, c, c' , A_c, A_{c'}) = \min\set{|D_c|, |D_{c'}|} \times \\
            &\qquad\begin{cases}
                1 & A_c \neq A_c \\
                \frac{1}{2}\sum_{a\in \dom(A)} \abs{ \frac{\cnt_{A=a}(D_c)}{\max\set{\abs{ D_c},1}} - \frac{\cnt_{A=a}(D_{c'})}{\max\set{|D_{c'}|,1}}} & A_c=A_{c'}    
            \end{cases}
\end{align*}
\end{definition}
To interpret \Cref{def:pairwise_diversity}, note that when both clusters are not empty and are explained by the same attribute $A$, we have $d(D, c, c', A, A)= \min\set{|D_c|, |D_{c'}|}\cdot \tvd( \proj_A(D_c),\proj_A(D_{c'}))$,  where $\tvd$ is defined in \Cref{eq:tvd}.
%\ag{TVD?}
Thus, for a given pair of clusters, the low-sensitivity pairwise diversity ranks attributes identically to the sensitive TVD deviation, and is  maximized when the clusters are explained by different attributes.

Following previous work on result diversification (e.g., \cite{borodin2017max, vieira2011query}), we define global diversity as the average of all pairwise diversities.
Note that achieving low sensitivity requires that pairs from smaller clusters have a reduced impact on the global diversity function, as is evident in \Cref{def:glob_diversity}. The sensitivity bound for this function is provided in \Cref{prop:sens-div}, leveraging the fact that a convex combination of sensitivity-1 functions has a sensitivity bounded by 1 (see 
\ifpaper
\cite{full_version} for proof).
\else
\Cref{lemma:convex-comb}).
\fi

\begin{definition}[Global Diversity]\label{def:glob_diversity}
For a dataset $D$ and  a clustering function $f:\dom(R)\to C$, define the diversity score of an attribute combination $\AC$ as
\begin{align*}
    \diversity(D, f, \AC) = \frac{1}{\binom{\numclust}{2}} \sum_{\set{c,c'}\subseteq C} d(D, f,c,c',\AC(c),\AC(c'))
\end{align*}
% \ag{Are we supposed to see $E$ on the right side of the equation?}\ron{TODO: change $E$ to macro. clarify equation.}
where the sum is over all distinct pairs of cluster labels.
\end{definition}
% Note that achieving low sensitivity requires that pairs from smaller clusters have less influence on the overall diversity function, as in our definition. Since a convex combination of sensitivity $1$ functions has sensitivity bounded by $1$ (See Lemma \Cref{??} in \Cref{fullversion}), we have
\begin{restatable}{proposition}{DivLowSens}\label{prop:sens-div}
   The sensitivity of $\diversity$ is bounded by $1$. Moreover, its range is $[0,R_{\diversityTabEE}]$ where
   $
        R_{\diversityTabEE} = \frac{1}{\binom{\numclust}{2}} \sum_{i=1}^{\numclust} (\numclust-i)\abs{D_{c_{i}}}
   $
   is a weighted average of the cluster sizes, and 
   % $|D_{c_1}| \le |D_{c_2}| \le \ldots \le |D_{c_{|C|}}|$.
   $|D_{c_i}| \le |D_{c_{i+1}}|$.
   % \ag{What is the role of $R_{\diversityTabEE}$? What is the connection to $Div$?}\ron{just notation for the range bound}
\end{restatable}
% Next, we combine the three score functions into a single objective and derive a sensitivity bound.
\subsection{Combining All Quality Functions}
\label{subsec:combining_scores}
Following prior work on explainability \cite{lakkaraju2016interpretable, mothilal2020explaining, rawal2020beyond, lv2023data, TabEE}, we combine  the different measures into a weighted sum, with weights that may be user-defined or preference-driven. We first define the single-cluster score function, which assesses the quality of an attribute $A$ in explaining a given cluster $D_c$.
  % We apply a common approach taken by multiple previous works on explainability (e.g., \cite{lakkaraju2016interpretable,mothilal2020explaining, rawal2020beyond,lv2023data, TabEE}). Specifically, we combine the different measures into a weighted sum, where the weights can be user-defined or adjusted based on specific requirements or preference.
  % \cite{TabEE} assigns equal weight to each score by default, a choice we adopt in our framework.

   % , and used in \Cref{alg:topk-single}. 
   % In~\cite{dpclustex_github} we prove the sensitivity bound given in \Cref{prop:single_score-sensitivity}.
  
\begin{definition}[Single-Cluster Score]\label{def:single_score}
Let $\gamma = (\gamma_{\intTabEE},\gamma_{\sufTabEE})$ be a pair of non-negative parameters that sum to $1$.
For a dataset $D$ and a clustering function $f:\dom(R)\to C$, define the quality score  $A\in \cA$ as a candidate attribute for a cluster $c\in C$, as
\begin{align*}
    \score_{\gamma}(D,f,c,A)  = \gamma_{\intTabEE}\cdot \interest(D,f,c,A) 
                             +\gamma_{\sufTabEE}\cdot \suf(D,f,c,A) .
\end{align*}
\end{definition}
% Since a convex combination of sensitivity $1$ functions has sensitivity bounded by $1$ (See Lemma \Cref{??} in \Cref{fullversion} \ag{Seems very general. Did you prove this or is this a result of prior work?}), and
% since $\score{\gamma}$ is a convex combination of function with sensitivity bounded by $1$ (\Cref{prop:interest-sensitivity}, \Cref{prop:suff-sensitivity}), applying  \Cref{lem:convex-low-sens} we obtain \Cref{prop:single_score-sensitivity}. Being a weighted average of $\interest$ and $\suf$, $\score_{\gamma}$ also ranges within $[0, \abs{D_c}]$, and hence its sensitivity is intuitively low. 
Importantly, we can bound the sensitivity of the score function.  
% (proof in~\cite{dpclustex_github}).  
\begin{restatable}{proposition}{SingLowSens}
\label{prop:single_score-sensitivity}
   $ \score_{\gamma}(D,f,c,A)$ has sensitivity bounded by $1$ and its range is  $[0, \abs{D_c}]$.
   % \ag{Why is this good? What is the range of $\score_{\gamma}$?}
\end{restatable}
The global score combines individual measures to assess the quality, with  smaller clusters contributing less than larger ones for the same distribution distance, addressing the high sensitivity of the original global score. As shown in \Cref{claim:interest-high-sens}, removing a single point from a small cluster can significantly alter its column distribution, leading to a large score change. In 
\ifpaper
\cite{full_version},
\else
\Cref{appendix:all},
\fi
we prove the bound \Cref{prop:global_score-sensitivity}.
\begin{definition}[Global Score]\label{def:global_score}
Let  $ \lambda = (\lambda_{\diversityTabEE}, \lambda_{\intTabEE}, \lambda_{\sufTabEE})$ be  non-negative parameters that sum to $1$.
For a dataset $D$ and a  clustering function $f$, define the overall quality score of an attribute combination $\AC$  as
\begin{align*}
    \globscore_{\lambda}(D,f, \AC) &=   \lambda_{\intTabEE}\cdot \interest(D, f, \AC)+\lambda_{\sufTabEE}\cdot \suf(D, f, \AC)  \\
     &\qquad \qquad +  \lambda_{\diversityTabEE}\cdot \diversity(D, f, \AC)
\end{align*}
where we extend $ \interest(D, f, \AC) = \frac{1}{\numclust} \sum_{c\in C} \interest( D, f, c, \AC(c))$ and $\suf(D, f, \AC) = \frac{1}{\numclust} \sum_{c\in C} \suf( D, f, c, \AC(c)). $ 

\end{definition}

\begin{restatable}{proposition}{GlobLowSens}
\label{prop:global_score-sensitivity}
   $\globscore_{\lambda}$ has sensitivity bounded by $1$. Moreover, its range is $[0,R_{\globscore_{\lambda}}]$ where
   \[
    R_{\globscore_{\lambda}} = (\lambda_{\intTabEE} + \lambda_{\sufTabEE}) \cdot  \frac{1}{\numclust} \sum_{c\in C}|D_c|+ \lambda_{\diversityTabEE}\cdot R_{\diversityTabEE}
   \]
   is a weighted average of the cluster sizes, and $R_{\diversityTabEE}$ is defined as in \Cref{prop:sens-div}. 
   % \ag{Why $R_{\globscore_{\lambda}}$ and not just the score function?} \ron{It's just notation for range upperbound}
\end{restatable}

% Our experiments show that, despite differences in objective functions, and even for small privacy budgets, \ourframework\ either selects the 
% same attributes as the non-private baseline or one with a comparable score, even when the score is measured using sensitive the functions.

\paratitle{Selecting the weight parameters}
%$\lambda_{\diversityTabEE}, \lambda_{\intTabEE}, \lambda_{\sufTabEE}$}}
Previous work on HBEs in the non-private setting \cite{TabEE} has shown through user studies and quantitative evaluations that equal weight distribution $\lambda_{\intTabEE}  = {\lambda_{\sufTabEE} = \lambda_{\diversityTabEE}} = 1/3$ produces high quality and informative explanations. We adopt this default parameter setting in our framework. However, weights can be adjusted based on preference. Our experiments show that  \ourframework\ maintains high explanation quality compared to the non-private baseline across different weight distributions.

\section{The \ourframework\ Framework}
\label{sec: alg}
In this section, we introduce our algorithms for computing explanations (depicted in \Cref{fig:system_architecture}), addressing \Cref{prob:mult}. 
Since the space of all possible attribute combinations is generally too large to analyze, we propose a novel DP-tailored manner of pruning the search space. 
Following previous work~\cite{TabEE}, we  construct a small set of high-quality candidate \emph{attributes} for each cluster (Stage-1), which  serve as the candidate pool for a \emph{global} explanation of all clusters (Stage-2). 
To optimize our privacy budget usage, we avoid generating noisy histograms at Stage-1. 

\subsection{Stage-1: Construct Private Candidate Sets}\label{subsec:stage-1}

We now present our candidate set construction algorithm (\Cref{fig:system_architecture}).

In \ourframework, to satisfy DP, we privately select top-$k$ candidate \emph{attributes} for each cluster based on our single-cluster score function.
Since doing so by iteratively applying the exponential mechanism $k$ times can be computationally expensive, we adopt the \textit{One-shot Top-k mechanism} (\Cref{subsec:prelim-dp}) to privately select top-$k$.

% \paragraph*{\Cref{alg:topk-single}: Select-Candidates}
The pseudo code in \Cref{alg:topk-single} provides the candidate selection procedure. 
we first set the privacy budget $\eps_{Topk} = \eps_{CandSet}/\numclust$ allocated  for each Top-$k$ selection, and in \Cref{topk-single: line 2} we set the noise scale parameter $\sigma = 2 k/\eps_{Topk}$. 
Then, we run the Top-$k$ attributes selection procedure for each cluster $c\in  C$. Specifically, in \Cref{topk-single: line 5} we compute the noisy single-cluster score (\Cref{def:single_score}) for each cluster and attribute. In  \Cref{topk-single: line 7}, we sort the attributes based on the noisy scores for each cluster. In  \Cref{topk-single: line 9}, we define the set $S_c$.
The output returned in \Cref{topk-single: line 11} consists of the sets $S_c$ for every $c\in C$.

Note that $1/\numclust$-fraction of the total privacy budget $\eps_{CandSet}$ is allocated for selecting top candidate attributes for each cluster. While one  might have hoped to use parallel composition, this is typically not possible. The quality score of an attribute for a given cluster depends on the entire dataset, not just the cluster itself. A high-scoring attribute exhibits large distributional deviation between the cluster and the full dataset, requiring consideration of tuples outside the cluster.

\begin{algorithm} \caption{Select-Candidates: Generate all single-cluster top-$k$ candidate attributes 
} 
\label{alg:topk-single}
\begin{algorithmic}[1]

\Require
Dataset $D$, clustering function $f:\dom(R)\to C$, 
hyperparameters $\gamma = (\gamma_{\intTabEE}, \gamma_{\sufTabEE})$
Attribute set $\cA$,  
privacy parameter $\eps_{CandSet}$,
cardinality $k$.

\Ensure Sets $S_{c_1}\dots,S_{c_{|C|}}$ where $S_{c_i}$ contains noisy Top-$k$ candidate explanation attributes for $D_{c_i}$.
\vspace{2mm}
\State $\eps_{Topk} \gets \eps_{CandSet}/\numclust$ \plabel{topk-single: line 1}
\State Set $\sigma \gets 2 k/\eps_{Topk}$ \plabel{topk-single: line 2}
\For{$c\in C$} \plabel{topk-single: line 3}
    \For{$ A\in \cA$} \plabel{topk-single: line 4}
    \State $s_A\gets  \score_\gamma( D, f, c,  A) + \Gumb{\sigma}$ \plabel{topk-single: line 5}
    \EndFor \plabel{topk-single: line 6}
    \State Sort $\set{ s_A \mid A\in \cA}$ in descending order. \plabel{topk-single: line 7}
    \State Let $A_1,\ldots,A_k$ be the attributes corresponding to the top-$k$ noisy scores.  \plabel{topk-single: line 8}
    \State Set $S_c \gets \set{A_1,\ldots,A_k}$ \plabel{topk-single: line 9}
\EndFor \plabel{topk-single: line 10}

\State\Return Candidate sets $S_{c_1}\dots,S_{c_{|C|}}$. \plabel{topk-single: line 11}

\end{algorithmic}
\end{algorithm}

The privacy and utility guarantees of \Cref{alg:topk-single} are given by the following proposition:
\begin{restatable}{proposition}{CandPrivUtility}
\label{prop:topk-single privacy}
Given a clustering function $f$,
a set of attributes $\cA$, a privacy parameter $\eps_{CandSet}$, non-negative hyperparameters $\gamma_{\intTabEE}, \gamma_{\sufTabEE}$ that sum to $1$, and a size parameter $k$, the following holds:
\begin{enumerate}
    \item  \Cref{alg:topk-single} satisfies  $\eps_{CandSet}$-DP.
    \item For $c\in  C$, denote by $\OPT_c^{(\ell)}$ the $\ell$-th highest (true) score, and by $A_c^{(\ell)}$
the $\ell$-th explanation attribute selected by \Cref{alg:topk-single} to $S_c$. For all $c$ and $\ell=1,2,\dots,k$, we have
\footnotesize{
\[
    \Pr\left[ \score(c,A_c^{(\ell)}) \le \OPT_c^{(\ell)} - \frac{2\numclust\cdot k}{\eps_{CandSet}}(\ln{\abs{\cA}}+t)\right] \le e^{-t}.
\]}
\end{enumerate}
where we denote $ \score(c, A_c^{(\ell)}) =\score( D, f, c, A_c^{(\ell)}) $
\end{restatable}

The proof of item 1 follows from \Cref{prop:composition} and
the analysis of the one-shot Top-$k$ mechanism \cite{durfee2019practical, durfee2021oneshot}. 
The proof of item 2 is based on the utility proposition of EM (Theorem 3.11  in \cite{dwork2014algorithmic}). 
Both proofs can be found in
 \ifpaper
    ~\cite{full_version}.
    \else
    \Cref{appendix:proofs}.
    \fi

\begin{example}\label{example:stage-1}
    Reconsider \Cref{example:top_exp_diabates}. \Cref{alg:topk-single} outputs the set 
    $S_1 = \set{\mathtt{lab\_proc},\; \mathtt{time\_in\_hospital},$ $\mathtt{num\_medications}}$, which indeed comprises the top-3 attributes for Cluster 1 obtained from the ranked list partly presented in \Cref{fig:hist_clust_example_diabetes_ranked}.
    \Cref{alg:topk-single} outputs one such set for each cluster.
\end{example}

\subsection{Stage-2: A Private Global Explanation}\label{subsec: stage_2}
We now present a privacy-preserving mechanism that, given a dataset $D$ and a clustering function $f$,  outputs an HBE for the given clustering while selecting the explanation attribute for each cluster $D_c$ from its corresponding set $S_c$ (\Cref{subsec:stage-1}), as depicted in \Cref{fig:system_architecture}.  By restricting the search space to the candidate sets, \ourframework\ evaluates only $k^{\abs{C}}$ attribute combinations, instead of the full space of $\abs{\cA}^{\abs{C}}$ combinations.

As in Stage-1, pre-computing noisy histograms and selecting the best candidate combination afterward incurs a waste of privacy budget, and evaluating the global score based on noisy histograms introduces more noise than necessary due to the injection of noise into \emph{each count} of the private histograms. To address these issues, we propose the following approach.

We begin by computing the candidate sets $S_c$ using \Cref{alg:topk-single} with privacy budget $\eps_{CandSet}$.
Then, we apply the EM (\Cref{def:exp-mech}) to select the noisy-best \emph{attribute combination} drawn from each cluster's candidate set using our global score function (\Cref{def:global_score}). Subsequently, we generate the noisy histograms \emph{only for the selected attribute combination}, employing a black-box mechanism for histogram generation while leveraging the parallel composition property for efficient privacy budget allocation.

The pseudo code describing our approach can be found in \Cref{alg:gen_global_explanation}. 
We first set the marginal weights for interestingness and sufficiency in the single-cluster score function. In \Cref{gen_global_explanation:line 2}, \Cref{alg:topk-single} is invoked to obtain candidate sets $S_c$ for each cluster. In \Cref{gen_global_explanation:line 5}, the EM is run using the global score function (\Cref{def:global_score}), privacy parameter $\eps_{TopComb}$, and the set of possible candidate attribute combinations, assigning each cluster $c$ an attribute from $S_c$. \Cref{gen_global_explanation:line 6} extracts all attributes selected at least once. In \Cref{gen_global_explanation:line 7}, we allocate $\eps_{hist,all} = \eps_{Hist}/(2|\cA'|)$ and $\eps_{hist,cluster} = \eps_{Hist}/2$ as privacy budgets for full-dataset and cluster histograms, respectively. Lines \ref{gen_global_explanation:line 8}–\ref{gen_global_explanation:line 10} compute full-dataset histograms. \Cref{gen_global_explanation:line 12} computes cluster histograms. In \Cref{gen_global_explanation:line 13}, we derive out-of-cluster histograms by subtracting cluster from full-dataset histograms, replacing negatives with 0. \Cref{gen_global_explanation:line 14} constructs explanations $e_c$ using these histograms, and \Cref{gen_global_explanation:line 16} outputs the global explanation $\set{e_c \mid c \in C}$.

\normalfont
\begin{algorithm}
\caption{ \ourframework\ Algorithm - Generate Global Explanation}\label{alg:gen_global_explanation}
\begin{algorithmic}[1]
\Require  Dataset $D$, clustering function $f$,
number of candidates $k$,
privacy parameters
$\eps_{CandSet},
\eps_{TopComb},
\eps_{Hist}$,
hyperparameters $\lambda_{\diversityTabEE}, \lambda_{\intTabEE}, \lambda_{\sufTabEE}$
\Ensure Global explanation $\set{e_c\mid c\in C}$ (\Cref{def:global-candidate})

\vspace{2mm}
\State Let $\gamma_{\sufTabEE} \gets \lambda_{\sufTabEE} / (\lambda_{\sufTabEE}+\lambda_{\intTabEE})$ and $\gamma_{\intTabEE} \gets \lambda_{\intTabEE} / (\lambda_{\sufTabEE}+\lambda_{\intTabEE})$ \plabel{gen_global_explanation:line 1}
\LineComment{Invoke \Cref{alg:topk-single}}
\State Let $S_{c_1},\dots S_{c_{\abs{C}}} \gets \text{Select-Candidates}(D, f, \gamma, \cA, \eps_{CandSet}, k)$ \plabel{gen_global_explanation:line 2}
% \Comment{\Cref{alg:topk-single}}

\LineComment{Select attribute combination}

% \State   $\eps \gets \sqrt{8\rho_{TopComb}}$ \plabel{gen_global_explanation:line 3}
\State Let  $\cR \gets \set{\AC \mid \forall c.\; \AC(c) \in S_c }$ be the set of candidate combinations, mapping each $c\in C$ to an attribute in $S_c$. \plabel{gen_global_explanation:line 4}
  \State $\AC \gets \cM_E(D,\globscore_\lambda, \Delta_{\globscore},\cR ,\eps_{TopComb})$  \plabel{gen_global_explanation:line 5}

\LineComment{Generate full-dataset histograms}
\State  Let $\cA' \gets \set{\AC(c)\mid c\in C}$ be the set of attributes appearing in at least once in the combination. \plabel{gen_global_explanation:line 6}
  \State $\eps_{hist,all} \gets \eps_{Hist} / (2\abs{\cA'})$,   $\eps_{hist,cluster} \gets \eps_{Hist}/2$ \plabel{gen_global_explanation:line 7}
    \For{every  $A\in\cA'$} \plabel{gen_global_explanation:line 8}
    \State $\widetilde{h}_A \gets \cM_{hist}(\proj_A(D), \eps_{hist,all})$  \plabel{gen_global_explanation:line 9}
    \EndFor  \plabel{gen_global_explanation:line 10}
    \LineComment{Compute single-cluster explanations}
\For{every  $c\in C$} \plabel{gen_global_explanation:line 11}
        \State Let $A_c\gets \AC(c)$ be the attribute selected for $D_c$.
         \State $\widetilde{h}^c \gets \cM_{hist}(\proj_{A_c}(D_c), \eps_{hist,cluster})$ \plabel{gen_global_explanation:line 12}
         \State $\widetilde{h}^{-c} \gets \max(\widetilde{h}_{A_c} - \widetilde{h}^c, 0)  $ \plabel{gen_global_explanation:line 13}
         \State $e_c \gets (c, A_c,\widetilde{h}^{-c} ,\widetilde{h}^c)$ \plabel{gen_global_explanation:line 14}
    \EndFor \plabel{gen_global_explanation:line 15}
   
\State\Return  global explanation $\set{e_c\mid c\in C}$. \plabel{gen_global_explanation:line 16}

\end{algorithmic}
\end{algorithm}
\normalsize
\normalfont

The following theorem states the privacy guarantee of \ourframework.
\begin{theorem}\label{thm:gen_global_explanation-privacy}
 Given a clustering function $f:\dom(R)\to C$, number of candidates $k$, privacy parameters
$\eps_{CandSet},
\eps_{TopComb},
\eps_{Hist}$,
% the \ourframework\ framework (\Cref{alg:gen_global_explanation}) 
\Cref{alg:gen_global_explanation} is 
$(\eps_{CandSet} +  \eps_{TopComb} + \eps_{Hist})$-DP.
\end{theorem}

\begin{proof}[Proof Sketch]
     The execution of \Cref{alg:topk-single} satisfies $\eps_{CandSet}$-DP by \Cref{prop:topk-single privacy}.
     Since the function  $\globscore_{\lambda}$ has sensitivity $1$ (\Cref{prop:global_score-sensitivity}), the execution of the exponential mechanism in \Cref{gen_global_explanation:line 5} guarantees $\eps_{TopComb}$-DP, applying  \Cref{thm:exponential_mech}.
        We assume that each execution of $\cM_{hist}(\cdot,\eps)$ satisfies $\eps$-DP. Therefore, by sequential composition, the computation of $\widetilde{h}_A$ for all $A\in \cA'$ satisfies overall $\eps_{Hist}/2$-DP. Since the clusters are disjoint, the computation of all noisy histograms $\widetilde{h}^c$ satisfies overall $\eps_{Hist}/2$-DP by parallel composition. Note that the computation of each $\widetilde{h^{-c}}$ in \Cref{gen_global_explanation:line 13} is post-processing, and therefore incurs no additional privacy loss. Overall, by sequential composition, the entire algorithm guarantees $(\eps_{CandSet} + \eps_{TopComb} + \eps_{Hist})$-DP. 
\end{proof}

\begin{example}
Consider the setting in \Cref{example:top_exp_diabates}, and suppose the input contains 3 clusters of the Diabetes dataset. First, a candidate set is obtained for each cluster, with the set for Cluster 1 shown in \Cref{example:stage-1}. In \Cref{gen_global_explanation:line 5}, the selected attribute combination is: Cluster 1: \texttt{lab\_proc}, Cluster 2: \texttt{lab\_proc}, and Cluster 3: \texttt{discharge\_disp}. DP noisy histograms are generated for these attributes.  \Cref{fig:hist_clust_example_diabetes} depicts a part of the output , showcasing the explanation for Cluster 1.
The full output of \Cref{alg:gen_global_explanation} contains such an explanation for each cluster.
\end{example}

\paratitle{Time complexity} 
The time complexity \ourframework\ is proportional to $O(\abs{\cA} \cdot |C| + k^{|C|})$, where the first term is due to Stage-1 (\Cref{alg:topk-single}) and the second to Stage-2 (\Cref{alg:gen_global_explanation}). Stage-1 constructs a candidate set for each cluster, requiring $O(\abs{\cA} \cdot |C|)$ evaluations of the single-cluster score function, each involving two count group-by queries. The noisy scores are computed only once using the one-shot top-$k$ mechanism, instead of $k$ times using repeated applications of the EM that require overall $O(k \abs{\cA} |C|)$ noisy scores evaluations. Stage-2 performs $O(k^{|C|})$ evaluations of the global score function, corresponding to the number of global explanation candidates. The complexity of each global score evaluation is as follows. The average interestingness and sufficiency across all clusters require $O(|C|)$ count group-by  queries. Computing global diversity requires $O(|C|^2)$ count group-by queries, as it is defined as average of pairwise diversities, with each pair requiring two count queries.

\section{Experiments}\label{sec: exp}

In this section, we evaluate the quality and efficiency of the explanations generated by \ourframework\ with the following questions:
\begin{enumerate}[leftmargin=*]
    \item With respect to the quality measures for HBEs, how does \ourframework\ perform compared to the non private method and a naive approach which computes all histograms in advance? 
    \item How close is the attribute combination selected by \ourframework\ to that of the non-private baseline?
    \item How efficient is \ourframework\ in computing the explanations? 
\end{enumerate}

\paratitle{Summary of our findings}
With a total privacy budget of $\eps=0.1$, \ourframework\ selects attributes of comparable quality to those chosen by the non-private baseline in all datasets and clustering methods. Moreover, at $\eps=1$, \ourframework\ consistently selects the same attributes as the non-private baseline across all runs and clustering methods for the Diabetes dataset. The execution time of \ourframework\ for generating explanations averages under 6.6 seconds across all datasets and clustering methods with at most 9 clusters.

\subsection{Experimental Settings}\label{sec:settings}
We next present our settings for the experiments. 
We have implemented \ourframework\  in Python 3.9.19 using the Pandas and NumPy libraries.
All experiments were run on an Intel Xeon CPU-based server with 24 cores
and 96 GB of RAM. We use the Geometric mechanism \cite{ghosh2009universally} for DP histogram generation, implemented by DiffPrivLib \cite{holohan2019diffprivlib}. The source code is available in \cite{dpclustex_github}.

\paratitle{Default parameters}
 Unless mentioned otherwise, the following default parameters are used. We set $\eps_{CandSet} = \eps_{TopComb} = \eps_{Hist} = 0.1$. Thus, the combined privacy budget for attribute selection is $\eps=0.2$, which we vary from $0.001$ to $1$ when evaluating its effect on the selected attributes.
 The number of candidate attributes per cluster selected at Stage-1 (denoted by  $k$ in  \Cref{subsec:stage-1}) is set to $3$, a choice supported by ablation studies presented in \Cref{fig:Num_candidates_ablation}, where it is varied from $1$ to $5$.
Following the discussion in \Cref{subsec:combining_scores}, we set the default values $\lambda_{\intTabEE} 
 = {\lambda_{\sufTabEE} = \lambda_{\diversityTabEE}} = 1/3$. We also evaluate alternative weight distributions,
 where we set one weight to zero and each of the remaining two to $1/2$. Unless otherwise stated, we use 5 clusters for evaluation.

\paratitle{Datasets}
We evaluate \ourframework\ on the following two  datasets:
\begin{itemize}[leftmargin=*]
    \item \textbf{US Census Data} \cite{census_1990}: 2,458,285 tuples and 68 attributes. This dataset contains a one percent sample of the Public Use Microdata Samples (PUMS) person records drawn from the 1990 census.
    \item \textbf{Diabetes} \cite{diabetes}:
    101,766 tuples and 47 attributes. This dataset comprises ten years (1999–2008) of clinical care data. Each tuple represents a hospital record of a diabetic patient. Numerical and large-domain categorical attributes are binned in accordance with \cite{diabetes_paper} to ensure interpretable histograms, following prior work on HBEs \cite{FEDEX, TabEE}. Domain sizes vary from 2 to 39. Further details can be found in 
    \ifpaper
    ~\cite{full_version}.
    \else
    \Cref{appendix-exp}.
    \fi
    \item \textbf{Stack Overflow Developer Survey} \cite{stackoverflow_survey}: 98,855 tuples and 60 attributes. This dataset is obtained from the 2018 Stack Overflow Developer Survey. We consider 60 attributes, which include demographic information, professional background, and work habits of the respondents. Numerical and large-domain categorical attributes are binned. Domain sizes vary from 2 to 22. The preprocessing of this dataset is detailed in 
    \ifpaper
    ~\cite{full_version}.
    \else
    \Cref{appendix-exp}. 
    \fi
\end{itemize}

\paratitle{Clustering methods}
To demonstrate the effectiveness and versatility of \ourframework, we evaluate it across diverse scenarios, using both private and non-private clustering methods, following prior work on DP classifier explanations \cite{patel2022model, harder2020interpretable, mochaourab2021robust}. 
In real-world deployment, to protect data privacy, the clustering function must be either privately computed or data-independent.
This evaluation highlights the robustness of our method in providing meaningful explanations for different clustering tasks, including:
\textit{(i)} $k$-means, \textit{(ii)} DP-$k$-means \cite{su2016differentially} implemented by DiffPrivLib \cite{holohan2019diffprivlib}, \textit{(iii)}  $k$-modes, \textit{(iv)}  Agglomerative Clustering, \textit{(v)}  Gaussian Mixture Models (GMMs).
The budget for DP-$k$-means is set to $\varepsilon = 1$, as commonly used for clustering in experimental settings  \cite{su2016differentially, su2017differentially, nguyen2021differentially, nguyen2018privacy}.
Categorical attributes are transformed into equivalent numerical data by mapping each domain value to
a unique integer. Due to its scalability limitations, Agglomerative clustering is not applied to the Census dataset. 

\paratitle{Baselines}
Despite the importance of explainability, to the best of our knowledge, no attempts have been made to develop explanations for clustering results under DP. Nevertheless, we consider one non-private algorithm and two devised DP adaptations of that baseline for comparative evaluation.
We compare the performance of \ourframework\ with the following approaches for generating explanations:
\begin{itemize}[leftmargin=*]
    \item \textbf{TabEE (non-private)}: The non-private algorithm in \cite{TabEE} for finding a high-scoring attribute combination. 
    The algorithm selects the top attribute combination from a pre-constructed candidate pool based on the original, sensitive definition of the quality functions, as we discussed in \Cref{sec:metrics}.
    \item \textbf{\dptabee}: We implement a direct adaptation of the TabEE algorithm to satisfy DP. This baseline uses the original, sensitive quality functions for attribute selection, but injects the required noise to satisfy DP, according to \Cref{thm:exponential_mech} and the sensitivity of the quality functions (i.e., \Cref{claim:interest-high-sens,claim:suf-high-sens}).
    \item \textbf{\dpnaive}: In  \Cref{sec: alg}  we discussed a naive approach for computing HBEs under DP. We implement this simple DP baseline as follows. Given a privacy budget $\eps$, we compute each of the full-dataset histograms using a budget $\eps/(2\abs{\cA})$ for each attribute. We compute the histogram of each cluster for each attribute using a budget of $\eps/(2\abs{\cA})$ per cluster. Then,  as a post-processing step,  we run the TabEE-based algorithm on the noisy histograms. By the composition and post-processing theorems, the entire algorithm satisfies $\eps$-DP. 
\end{itemize}

\paratitle{Evaluation measures}
While in this work we introduce low-sensitivity variants of \emph{interestingness}, \emph{sufficiency} and \emph{diversity}, which are incorporated into our algorithm, the \emph{original}, sensitive variants of these functions can still be used for evaluating the quality of the selected attribute combination. Hence, we let $Quality = \lambda_{\intTabEE}\cdot \intTabEE + \lambda_{\sufTabEE}\cdot \sufTabEE + \lambda_{\diversityTabEE}\cdot \diversityTabEE$. 
be the global score function from \cite{TabEE}, using the sensitive quality functions $\intTabEE,\sufTabEE, \diversityTabEE$
(\Cref{sec:metrics}).

We also evaluate the similarity between the attribute combination selected by  non-private baseline, denoted $\AC^*$, and that of \ourframework\ or the \dpnaive\ 
baseline. To this end, we adapt the mean absolute error (MAE) \cite{MAE} to our discrete setting. 
The MAE score for a combination $\AC$  is given by $\mathrm{MAE}(E) = \frac{1}{\numclust} \sum_{c\in C} \one_{\set{ \AC(c) \neq \AC^*(c)}}$.

\subsection{Quality Analysis}
\label{subsec: acc}
We detail our experimental analysis for the different configurations of our framework. All results are averaged over 10 runs.

\begin{figure*}
    \centering
    \includegraphics[width=\linewidth]{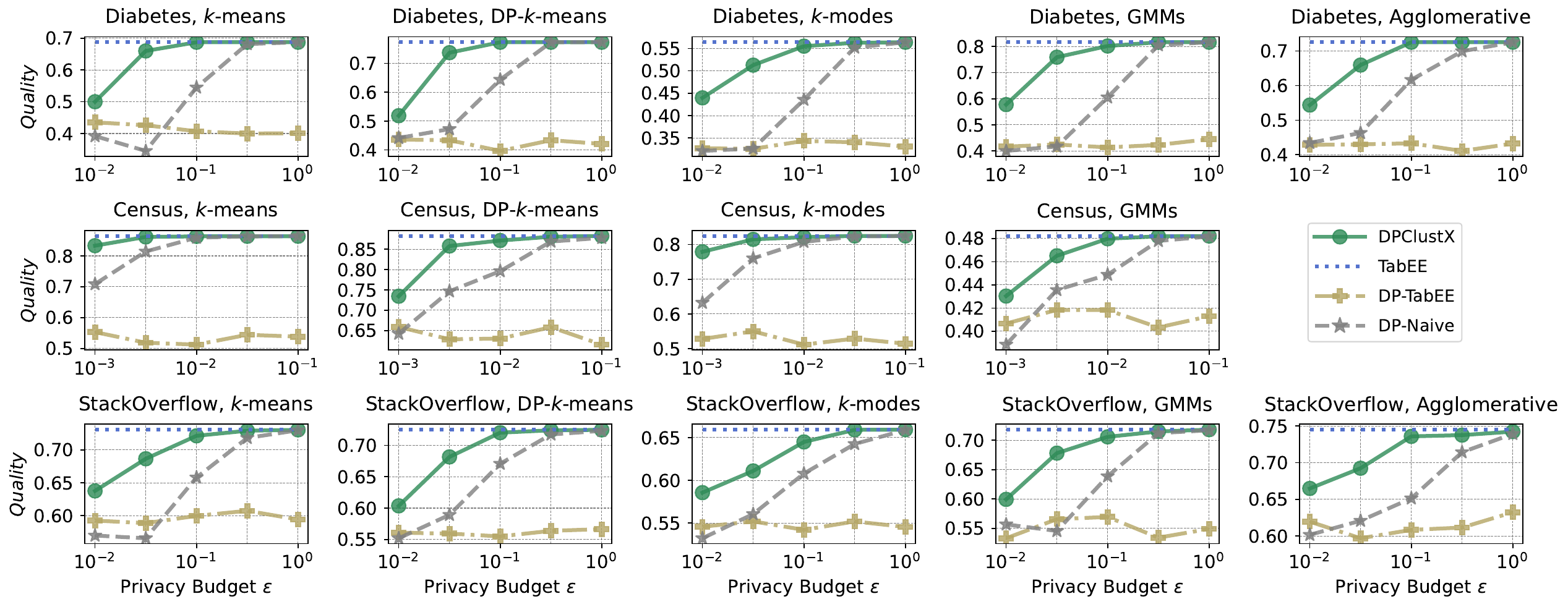}
\caption{$Quality$ values of the selected attribute combination as the total privacy budget $\eps$ varies. Note that the range of the $Quality$ axis differs across methods, reflecting the substantial variation in explanation quality between clustering approaches.}
    \label{fig:quality_scores}
\end{figure*}

\paratitle{Selected attributes quality score}
In this experiment (\Cref{fig:quality_scores}), we evaluate the $Quality$ score of the selected attribute combination with default parameters for different privacy budgets $\eps$, where $\eps_{CandSet} = \eps_{TopComb} = \eps/2$. Note that this experiment examines the attribute choice, hence histogram generation is not needed. 
Results for additional cluster numbers are provided in
\ifpaper
~\cite{full_version}, 
\else
\Cref{appendix-exp}, 
\fi
showing similar trends.
We find that increasing the total privacy budget $\eps$ improves the quality scores achieved by \ourframework\ in all cases, and that it consistently outperforms the other DP baselines.
Moreover, \dptabee\ failed to improve in the examined range.
For the Diabetes dataset, at $\eps = 0.1$, \ourframework\ scores are  only $0.66\%$ lower than TabEE on average, while \dpnaive\ scores $20.22\%$ lower.
At $\eps = 1$, \ourframework\ matches TabEE across all methods. For the Census dataset, at $\eps = 10^{-2.5}$, \ourframework\ attains scores $1.6\%$ lower on average, while \dpnaive\ scores are $9.8\%$ lower. At $\eps = 0.1$, the difference for \ourframework\ is only $0.003\%$. 
For the Stack Overflow dataset, the scores of \ourframework\ are only $1.3\%$ lower than TabEE on average at $\eps=0.1$, while \dpnaive\ scores $9.8\%$ lower. The scores of \dptabee\ are lower by $19\%$ even at $\eps=1$.

\begin{figure}
    \centering
    \includegraphics[width=0.9\linewidth]{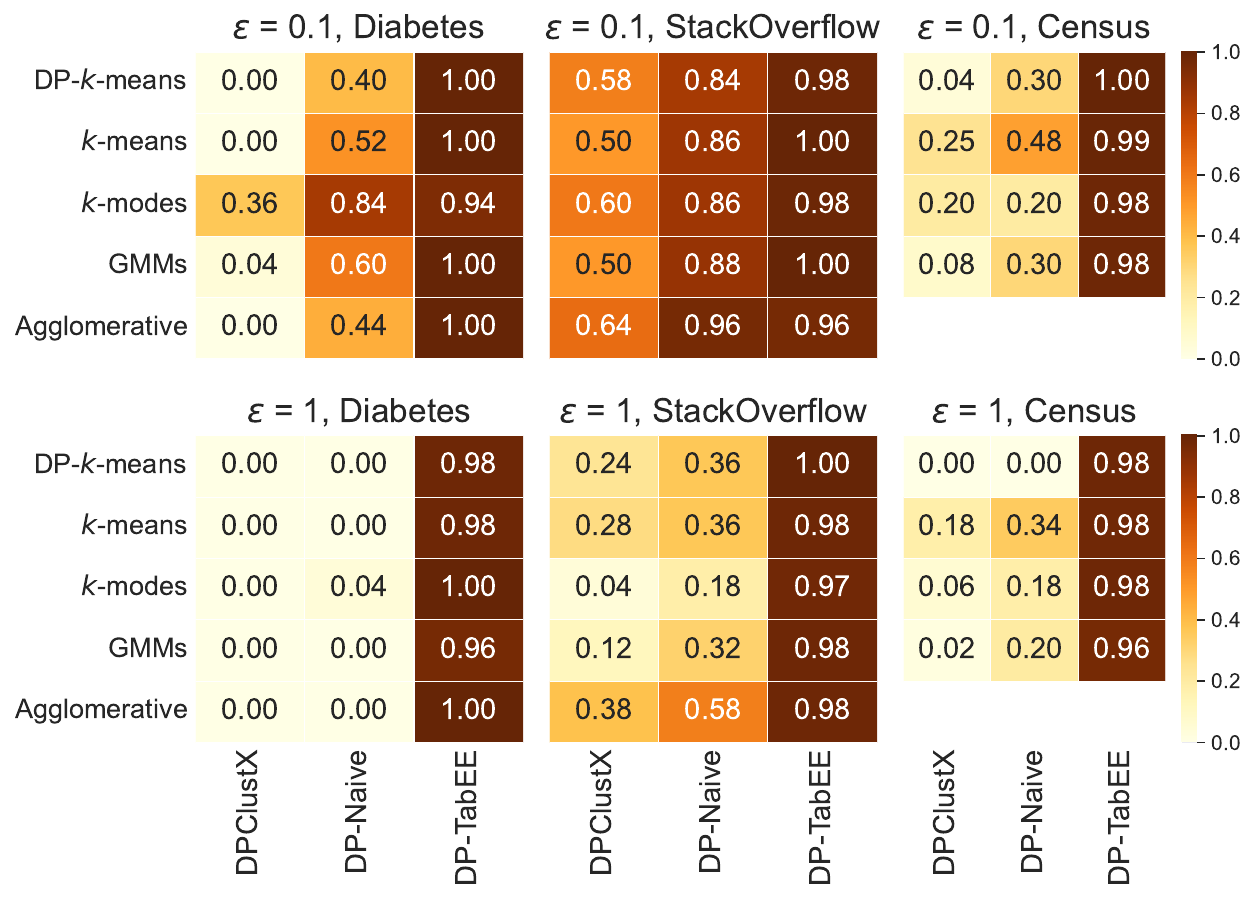}
    \caption{Mean Absolute Error (MAE) values of the selected attribute combination compared to the non-private TabEE baseline~\cite{TabEE}, as the total privacy budget $\eps$ varies.}
    \label{fig:similarity_scores}
\end{figure}

\paratitle{Selected attributes error}\label{para: attr_err}
This experiment (\Cref{fig:similarity_scores}) examines the MAE of the selected attribute combination with varying privacy budgets $\eps$, where $\eps_{CandSet} = \eps_{TopComb} = \eps/2$. 
Note that all attributes are considered distinct, despite possible correlations. Hence, an MAE score of $0$ implies identical attribute choice to that of  the non-private TabEE baseline.
The results show that \ourframework\ outperforms the DP baselines in all cases. For the Diabetes dataset, \ourframework\ selects the same combination as TabEE for all methods at $\eps = 1$, and for most methods at $\eps = 0.1$, with an MAE of $0.04$ for GMMs and $0.36$ for k-modes.
 For the Census dataset, the MAE values are below $0.25$ for all methods at $\eps=0.1$ and below  $0.18$ at $\eps=1$. 
For the Stack Overflow dataset at $\eps=1$, \ourframework\ obtains average $\mathrm{MAE}$ scores below $0.4$ for all methods, and below $0.12$ for GMMs and k-modes. 
Notably, with the Stack Overflow and Census datasets, multiple choices achieve nearly optimal scores due to attribute correlations. Consequently, under DP randomization, it is expected that the selection will not always favor the top solution, leading to relatively higher MAE for all DP methods, despite the nearly-optimal quality scores (\Cref{fig:quality_scores}).

\begin{figure}
    \centering
    \includegraphics[width=\linewidth]{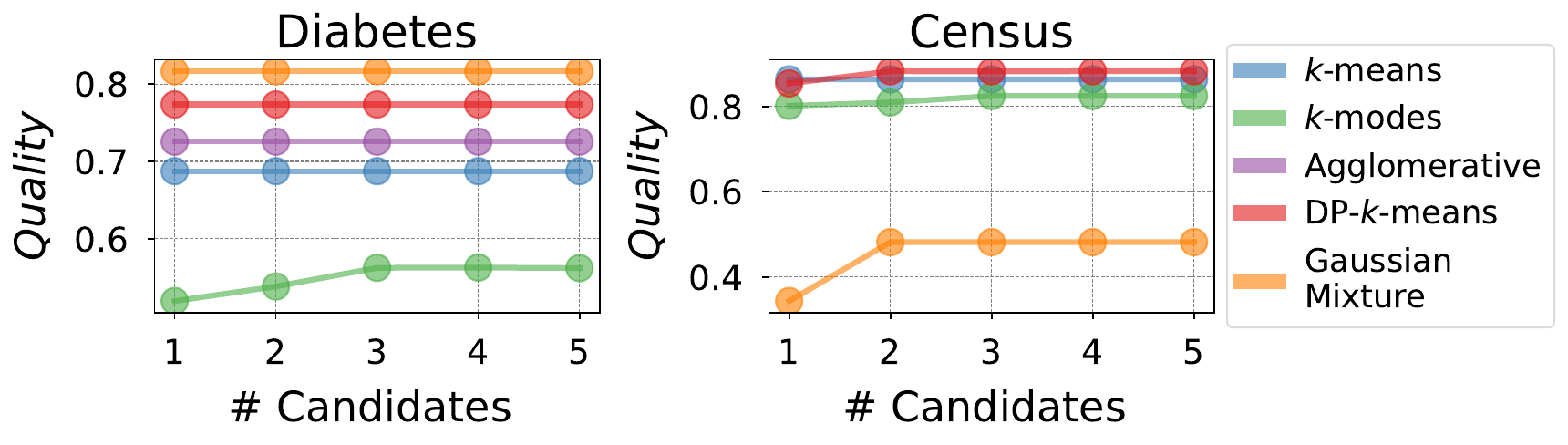}
    \caption{ Explanation $Quality$ values for \ourframework\ with a varying number of candidate attributes per cluster selected at Stage-1 (\Cref{subsec:stage-1}).}
    \label{fig:Num_candidates_ablation}
\end{figure}

\paratitle{Quality for different candidate set sizes}
In this experiment (\Cref{fig:Num_candidates_ablation}) We evaluate the score of the selected attribute combination with varying candidate set sizes from Stage-1 of our algorithm. We focus on the Census and Diabetes datasets, as the Stack Overflow dataset exhibited similar trends. 
For the Diabetes dataset and all methods except $k$-modes, we find that increasing the candidate size from $1$ to $5$ had no effect, with the same attributes selected in all runs. For $k$-modes, the score increases by $8\%$ between $1$ and $3$ candidates, and stabilizes. For the Census dataset, a positive trend is observed for all methods between $1$ and $3$, peaking at $3$ and stabilizing. GMMs shows a $40\%$ score improvement between $1$ and $2$ candidates, with the same attributes selected for $3$. 
We set the default size to $3$, as further increase did not improve quality in our experiments, but considerably increased the running time, as shown in \Cref{fig:time_by_num_cand}.

\paratitle{Quality for different choices of weights}
We measure the $Quality$ scores of the selected attribute combination for different weight parameters $\lambda_{\intTabEE}$, $\lambda_{\sufTabEE}$, and $\lambda_{\diversityTabEE}$, setting one to zero and the remaining two to $1/2$. 
The results,
detailed in
\ifpaper
~\cite{full_version}, 
\else
\Cref{table:weight_exp}, 
\fi
show zero or negligible score difference on the Census dataset for 3 clusters. For the Diabetes dataset, scores are lower by merely $0.11\%$ on average across all clustering methods and weight configurations. For 5 clusters, the results show a minor difference of $0.06\%$ on average on the Diabetes dataset, and of $0.13\%$ on the Census dataset. 
For 7 clusters, we find that the scores are lower by  $0.4\%$ on average on the Diabetes dataset. and by only $0.08\%$ for Census dataset.
These minor differences indicate that \ourframework\ selects an attribute combination with a quality comparable to that of the non-private baseline across different weights configurations, offering the same flexibility in parameter selection.

\begin{figure}
       \begin{minipage}{\linewidth}
    \begin{subfigure}{\linewidth}
    \centering
         \includegraphics[width=0.9\textwidth]{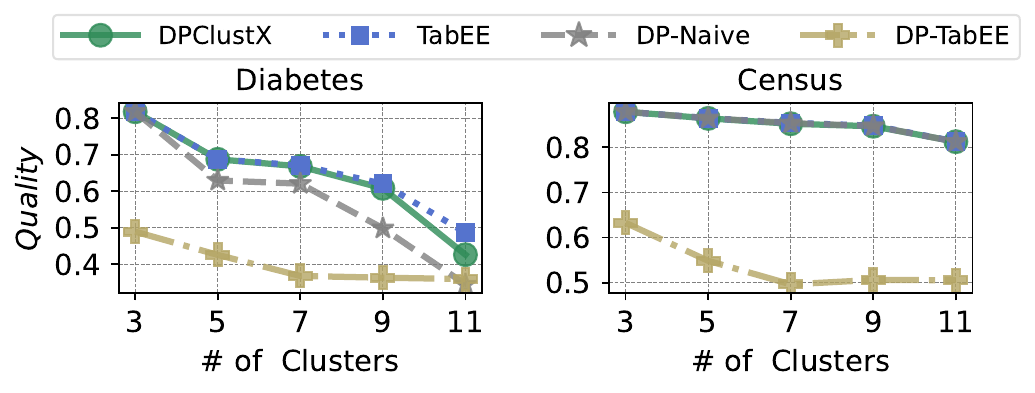}  
         \caption{Trend  by number of clusters.}
          \label{fig:quality_by_num_clust}
    \end{subfigure}
    \hfill
        \begin{subfigure}{\linewidth}
        \centering
   \includegraphics[width=0.9\textwidth]{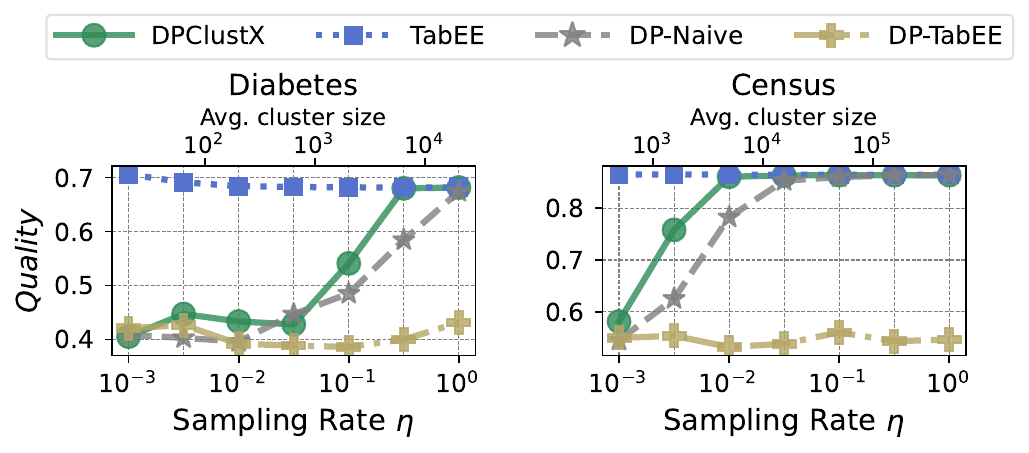}
        \caption{Trend by sample size. The bottom x-axis indicates the sampling rate, and the top x-axis shows the corresponding average cluster size.}
        \label{fig:quality_by_clust_size}
    \end{subfigure}
        % \vspace{0.2cm}
    \end{minipage}
    \vspace{2mm}
       \caption{$Quality$ values for the selected attribute combination as the number of clusters varies (top) and as the sample size per cluster varies (bottom). Note the differing $Quality$ axis ranges.}
        \label{fig:quality_by_clust_size_num_clust}
        \vspace{2mm}
\end{figure}

\paratitle{Quality for different numbers of clusters}
We examine the impact of varying the number of clusters on the quality of explanations generated by \ourframework\ compared to the baselines. \Cref{fig:quality_by_num_clust} presents the results for the Census and Diabetes datasets using $k$-means clustering, with other methods exhibiting similar trends.
The results indicate that explanation quality decreases as the number of clusters increases, even without privacy constraints. 
%For the smaller Diabetes dataset
%, the $Quality$ values of the non-private TabEE baseline decreased by $38\%$ when the number of clusters increased from 3 to 11. For the larger Census dataset, the decrease was  $7.5\%$. 
In all cases, \ourframework\ outperforms the DP baselines. For  Census, both \ourframework\ and \dpnaive\ achieve scores comparable to TabEE, whereas \dptabee\ obtains significantly lower scores ($36.7\%$ difference on average).
For  Diabetes, \ourframework\ outperforms the DP baselines, and maintains a score close to TabEE (lower by $2\%$ on average, and by $0.4\%$ when the number of clusters is at most $9$).
Note that the presence of small clusters, which tends to occur with a larger number of clusters, inevitably leads to some degradation in utility of DP methods.
%(e.g., see the discussion preceding \Cref{claim:interest-high-sens}). 
 Intuitively, the
small count differences are masked by DP noise, leading
to inaccurate quality evaluations of the histograms.
However, this effect is not observed in \Cref{fig:quality_by_num_clust} for the Census dataset as it is larger. 
%The impact of cluster sizes is studied next.
% This effect is observed in \Cref{fig:quality_by_num_clust} for the smaller Diabetes dataset.

%The impact of cluster sizes is studied next.

 \paratitle{Quality for different cluster sizes}
\label{subsec:quality_by_clust_size}
We study the impact of varying the average cluster size on the quality scores. (\Cref{fig:quality_by_clust_size}). A subset comprising $\eta$ fraction of the tuples in each cluster is sampled, where $\eta$ ranges from $10^{-3}$ to $1$ (full dataset), and an explanation is generated for the sampled data. \Cref{fig:quality_by_clust_size} shows the results for the Census and Diabetes datasets under $k$-means clustering, while for other cases the results  exhibited similar trends.
We find that the non-private TabEE baseline maintains a stable explanation quality, while the DP methods exhibit a decrease in quality with smaller cluster sizes.
For the Diabetes dataset, at $\eta = 10^{-0.5}$, the average cluster size is $6436$, and \ourframework\ performs comparably to TabEE, with a minor difference of $0.2\%$, while \dpnaive\ scores $15\%$ lower, and \dptabee\ consistently performs poorly with a $42\%$ difference. At $\eta = 10^{-1}$, the average cluster size is $2035$, and \ourframework's score decreases by $20\%$. 
For the Census dataset, \ourframework\ performs comparably to the non-private TabEE, with an average difference of $0.09\%$ 
at $\eta \ge 10^{-2}$ and an average cluster size of $\ge 4917$. At $\eta = 10^{-2.5}$, the average cluster size is $1555$, and the score decreases by $12\%$. In all cases, \ourframework\ consistently outperforms or matches the two DP baselines with smaller cluster sizes.

\paratitle{Impact of attribute correlations on quality}
We assess the stability of \ourframework\ in the presence of attribute correlations by adopting the experimental setting of \cite{TabEE}. We cluster the datasets after adding attributes that are highly correlated with the original ones. Specifically, for each original attribute, we generate a corresponding correlated attribute by randomly perturbing a small portion of the records, while maintaining a Cramer's V value of  0.85, a standard association measure \cite{cramer1999mathematical} between attributes.
Then, we run \ourframework\ twice, once with the extended set of attributes and once with the original set,
and compare the $Quality$ scores of the selected attributes in both scenarios.
We find that, for all datasets, the difference is below $2\%$ on average, indicating a minor change. 
This difference is mostly due to the diversity measure, as an attribute and its noisy counterpart are considered different, hence selecting both maximizes diversity. In contrast, while running with the original attribute set, \ourframework\  may select the same attribute twice, potentially contributing less to diversity. Considering only interestingness and sufficiency, the difference is below $0.1\%$ in all cases.

\subsection{Performance Analysis}
\label{subsec: pref}
% \ag{Some are still missing: DB size, size of $k$, etc.}\ron{Added}
We measure the execution time of our algorithm with various parameter settings for both datasets. We focus on $k$-means and GMMs clustering methods, as other methods were unable to scale to a large number of clusters or returned mostly empty clusters when the number of clusters exceeded $9$. 
The default settings for these experiments are $9$ clusters, $3$ candidate attributes per cluster, and the entire dataset with all attributes used. 
Each experiment varies one of these parameters and is averaged over 10 runs.

\begin{figure*}[t]
    \begin{subfigure}[t]{0.23\linewidth}
        \centering
        \includegraphics[height=3cm]{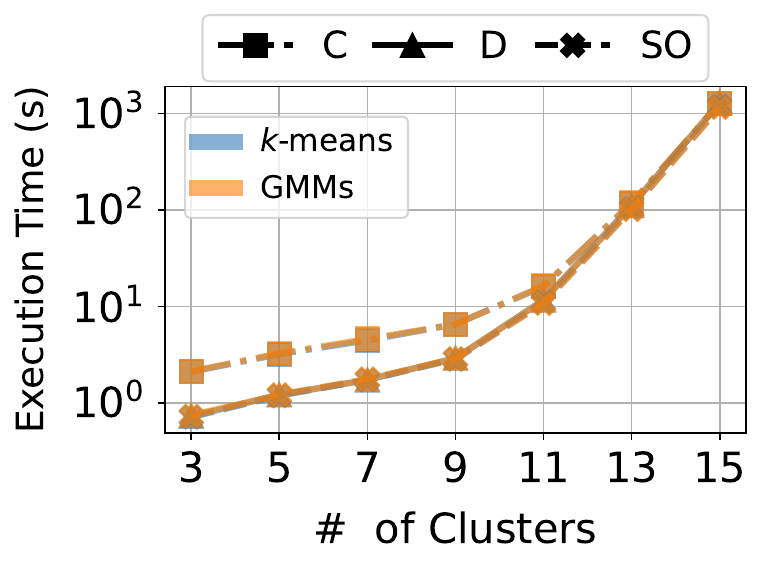}
        \caption{Trend  by number of clusters (log-scale).}
        \label{fig:time_by_num_clust}
        \vspace{0pt}
    \end{subfigure}
    \hfill
    % Subfigure 2
    \begin{subfigure}[t]{0.23\linewidth}
        \centering
        \includegraphics[height=3cm]{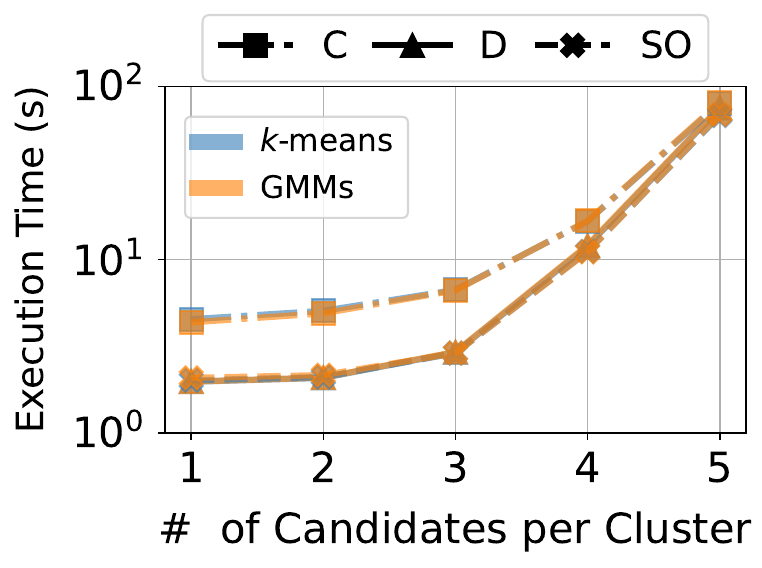}
        \caption{Trend by size of a single cluster candidate set (log-scale).}
        \label{fig:time_by_num_cand}
        \vspace{0pt}
    \end{subfigure}
    \hfill
    % Subfigure 3
    \begin{subfigure}[t]{0.23\linewidth}
        \centering
        \includegraphics[height=3cm]{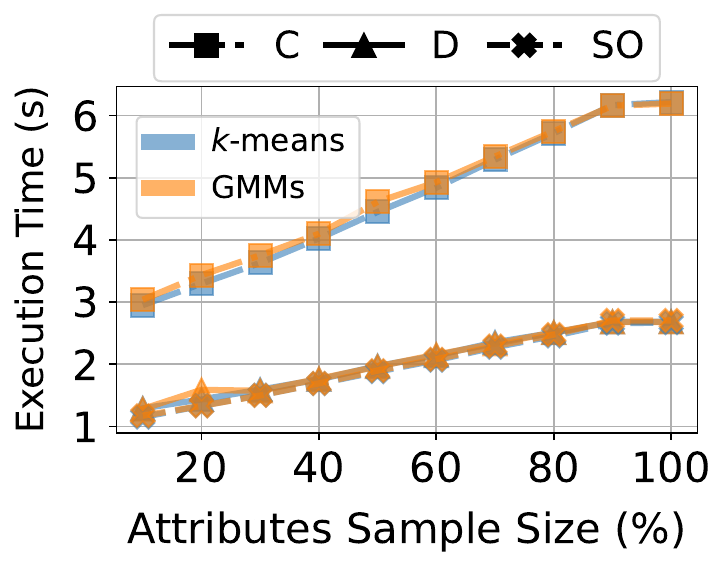}
        \caption{Trend by attributes sample size.}
        \label{fig:time_by_num_attr}
        \vspace{0pt}
    \end{subfigure}
    \hfill
    % Subfigure 3
    \begin{subfigure}[t]{0.23\linewidth}
        \centering
        \includegraphics[height=3cm]{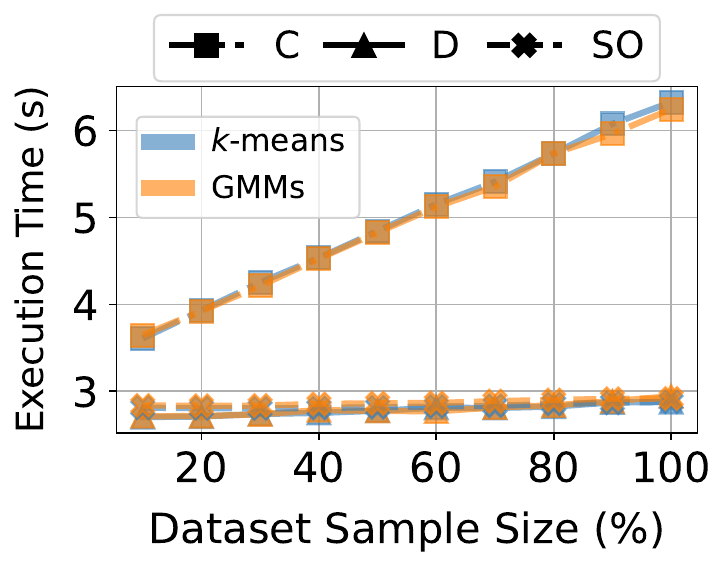}
        \caption{Trend by dataset sample size.}
        \label{fig:time_by_sample_size}
        \vspace{0pt}
    \end{subfigure}
    \caption{\ourframework's execution time trends for the Census (C), Diabetes (D), and Stack Overflow (SO) datasets by different parameters.}
    
    \label{fig:time_experiments}
\end{figure*}

\paratitle{Number of clusters}
\Cref{fig:time_by_num_clust} illustrates the relationship between the execution time (in seconds, log scale) and the number of clusters for which an explanation is generated.
We observe that the running time increases exponentially with the number of clusters, yet remains reasonably low across both methods when the number of clusters is at most $11$. For the larger Census dataset, the runtimes for up to $9$ clusters across both methods are less than $6.6$ seconds.
For the Diabetes and Stack Overflow datasets, all runtimes are below $3$ seconds up to $9$ clusters, exhibiting a similar trend. For $15$ clusters, the runtimes are up to $1275$ seconds for Stack Overflow and Diabetes, and $1314$ seconds for Census.
% As most applications \cite{TabEE, FEDEX}  consider a number of clusters bounded by $10$, these results indicate that our algorithm maintains short execution time, making it well-suited for interactive user experience.
% \paratitle{Candidate set size} \Cref{fig:time_by_num_clust} illustrates the relationship between the running time (in seconds, log scale) and the size of the candidate set constructed in Stage-1 of our algorithm for each cluster. The number of clusters is fixed at $9$. In all cases, running times stay below $10$ seconds with $3$ candidates and increase significantly with more candidates, reaching $80$ seconds with 5 candidates for the Census dataset and $75$ seconds for the Diabetes dataset.

\paratitle{Number of candidate attributes per cluster} \Cref{fig:time_by_num_cand} illustrates the relationship between the execution time (in seconds, log scale) and the size of the candidate set constructed by \Cref{alg:topk-single} for each cluster. The results indicate a significant execution time increase with larger sizes, highlighting the benefit of selecting a smaller size.
The execution times remain below $3$ seconds with at most  $3$ candidates for the Diabetes and Stack Overflow datasets, and under $7$ seconds for the Census dataset. 
 However, with $5$ candidates, execution times rise up to $80$ seconds for Census, $75$ seconds for Diabetes, and $69$ seconds for Stack Overflow.

\paratitle{Number of attributes} In this experiment (\Cref{fig:time_by_num_attr}), we randomly sample a subset of the attributes for each dataset, and generate an explanation using only the sampled attributes.
\Cref{fig:time_by_num_attr} depicts the relationship between running time (in seconds) and the percentage of attributes selected from the dataset. The results indicate a linear increase in running time, implying that while the number of attributes has an effect, it is relatively small. For instance, with a $50\%$ sample, the execution times are at most $4.6$ seconds for the Census dataset and $2$ seconds for Diabetes and Stack Overflow, while with a $100\%$ sample (full dataset), the execution times increase to a maximum of $6.3$ seconds for Census and at most $2.9$ seconds for Diabetes and Stack Overflow.

\paratitle{Dataset size} 
 In this experiment (\Cref{fig:time_by_sample_size}), we randomly sample a portion of the tuples for each dataset, and generate an explanation using the sampled data.
\Cref{fig:time_by_sample_size} shows the relationship between execution time (in seconds) and the percentage of tuples sampled from each dataset. The results indicate a linear increase in running time, suggesting that while an increase in the number of tuples leads to longer execution times, the effect remains relatively small.
With a $50\%$ sample, the execution times are at most $4.8$ seconds for the Census dataset and below $2.6$ seconds for Diabetes and Stack Overflow. With a $100\%$ sample (full dataset), the execution times are at most $6.3$ seconds for Census and below $2.9$ seconds for Diabetes and Stack Overflow.

\subsection{Case Study}
\label{subsec:case-studies}
We now present a case study over the Census dataset with default parameters.
The dataset is clustered into $3$ clusters using $k$-means.

\begin{figure}
    \centering
       \begin{minipage}{\linewidth}
    \begin{subfigure}{\linewidth}
       \centering
         \includegraphics[width=\textwidth]{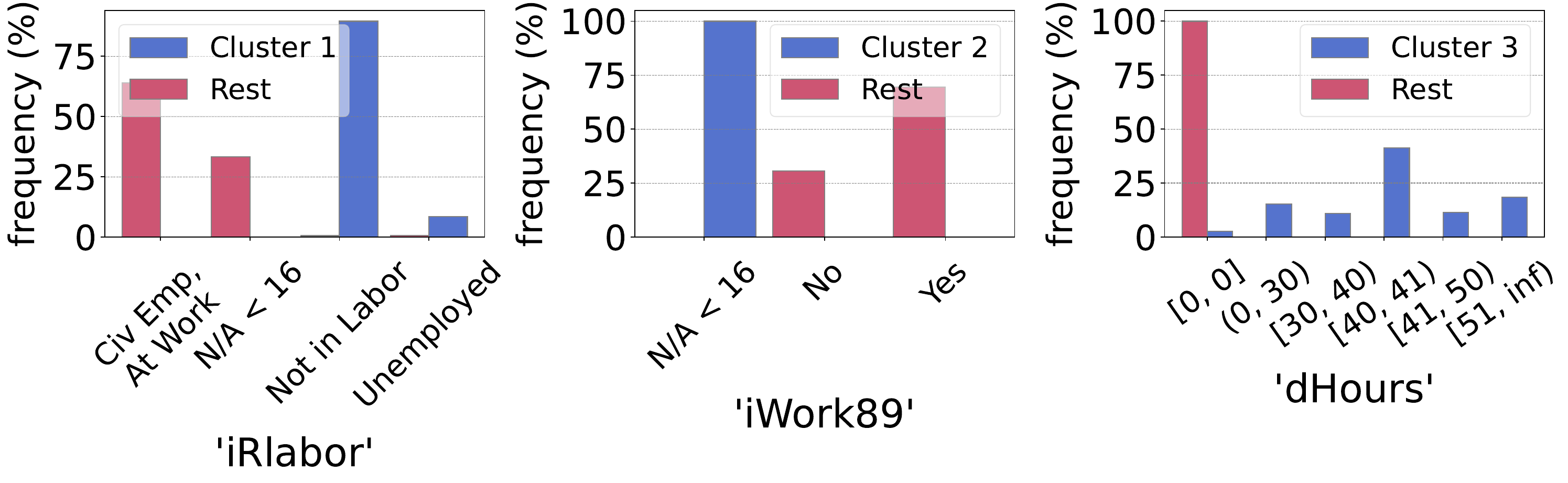}  
         \vspace{-4mm}
        \caption{\ourframework\ explanation.}
      \label{fig:census-private-case1}
    \end{subfigure}
    \hfill
    
    \begin{subfigure}{\linewidth}
       \centering
   \includegraphics[width=\textwidth]{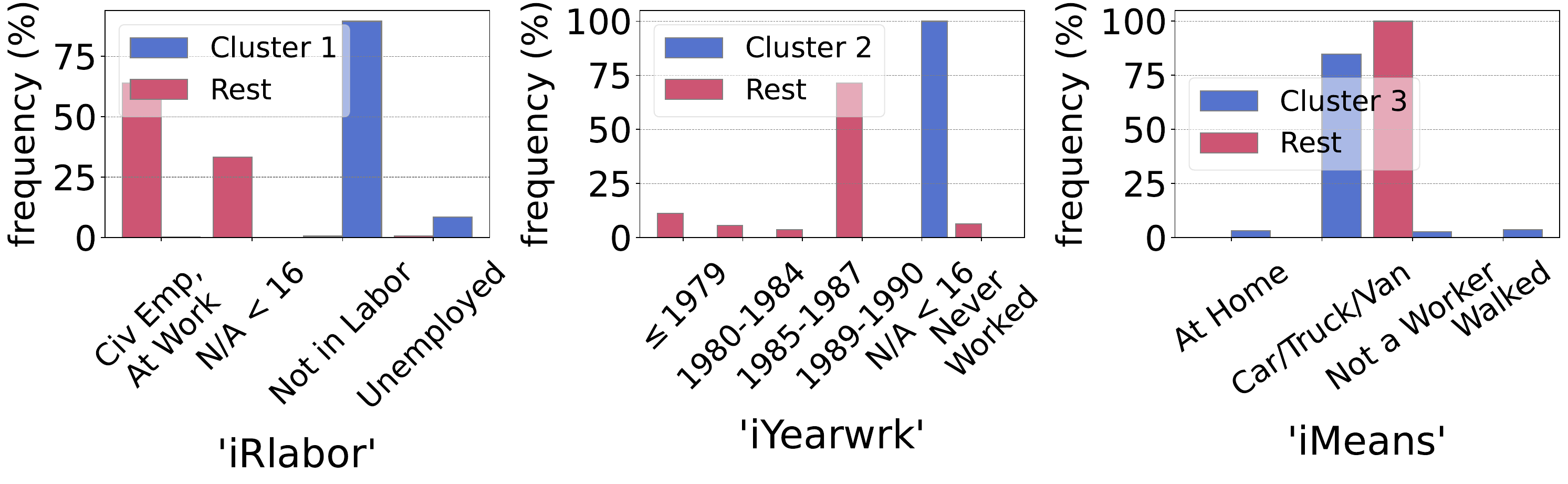}    
   % \vspace{-2mm}
 \caption{Non-private TabEE~\cite{TabEE} explanation.}
      \label{fig:census-tabee-case1}
    \end{subfigure}
        % \vspace{0.2cm}
    \end{minipage}
    \caption{Explanations for the US Census dataset case study.}
\end{figure}

\ourframework\ selects the attributes $\mathtt{iRlabor}$, which represents employment status,  $\mathtt{iWork89}$, which indicates whether the individual worked in 1989, and  $\mathtt{dHours}$, which denotes the number of working hours in the previous week.
The final output is shown in \Cref{fig:census-private-case1}. 
\Cref{fig:census-tabee-case1} shows the non-private explanation generated by TabEE. In this case, the selected attributes are  
$\mathtt{iRlabor}$, $\mathtt{iYearwrk}$, which specifies the last year the individual worked, and  $\mathtt{iMeans}$, which describes the means of transportation to work.
Since the two explanations agree on $1$ out of $3$ attributes, \ourframework\ achieves $\mathrm{MAE}=2/3$. However, the $Quality$ of the non-private baseline is only $0.04\%$ higher, and the explanation generated by \ourframework\ conveys similar insights.

Both explanations indicate that Cluster 1 predominantly consists of adults who are currently not working, and Cluster 2 of individuals under age 16, for whom data is unavailable. In this group, the attributes $\mathtt{iWork89}$ and $\mathtt{iYearwrk}$ are correlated.
Both \ourframework\ and TabEE reveal that Cluster 3 primarily consists of working individuals, as they worked more than 0 hours last week and have means of transportation to work. Thus, a similar conclusion is reached, though different attributes are used, which are correlated among non-working individuals.
Note that \ourframework\ is not guaranteed to select attributes most correlated with those chosen by TabEE in general.

\section{Related Work}
\label{sec:related}
We next survey related work in relevant fields. 

\paratitle{Differentially private explanations}
Several works have explored integrating DP with ML models and query result explanation methods to address the dual challenge of transparency and privacy. The work of \cite{patel2022model} introduced differentially private mechanisms for model explanations, providing interpretable insights into model behavior while satisfying DP. Their approach adapts traditional explanation methods such as feature importance scores and local interpretable model-agnostic explanations (LIME) to operate under DP constraints, thereby ensuring that the explanations themselves do not leak sensitive information. Additionally, significant contributions have been made in creating a framework for generating differentially private Shapley values, enabling the interpretation of model predictions with strong privacy guarantees \cite{jordon2019differentially}. Further approaches include a DP SVM mechanism for robust counterfactual explanations \citep{mochaourab2021robust} and DP Locally Linear Maps, which provide both local and global model explanations while enabling a favorable privacy-accuracy trade-off by efficiently managing the number of parameters \citep{harder2020interpretable}.
See \cite{nguyen2024survey} for a survey on privacy-preserving model explanations. In the context of query result explanation, DPXPlain \cite{tao2022dpxplain} is a framework that generates differentially private explanations for aggregate queries based on the notion of intervention.
These efforts represent crucial steps towards creating trustworthy ML and data analysis systems that offer both transparency and robust privacy guarantees, thus addressing the critical need for ethical AI deployment in sensitive applications. 

\paratitle{Clustering explanations}
In the non-private setting, clustering explanations are well-studied. For instance, \cite{moshkovitz2020explainable, makarychev2021near, esfandiari2022almost, makarychev2022explainable, gamlath2021nearly}  study tree-based interpretable clustering algorithms specifically for $k$-means or $k$-medians.
\citet{Cluster-Explorer} provides black-box clustering explanations using filter predicates. 
\citet{TabEE} use histogram-based explanations for clusters in tabular datasets, induced by clusters formed in the tabular-embedding space. However, \emph{differentially-private} explanation algorithms for clustering have not been explored beyond our work. 

\paratitle{Private data summarization and exploration tools}
In the realm of databases, privacy-preserving techniques have been a focal point of research, particularly with the advent of differential privacy (DP). In recent years, DP has been increasingly adopted to 
practical systems for interactive data analysis, such as PINQ \cite{mcsherry2009privacy}, PrivateSQL \cite{privatesql}, FLEX  \cite{FLEX}, and Chorus \cite{johnson2020chorus}.
Another significant contribution is the DPCube framework by Xiao et al., which enables efficient and private OLAP data cube construction, by differentially private histogram release through multidimensional partitioning, thereby enhancing the usability of summarized data under privacy constraints \cite{xiao2012dpcube}. The RONA algorithm by McKenna et al. improves the accuracy of DP synthetic data generation through an iterative refinement process, supporting private data release \cite{mckenna2021winning}. Additionally, Zhang et al. introduced PrivBayes, a method for generating differentially private synthetic data using Bayesian networks to capture correlations in the data while preserving privacy \cite{zhang2017privbayes}. In the context of data exploration, APEx \cite{ge2019apex} is a system that allows data analysts to pose adaptively chosen queries along with
required accuracy bounds, identifying algorithms with the least privacy loss to answer these queries accurately under DP.

These advancements collectively highlight the progress and ongoing challenges in developing privacy-preserving database systems that balance data utility with stringent privacy requirements.

% \vspace{-1mm}
\section{Conclusion and Future Work}\label{sec:conc}
We proposed \ourframework, a framework for generating global, histogram based explanations for black-box clustering results while preserving differential privacy. These explanations consist of histograms on carefully selected attributes for each cluster and the remaining dataset, highlighting significant distributional shifts alongside additional high-quality characteristics.
We demonstrated through extensive experiments that the explanations generated by  \ourframework\  are comparable to the non-privately generated explanations even under tight privacy budgets.

There are several interesting future directions. First, the current
framework outputs one explanation per cluster, aligning with the non-private \cite{TabEE}.
It can be generalized to output multiple explanations per cluster for a more comprehensive understanding, but complexity may increase (see \Cref{appendix:multexp}). Extending the framework to efficiently output multiple explanations per cluster is important future work.
Second, \ourframework\ uses one-dimensional histograms due to their simplicity and interpretability, building upon existing work. One possible way to extend \ourframework\ to higher-dimensional histograms is by considering the Cartesian product of the domains. However, such an extension is not straightforward, as it comes at the cost of increased complexity, and may result in histograms where all counts are small, making it challenging to accurately compute them under DP.
Examining the quality and interpretability of such explanations is an interesting direction.
Third, it would be intriguing to examine the impact of different discretization and binning
approaches on the performance of our system. 
Fourth, the extension of \ourframework\ to different score functions that emphasize different facets of explainability would be an interesting direction.  
These directions will enhance the usability and interpretability of clustering methods under DP.
\begin{acks}
This research was supported by the Israel Science Foundation (ISF) under grant 2707/22 of the Breakthrough Research Grant (BRG) Program.
The work of Amir Gilad was funded by the Israel Science Foundation (ISF) under grant 1702/24, the Scharf-Ullman Endowment, and the Alon Scholarship.
\end{acks}

% \ifpaper
% \clearpage
% \balance
% \else
% \fi
% \clearpage
% \newpage
\balance
\bibliographystyle{ACM-Reference-Format}
\bibliography{sigproc}

%%% -*-BibTeX-*-
%%% Do NOT edit. File created by BibTeX with style
%%% ACM-Reference-Format-Journals [18-Jan-2012].

\begin{thebibliography}{76}

%%% ====================================================================
%%% NOTE TO THE USER: you can override these defaults by providing
%%% customized versions of any of these macros before the \bibliography
%%% command.  Each of them MUST provide its own final punctuation,
%%% except for \shownote{}, \showDOI{}, and \showURL{}.  The latter two
%%% do not use final punctuation, in order to avoid confusing it with
%%% the Web address.
%%%
%%% To suppress output of a particular field, define its macro to expand
%%% to an empty string, or better, \unskip, like this:
%%%
%%% \newcommand{\showDOI}[1]{\unskip}   % LaTeX syntax
%%%
%%% \def \showDOI #1{\unskip}           % plain TeX syntax
%%%
%%% ====================================================================

\ifx \showCODEN    \undefined \def \showCODEN     #1{\unskip}     \fi
\ifx \showDOI      \undefined \def \showDOI       #1{#1}\fi
\ifx \showISBNx    \undefined \def \showISBNx     #1{\unskip}     \fi
\ifx \showISBNxiii \undefined \def \showISBNxiii  #1{\unskip}     \fi
\ifx \showISSN     \undefined \def \showISSN      #1{\unskip}     \fi
\ifx \showLCCN     \undefined \def \showLCCN      #1{\unskip}     \fi
\ifx \shownote     \undefined \def \shownote      #1{#1}          \fi
\ifx \showarticletitle \undefined \def \showarticletitle #1{#1}   \fi
\ifx \showURL      \undefined \def \showURL       {\relax}        \fi
% The following commands are used for tagged output and should be
% invisible to TeX
\providecommand\bibfield[2]{#2}
\providecommand\bibinfo[2]{#2}
\providecommand\natexlab[1]{#1}
\providecommand\showeprint[2][]{arXiv:#2}

\bibitem[dpc(2024)]%
        {dpclustex_github}
 \bibinfo{year}{2024}\natexlab{}.
\newblock \bibinfo{title}{DPClustX Git Repository}.
\newblock
\newblock
\urldef\tempurl%
\url{https://github.com/ronzadi/DPClustX}
\showURL{%
\tempurl}


\bibitem[Abouelmehdi et~al\mbox{.}(2018)]%
        {AbouelmehdiHK18}
\bibfield{author}{\bibinfo{person}{Karim Abouelmehdi}, \bibinfo{person}{Abderrahim~Beni Hssane}, {and} \bibinfo{person}{Hayat Khaloufi}.} \bibinfo{year}{2018}\natexlab{}.
\newblock \showarticletitle{Big healthcare data: preserving security and privacy}.
\newblock \bibinfo{journal}{\emph{J. Big Data}}  \bibinfo{volume}{5} (\bibinfo{year}{2018}), \bibinfo{pages}{1}.
\newblock
\urldef\tempurl%
\url{https://doi.org/10.1186/S40537-017-0110-7}
\showDOI{\tempurl}


\bibitem[Abowd(2018)]%
        {Abowd18}
\bibfield{author}{\bibinfo{person}{John~M. Abowd}.} \bibinfo{year}{2018}\natexlab{}.
\newblock \showarticletitle{The {U.S.} Census Bureau Adopts Differential Privacy}. In \bibinfo{booktitle}{\emph{Proceedings of the 24th {ACM} {SIGKDD} International Conference on Knowledge Discovery {\&} Data Mining, {KDD} 2018, London, UK, August 19-23, 2018}}, \bibfield{editor}{\bibinfo{person}{Yike Guo} {and} \bibinfo{person}{Faisal Farooq}} (Eds.). \bibinfo{publisher}{{ACM}}, \bibinfo{pages}{2867}.
\newblock
\urldef\tempurl%
\url{https://doi.org/10.1145/3219819.3226070}
\showDOI{\tempurl}


\bibitem[Acs et~al\mbox{.}(2012)]%
        {acs2012differentially}
\bibfield{author}{\bibinfo{person}{Gergely Acs}, \bibinfo{person}{Claude Castelluccia}, {and} \bibinfo{person}{Rui Chen}.} \bibinfo{year}{2012}\natexlab{}.
\newblock \showarticletitle{Differentially private histogram publishing through lossy compression}. In \bibinfo{booktitle}{\emph{2012 IEEE 12th International Conference on Data Mining}}. IEEE, \bibinfo{pages}{1--10}.
\newblock


\bibitem[Amer-Yahia et~al\mbox{.}(2021)]%
        {amer2021exploring}
\bibfield{author}{\bibinfo{person}{Sihem Amer-Yahia}, \bibinfo{person}{Tova Milo}, {and} \bibinfo{person}{Brit Youngmann}.} \bibinfo{year}{2021}\natexlab{}.
\newblock \showarticletitle{Exploring ratings in subjective databases}. In \bibinfo{booktitle}{\emph{Proceedings of the 2021 International Conference on Management of Data}}. \bibinfo{pages}{62--75}.
\newblock


\bibitem[Borodin et~al\mbox{.}(2017)]%
        {borodin2017max}
\bibfield{author}{\bibinfo{person}{Allan Borodin}, \bibinfo{person}{Aadhar Jain}, \bibinfo{person}{Hyun~Chul Lee}, {and} \bibinfo{person}{Yuli Ye}.} \bibinfo{year}{2017}\natexlab{}.
\newblock \showarticletitle{Max-sum diversification, monotone submodular functions, and dynamic updates}.
\newblock \bibinfo{journal}{\emph{ACM Transactions on Algorithms (TALG)}} \bibinfo{volume}{13}, \bibinfo{number}{3} (\bibinfo{year}{2017}), \bibinfo{pages}{1--25}.
\newblock


\bibitem[Clore et~al\mbox{.}(2014)]%
        {diabetes}
\bibfield{author}{\bibinfo{person}{John Clore}, \bibinfo{person}{Krzysztof Cios}, \bibinfo{person}{Jon DeShazo}, {and} \bibinfo{person}{Beata Strack}.} \bibinfo{year}{2014}\natexlab{}.
\newblock \bibinfo{title}{Diabetes 130-US Hospitals for Years 1999–2008}.
\newblock \bibinfo{howpublished}{UCI Machine Learning Repository}.
\newblock
\newblock
\shownote{DOI: https://doi.org/10.24432/C5230J}.


\bibitem[Copul et~al\mbox{.}(2024)]%
        {TabEE}
\bibfield{author}{\bibinfo{person}{Roni Copul}, \bibinfo{person}{Nave Frost}, \bibinfo{person}{Tova Milo}, {and} \bibinfo{person}{Kathy Razmadze}.} \bibinfo{year}{2024}\natexlab{}.
\newblock \showarticletitle{TabEE: Tabular Embeddings Explanations}.
\newblock \bibinfo{journal}{\emph{Proceedings of the ACM on Management of Data}} \bibinfo{volume}{2}, \bibinfo{number}{1} (\bibinfo{year}{2024}), \bibinfo{pages}{1--26}.
\newblock


\bibitem[Cram{\'e}r(1999)]%
        {cramer1999mathematical}
\bibfield{author}{\bibinfo{person}{Harald Cram{\'e}r}.} \bibinfo{year}{1999}\natexlab{}.
\newblock \bibinfo{booktitle}{\emph{Mathematical methods of statistics}}. Vol.~\bibinfo{volume}{43}.
\newblock \bibinfo{publisher}{Princeton university press}.
\newblock


\bibitem[Dasgupta et~al\mbox{.}(2022)]%
        {dasgupta2022framework}
\bibfield{author}{\bibinfo{person}{Sanjoy Dasgupta}, \bibinfo{person}{Nave Frost}, {and} \bibinfo{person}{Michal Moshkovitz}.} \bibinfo{year}{2022}\natexlab{}.
\newblock \showarticletitle{Framework for evaluating faithfulness of local explanations}. In \bibinfo{booktitle}{\emph{International Conference on Machine Learning}}. PMLR, \bibinfo{pages}{4794--4815}.
\newblock


\bibitem[Deutch et~al\mbox{.}(2022)]%
        {FEDEX}
\bibfield{author}{\bibinfo{person}{Daniel Deutch}, \bibinfo{person}{Amir Gilad}, \bibinfo{person}{Tova Milo}, \bibinfo{person}{Amit Mualem}, {and} \bibinfo{person}{Amit Somech}.} \bibinfo{year}{2022}\natexlab{}.
\newblock \showarticletitle{FEDEX: An Explainability Framework for Data Exploration Steps}.
\newblock \bibinfo{journal}{\emph{Proceedings of the VLDB Endowment}} \bibinfo{volume}{15}, \bibinfo{number}{13} (\bibinfo{year}{2022}), \bibinfo{pages}{3854--3868}.
\newblock


\bibitem[Ding et~al\mbox{.}(2017)]%
        {ding2017collecting}
\bibfield{author}{\bibinfo{person}{Bolin Ding}, \bibinfo{person}{Janardhan Kulkarni}, {and} \bibinfo{person}{Sergey Yekhanin}.} \bibinfo{year}{2017}\natexlab{}.
\newblock \showarticletitle{Collecting Telemetry Data Privately}. In \bibinfo{booktitle}{\emph{Proceedings of the 31st International Conference on Neural Information Processing Systems}} (Long Beach, California, USA) \emph{(\bibinfo{series}{NIPS'17})}. \bibinfo{publisher}{Curran Associates Inc.}, \bibinfo{address}{Red Hook, NY, USA}, \bibinfo{pages}{3574–3583}.
\newblock
\showISBNx{9781510860964}


\bibitem[Dong et~al\mbox{.}(2020)]%
        {dong2020optimal}
\bibfield{author}{\bibinfo{person}{Jinshuo Dong}, \bibinfo{person}{David Durfee}, {and} \bibinfo{person}{Ryan Rogers}.} \bibinfo{year}{2020}\natexlab{}.
\newblock \showarticletitle{Optimal differential privacy composition for exponential mechanisms}. In \bibinfo{booktitle}{\emph{International Conference on Machine Learning}}. PMLR, \bibinfo{pages}{2597--2606}.
\newblock


\bibitem[Durfee and Rogers(2021)]%
        {durfee2021oneshot}
\bibfield{author}{\bibinfo{person}{D. Durfee} {and} \bibinfo{person}{R. Rogers}.} \bibinfo{year}{2021}\natexlab{}.
\newblock \bibinfo{title}{One-shot DP top-k mechanisms}.
\newblock \bibinfo{howpublished}{DifferentialPrivacy.org}.
\newblock
\newblock
\shownote{\url{https://differentialprivacy.org/one-shot-top-k/}}.


\bibitem[Durfee and Rogers(2019)]%
        {durfee2019practical}
\bibfield{author}{\bibinfo{person}{David Durfee} {and} \bibinfo{person}{Ryan~M Rogers}.} \bibinfo{year}{2019}\natexlab{}.
\newblock \showarticletitle{Practical differentially private top-k selection with pay-what-you-get composition}.
\newblock \bibinfo{journal}{\emph{Advances in Neural Information Processing Systems}}  \bibinfo{volume}{32} (\bibinfo{year}{2019}).
\newblock


\bibitem[Dwork(2006)]%
        {dwork2006differential}
\bibfield{author}{\bibinfo{person}{Cynthia Dwork}.} \bibinfo{year}{2006}\natexlab{}.
\newblock \showarticletitle{Differential privacy}. In \bibinfo{booktitle}{\emph{International colloquium on automata, languages, and programming}}. Springer, \bibinfo{pages}{1--12}.
\newblock


\bibitem[Dwork(2019)]%
        {dwork2019differential}
\bibfield{author}{\bibinfo{person}{Cynthia Dwork}.} \bibinfo{year}{2019}\natexlab{}.
\newblock \showarticletitle{Differential Privacy and the US Census}. In \bibinfo{booktitle}{\emph{Proceedings of the 38th ACM SIGMOD-SIGACT-SIGAI Symposium on Principles of Database Systems}} (Amsterdam, Netherlands) \emph{(\bibinfo{series}{PODS '19})}. \bibinfo{publisher}{Association for Computing Machinery}, \bibinfo{address}{New York, NY, USA}, \bibinfo{pages}{1}.
\newblock
\showISBNx{9781450362276}
\urldef\tempurl%
\url{https://doi.org/10.1145/3294052.3322188}
\showDOI{\tempurl}


\bibitem[Dwork et~al\mbox{.}(2006)]%
        {dwork2006calibrating}
\bibfield{author}{\bibinfo{person}{Cynthia Dwork}, \bibinfo{person}{Frank McSherry}, \bibinfo{person}{Kobbi Nissim}, {and} \bibinfo{person}{Adam Smith}.} \bibinfo{year}{2006}\natexlab{}.
\newblock \showarticletitle{Calibrating noise to sensitivity in private data analysis}. In \bibinfo{booktitle}{\emph{Theory of Cryptography Conference}}. Springer, \bibinfo{pages}{265--284}.
\newblock


\bibitem[Dwork et~al\mbox{.}(2014)]%
        {dwork2014algorithmic}
\bibfield{author}{\bibinfo{person}{Cynthia Dwork}, \bibinfo{person}{Aaron Roth}, {et~al\mbox{.}}} \bibinfo{year}{2014}\natexlab{}.
\newblock \showarticletitle{The algorithmic foundations of differential privacy}.
\newblock \bibinfo{journal}{\emph{Foundations and Trends{\textregistered} in Theoretical Computer Science}} \bibinfo{volume}{9}, \bibinfo{number}{3--4} (\bibinfo{year}{2014}), \bibinfo{pages}{211--407}.
\newblock


\bibitem[Erlingsson et~al\mbox{.}(2014)]%
        {erlingsson2014rappor}
\bibfield{author}{\bibinfo{person}{\'{U}lfar Erlingsson}, \bibinfo{person}{Vasyl Pihur}, {and} \bibinfo{person}{Aleksandra Korolova}.} \bibinfo{year}{2014}\natexlab{}.
\newblock \showarticletitle{RAPPOR: Randomized Aggregatable Privacy-Preserving Ordinal Response}. In \bibinfo{booktitle}{\emph{Proceedings of the 2014 ACM SIGSAC Conference on Computer and Communications Security}} (Scottsdale, Arizona, USA) \emph{(\bibinfo{series}{CCS '14})}. \bibinfo{publisher}{Association for Computing Machinery}, \bibinfo{address}{New York, NY, USA}, \bibinfo{pages}{1054–1067}.
\newblock
\showISBNx{9781450329576}
\urldef\tempurl%
\url{https://doi.org/10.1145/2660267.2660348}
\showDOI{\tempurl}


\bibitem[Esfandiari et~al\mbox{.}(2022)]%
        {esfandiari2022almost}
\bibfield{author}{\bibinfo{person}{Hossein Esfandiari}, \bibinfo{person}{Vahab Mirrokni}, {and} \bibinfo{person}{Shyam Narayanan}.} \bibinfo{year}{2022}\natexlab{}.
\newblock \showarticletitle{Almost tight approximation algorithms for explainable clustering}. In \bibinfo{booktitle}{\emph{Proceedings of the 2022 Annual ACM-SIAM Symposium on Discrete Algorithms (SODA)}}. SIAM, \bibinfo{pages}{2641--2663}.
\newblock


\bibitem[Gabriele and Chiasson(2020)]%
        {GabrieleC20}
\bibfield{author}{\bibinfo{person}{Sandra Gabriele} {and} \bibinfo{person}{Sonia Chiasson}.} \bibinfo{year}{2020}\natexlab{}.
\newblock \showarticletitle{Understanding Fitness Tracker Users' Security and Privacy Knowledge, Attitudes and Behaviours}. In \bibinfo{booktitle}{\emph{{CHI} '20: {CHI} Conference on Human Factors in Computing Systems, Honolulu, HI, USA, April 25-30, 2020}}, \bibfield{editor}{\bibinfo{person}{Regina Bernhaupt}, \bibinfo{person}{Florian~'Floyd' Mueller}, \bibinfo{person}{David Verweij}, \bibinfo{person}{Josh Andres}, \bibinfo{person}{Joanna McGrenere}, \bibinfo{person}{Andy Cockburn}, \bibinfo{person}{Ignacio Avellino}, \bibinfo{person}{Alix Goguey}, \bibinfo{person}{Pernille Bj{\o}n}, \bibinfo{person}{Shengdong Zhao}, \bibinfo{person}{Briane~Paul Samson}, {and} \bibinfo{person}{Rafal Kocielnik}} (Eds.). \bibinfo{publisher}{{ACM}}, \bibinfo{pages}{1--12}.
\newblock
\urldef\tempurl%
\url{https://doi.org/10.1145/3313831.3376651}
\showDOI{\tempurl}


\bibitem[Gamlath et~al\mbox{.}(2021)]%
        {gamlath2021nearly}
\bibfield{author}{\bibinfo{person}{Buddhima Gamlath}, \bibinfo{person}{Xinrui Jia}, \bibinfo{person}{Adam Polak}, {and} \bibinfo{person}{Ola Svensson}.} \bibinfo{year}{2021}\natexlab{}.
\newblock \showarticletitle{Nearly-tight and oblivious algorithms for explainable clustering}.
\newblock \bibinfo{journal}{\emph{Advances in Neural Information Processing Systems}}  \bibinfo{volume}{34} (\bibinfo{year}{2021}), \bibinfo{pages}{28929--28939}.
\newblock


\bibitem[Ge et~al\mbox{.}(2019)]%
        {ge2019apex}
\bibfield{author}{\bibinfo{person}{Chang Ge}, \bibinfo{person}{Xi He}, \bibinfo{person}{Ihab~F Ilyas}, {and} \bibinfo{person}{Ashwin Machanavajjhala}.} \bibinfo{year}{2019}\natexlab{}.
\newblock \showarticletitle{Apex: Accuracy-aware differentially private data exploration}. In \bibinfo{booktitle}{\emph{Proceedings of the 2019 International Conference on Management of Data}}. \bibinfo{pages}{177--194}.
\newblock


\bibitem[Ghazi et~al\mbox{.}(2020)]%
        {ghazi2020differentially}
\bibfield{author}{\bibinfo{person}{Badih Ghazi}, \bibinfo{person}{Ravi Kumar}, {and} \bibinfo{person}{Pasin Manurangsi}.} \bibinfo{year}{2020}\natexlab{}.
\newblock \showarticletitle{Differentially private clustering: Tight approximation ratios}.
\newblock \bibinfo{journal}{\emph{Advances in Neural Information Processing Systems}}  \bibinfo{volume}{33} (\bibinfo{year}{2020}), \bibinfo{pages}{4040--4054}.
\newblock


\bibitem[Ghosh et~al\mbox{.}(2009)]%
        {ghosh2009universally}
\bibfield{author}{\bibinfo{person}{Arpita Ghosh}, \bibinfo{person}{Tim Roughgarden}, {and} \bibinfo{person}{Mukund Sundararajan}.} \bibinfo{year}{2009}\natexlab{}.
\newblock \showarticletitle{Universally utility-maximizing privacy mechanisms}. In \bibinfo{booktitle}{\emph{Proceedings of the forty-first annual ACM symposium on Theory of computing}}. \bibinfo{pages}{351--360}.
\newblock


\bibitem[Gupta et~al\mbox{.}(2010)]%
        {gupta2010differentially}
\bibfield{author}{\bibinfo{person}{Anupam Gupta}, \bibinfo{person}{Katrina Ligett}, \bibinfo{person}{Frank McSherry}, \bibinfo{person}{Aaron Roth}, {and} \bibinfo{person}{Kunal Talwar}.} \bibinfo{year}{2010}\natexlab{}.
\newblock \showarticletitle{Differentially private combinatorial optimization}. In \bibinfo{booktitle}{\emph{Proceedings of the twenty-first annual ACM-SIAM symposium on Discrete Algorithms}}. SIAM, \bibinfo{pages}{1106--1125}.
\newblock


\bibitem[Harder et~al\mbox{.}(2020)]%
        {harder2020interpretable}
\bibfield{author}{\bibinfo{person}{Frederik Harder}, \bibinfo{person}{Matthias Bauer}, {and} \bibinfo{person}{Mijung Park}.} \bibinfo{year}{2020}\natexlab{}.
\newblock \showarticletitle{Interpretable and differentially private predictions}. In \bibinfo{booktitle}{\emph{Proceedings of the AAAI Conference on Artificial Intelligence}}, Vol.~\bibinfo{volume}{34}. \bibinfo{pages}{4083--4090}.
\newblock


\bibitem[Hay et~al\mbox{.}(2010)]%
        {hay2010boosting}
\bibfield{author}{\bibinfo{person}{Michael Hay}, \bibinfo{person}{Vibhor Rastogi}, \bibinfo{person}{Gerome Miklau}, {and} \bibinfo{person}{Dan Suciu}.} \bibinfo{year}{2010}\natexlab{}.
\newblock \showarticletitle{Boosting the Accuracy of Differentially Private Histograms Through Consistency}.
\newblock \bibinfo{journal}{\emph{Proceedings of the VLDB Endowment}} \bibinfo{volume}{3}, \bibinfo{number}{1} (\bibinfo{year}{2010}).
\newblock


\bibitem[Hilderman and Hamilton(2013)]%
        {hilderman2013knowledge}
\bibfield{author}{\bibinfo{person}{Robert~J Hilderman} {and} \bibinfo{person}{Howard~J Hamilton}.} \bibinfo{year}{2013}\natexlab{}.
\newblock \bibinfo{booktitle}{\emph{Knowledge Discovery and Measures of Interest}}. Vol.~\bibinfo{volume}{638}.
\newblock \bibinfo{publisher}{Springer Science \& Business Media}.
\newblock


\bibitem[Holohan et~al\mbox{.}(2019)]%
        {holohan2019diffprivlib}
\bibfield{author}{\bibinfo{person}{Naoise Holohan}, \bibinfo{person}{Stefano Braghin}, \bibinfo{person}{P{\'o}l Mac~Aonghusa}, {and} \bibinfo{person}{Killian Levacher}.} \bibinfo{year}{2019}\natexlab{}.
\newblock \showarticletitle{Diffprivlib: the IBM differential privacy library}.
\newblock \bibinfo{journal}{\emph{arXiv preprint arXiv:1907.02444}} (\bibinfo{year}{2019}).
\newblock


\bibitem[Hu et~al\mbox{.}(2024)]%
        {hu2024interpretable}
\bibfield{author}{\bibinfo{person}{Lianyu Hu}, \bibinfo{person}{Mudi Jiang}, \bibinfo{person}{Junjie Dong}, \bibinfo{person}{Xinying Liu}, {and} \bibinfo{person}{Zengyou He}.} \bibinfo{year}{2024}\natexlab{}.
\newblock \showarticletitle{Interpretable Clustering: A Survey}.
\newblock \bibinfo{journal}{\emph{arXiv preprint arXiv:2409.00743}} (\bibinfo{year}{2024}).
\newblock


\bibitem[Johnson et~al\mbox{.}(2020)]%
        {johnson2020chorus}
\bibfield{author}{\bibinfo{person}{Noah Johnson}, \bibinfo{person}{Joseph~P Near}, \bibinfo{person}{Joseph~M Hellerstein}, {and} \bibinfo{person}{Dawn Song}.} \bibinfo{year}{2020}\natexlab{}.
\newblock \showarticletitle{Chorus: a programming framework for building scalable differential privacy mechanisms}. In \bibinfo{booktitle}{\emph{2020 IEEE European Symposium on Security and Privacy (EuroS\&P)}}. IEEE, \bibinfo{pages}{535--551}.
\newblock


\bibitem[Johnson et~al\mbox{.}(2018)]%
        {FLEX}
\bibfield{author}{\bibinfo{person}{Noah Johnson}, \bibinfo{person}{Joseph~P Near}, {and} \bibinfo{person}{Dawn Song}.} \bibinfo{year}{2018}\natexlab{}.
\newblock \showarticletitle{Towards practical differential privacy for SQL queries}.
\newblock \bibinfo{journal}{\emph{Proceedings of the VLDB Endowment}} \bibinfo{volume}{11}, \bibinfo{number}{5} (\bibinfo{year}{2018}), \bibinfo{pages}{526--539}.
\newblock


\bibitem[Jordon et~al\mbox{.}(2019)]%
        {jordon2019differentially}
\bibfield{author}{\bibinfo{person}{James Jordon}, \bibinfo{person}{Jinsung Yoon}, {and} \bibinfo{person}{Mihaela van~der Schaar}.} \bibinfo{year}{2019}\natexlab{}.
\newblock \showarticletitle{Differentially private model personalization}. In \bibinfo{booktitle}{\emph{Proceedings of the 36th International Conference on Machine Learning}}. PMLR.
\newblock


\bibitem[Kotsogiannis et~al\mbox{.}(2019)]%
        {privatesql}
\bibfield{author}{\bibinfo{person}{Ios Kotsogiannis}, \bibinfo{person}{Yuchao Tao}, \bibinfo{person}{Xi He}, \bibinfo{person}{Maryam Fanaeepour}, \bibinfo{person}{Ashwin Machanavajjhala}, \bibinfo{person}{Michael Hay}, {and} \bibinfo{person}{Gerome Miklau}.} \bibinfo{year}{2019}\natexlab{}.
\newblock \showarticletitle{Privatesql: a differentially private sql query engine}.
\newblock \bibinfo{journal}{\emph{Proceedings of the VLDB Endowment}} \bibinfo{volume}{12}, \bibinfo{number}{11} (\bibinfo{year}{2019}), \bibinfo{pages}{1371--1384}.
\newblock


\bibitem[Lakkaraju et~al\mbox{.}(2016)]%
        {lakkaraju2016interpretable}
\bibfield{author}{\bibinfo{person}{Himabindu Lakkaraju}, \bibinfo{person}{Stephen~H Bach}, {and} \bibinfo{person}{Jure Leskovec}.} \bibinfo{year}{2016}\natexlab{}.
\newblock \showarticletitle{Interpretable decision sets: A joint framework for description and prediction}. In \bibinfo{booktitle}{\emph{Proceedings of the 22nd ACM SIGKDD international conference on knowledge discovery and data mining}}. \bibinfo{pages}{1675--1684}.
\newblock


\bibitem[Lee et~al\mbox{.}(2021)]%
        {lee2021lux}
\bibfield{author}{\bibinfo{person}{Doris Jung-Lin Lee}, \bibinfo{person}{Dixin Tang}, \bibinfo{person}{Kunal Agarwal}, \bibinfo{person}{Thyne Boonmark}, \bibinfo{person}{Caitlyn Chen}, \bibinfo{person}{Jake Kang}, \bibinfo{person}{Ujjaini Mukhopadhyay}, \bibinfo{person}{Jerry Song}, \bibinfo{person}{Micah Yong}, \bibinfo{person}{Marti~A Hearst}, {et~al\mbox{.}}} \bibinfo{year}{2021}\natexlab{}.
\newblock \showarticletitle{Lux: always-on visualization recommendations for exploratory dataframe workflows}.
\newblock \bibinfo{journal}{\emph{PVLDB}} \bibinfo{volume}{15}, \bibinfo{number}{3} (\bibinfo{year}{2021}), \bibinfo{pages}{727--738}.
\newblock


\bibitem[Levin and Peres(2017)]%
        {levin2017markov}
\bibfield{author}{\bibinfo{person}{David~A Levin} {and} \bibinfo{person}{Yuval Peres}.} \bibinfo{year}{2017}\natexlab{}.
\newblock \bibinfo{booktitle}{\emph{Markov chains and mixing times}}. Vol.~\bibinfo{volume}{107}.
\newblock \bibinfo{publisher}{American Mathematical Soc.}
\newblock


\bibitem[Lin and Kifer(2013)]%
        {lin2013information}
\bibfield{author}{\bibinfo{person}{Bing-Rong Lin} {and} \bibinfo{person}{Daniel Kifer}.} \bibinfo{year}{2013}\natexlab{}.
\newblock \showarticletitle{Information preservation in statistical privacy and bayesian estimation of unattributed histograms}. In \bibinfo{booktitle}{\emph{Proceedings of the 2013 ACM SIGMOD International Conference on Management of Data}}. \bibinfo{pages}{677--688}.
\newblock


\bibitem[Lin(1991)]%
        {lin1991divergence}
\bibfield{author}{\bibinfo{person}{Jianhua Lin}.} \bibinfo{year}{1991}\natexlab{}.
\newblock \showarticletitle{Divergence measures based on the Shannon entropy}.
\newblock \bibinfo{journal}{\emph{IEEE Transactions on Information theory}} \bibinfo{volume}{37}, \bibinfo{number}{1} (\bibinfo{year}{1991}), \bibinfo{pages}{145--151}.
\newblock


\bibitem[Luo et~al\mbox{.}(2018)]%
        {luo2018deepeye}
\bibfield{author}{\bibinfo{person}{Yuyu Luo}, \bibinfo{person}{Xuedi Qin}, \bibinfo{person}{Nan Tang}, {and} \bibinfo{person}{Guoliang Li}.} \bibinfo{year}{2018}\natexlab{}.
\newblock \showarticletitle{DeepEye: Towards Automatic Data Visualization}. ICDE.
\newblock


\bibitem[Lv and Chen(2023)]%
        {lv2023data}
\bibfield{author}{\bibinfo{person}{Ge Lv} {and} \bibinfo{person}{Lei Chen}.} \bibinfo{year}{2023}\natexlab{}.
\newblock \showarticletitle{On data-aware global explainability of graph neural networks}.
\newblock \bibinfo{journal}{\emph{Proceedings of the VLDB Endowment}} \bibinfo{volume}{16}, \bibinfo{number}{11} (\bibinfo{year}{2023}), \bibinfo{pages}{3447--3460}.
\newblock


\bibitem[Makarychev and Shan(2021)]%
        {makarychev2021near}
\bibfield{author}{\bibinfo{person}{Konstantin Makarychev} {and} \bibinfo{person}{Liren Shan}.} \bibinfo{year}{2021}\natexlab{}.
\newblock \showarticletitle{Near-optimal algorithms for explainable k-medians and k-means}. In \bibinfo{booktitle}{\emph{International Conference on Machine Learning}}. PMLR, \bibinfo{pages}{7358--7367}.
\newblock


\bibitem[Makarychev and Shan(2022)]%
        {makarychev2022explainable}
\bibfield{author}{\bibinfo{person}{Konstantin Makarychev} {and} \bibinfo{person}{Liren Shan}.} \bibinfo{year}{2022}\natexlab{}.
\newblock \showarticletitle{Explainable k-means: don’t be greedy, plant bigger trees!}. In \bibinfo{booktitle}{\emph{Proceedings of the 54th Annual ACM SIGACT Symposium on Theory of Computing}}. \bibinfo{pages}{1629--1642}.
\newblock


\bibitem[McKenna et~al\mbox{.}(2021)]%
        {mckenna2021winning}
\bibfield{author}{\bibinfo{person}{Ryan McKenna}, \bibinfo{person}{Gerome Miklau}, {and} \bibinfo{person}{Daniel Sheldon}.} \bibinfo{year}{2021}\natexlab{}.
\newblock \showarticletitle{Winning the NIST Contest: A scalable and general approach to differentially private synthetic data}.
\newblock \bibinfo{journal}{\emph{arXiv preprint arXiv:2108.04978}} (\bibinfo{year}{2021}).
\newblock


\bibitem[McSherry and Talwar(2007)]%
        {mcsherry2007mechanism}
\bibfield{author}{\bibinfo{person}{Frank McSherry} {and} \bibinfo{person}{Kunal Talwar}.} \bibinfo{year}{2007}\natexlab{}.
\newblock \showarticletitle{Mechanism design via differential privacy}. In \bibinfo{booktitle}{\emph{48th Annual IEEE Symposium on Foundations of Computer Science (FOCS'07)}}. IEEE, \bibinfo{pages}{94--103}.
\newblock


\bibitem[McSherry(2009)]%
        {mcsherry2009privacy}
\bibfield{author}{\bibinfo{person}{Frank~D McSherry}.} \bibinfo{year}{2009}\natexlab{}.
\newblock \showarticletitle{Privacy integrated queries: an extensible platform for privacy-preserving data analysis}. In \bibinfo{booktitle}{\emph{Proceedings of the 2009 ACM SIGMOD International Conference on Management of data}}. \bibinfo{pages}{19--30}.
\newblock


\bibitem[Meek et~al\mbox{.}(2001)]%
        {census_1990}
\bibfield{author}{\bibinfo{person}{Chris Meek}, \bibinfo{person}{Bo Thiesson}, {and} \bibinfo{person}{David Heckerman}.} \bibinfo{year}{2001}\natexlab{}.
\newblock \bibinfo{title}{US Census Data (1990)}.
\newblock \bibinfo{howpublished}{UCI Machine Learning Repository}.
\newblock
\newblock
\shownote{DOI: https://doi.org/10.24432/C5VP42}.


\bibitem[Mochaourab et~al\mbox{.}(2021)]%
        {mochaourab2021robust}
\bibfield{author}{\bibinfo{person}{Rami Mochaourab}, \bibinfo{person}{Sugandh Sinha}, \bibinfo{person}{Stanley Greenstein}, {and} \bibinfo{person}{Panagiotis Papapetrou}.} \bibinfo{year}{2021}\natexlab{}.
\newblock \showarticletitle{Robust counterfactual explanations for privacy-preserving SVM}. In \bibinfo{booktitle}{\emph{International Conference on Machine Learning (ICML 2021), Workshop on Socially Responsible Machine Learning}}.
\newblock


\bibitem[Moshkovitz et~al\mbox{.}(2020)]%
        {moshkovitz2020explainable}
\bibfield{author}{\bibinfo{person}{Michal Moshkovitz}, \bibinfo{person}{Sanjoy Dasgupta}, \bibinfo{person}{Cyrus Rashtchian}, {and} \bibinfo{person}{Nave Frost}.} \bibinfo{year}{2020}\natexlab{}.
\newblock \showarticletitle{Explainable k-means and k-medians clustering}. In \bibinfo{booktitle}{\emph{International conference on machine learning}}. PMLR, \bibinfo{pages}{7055--7065}.
\newblock


\bibitem[Mothilal et~al\mbox{.}(2020)]%
        {mothilal2020explaining}
\bibfield{author}{\bibinfo{person}{Ramaravind~K Mothilal}, \bibinfo{person}{Amit Sharma}, {and} \bibinfo{person}{Chenhao Tan}.} \bibinfo{year}{2020}\natexlab{}.
\newblock \showarticletitle{Explaining machine learning classifiers through diverse counterfactual explanations}. In \bibinfo{booktitle}{\emph{Proceedings of the 2020 conference on fairness, accountability, and transparency}}. \bibinfo{pages}{607--617}.
\newblock


\bibitem[Nguyen(2018)]%
        {nguyen2018privacy}
\bibfield{author}{\bibinfo{person}{Huu~Hiep Nguyen}.} \bibinfo{year}{2018}\natexlab{}.
\newblock \showarticletitle{Privacy-preserving mechanisms for k-modes clustering}.
\newblock \bibinfo{journal}{\emph{Computers \& Security}}  \bibinfo{volume}{78} (\bibinfo{year}{2018}), \bibinfo{pages}{60--75}.
\newblock


\bibitem[Nguyen et~al\mbox{.}(2021)]%
        {nguyen2021differentially}
\bibfield{author}{\bibinfo{person}{Huy~L Nguyen}, \bibinfo{person}{Anamay Chaturvedi}, {and} \bibinfo{person}{Eric~Z Xu}.} \bibinfo{year}{2021}\natexlab{}.
\newblock \showarticletitle{Differentially private k-means via exponential mechanism and max cover}. In \bibinfo{booktitle}{\emph{Proceedings of the AAAI conference on artificial intelligence}}, Vol.~\bibinfo{volume}{35}. \bibinfo{pages}{9101--9108}.
\newblock


\bibitem[Nguyen et~al\mbox{.}(2024)]%
        {nguyen2024survey}
\bibfield{author}{\bibinfo{person}{Thanh~Tam Nguyen}, \bibinfo{person}{Thanh~Trung Huynh}, \bibinfo{person}{Zhao Ren}, \bibinfo{person}{Thanh~Toan Nguyen}, \bibinfo{person}{Phi~Le Nguyen}, \bibinfo{person}{Hongzhi Yin}, {and} \bibinfo{person}{Quoc Viet~Hung Nguyen}.} \bibinfo{year}{2024}\natexlab{}.
\newblock \showarticletitle{A survey of privacy-preserving model explanations: Privacy risks, attacks, and countermeasures}.
\newblock \bibinfo{journal}{\emph{arXiv preprint arXiv:2404.00673}} (\bibinfo{year}{2024}).
\newblock


\bibitem[Overflow(2018)]%
        {stackoverflow_survey}
\bibfield{author}{\bibinfo{person}{Stack Overflow}.} \bibinfo{year}{2018}\natexlab{}.
\newblock \bibinfo{title}{Stack Overflow Annual Developer Survey}.
\newblock \bibinfo{howpublished}{\url{https://survey.stackoverflow.co}}.
\newblock


\bibitem[Oyewole et~al\mbox{.}(2024)]%
        {oyewole2024data}
\bibfield{author}{\bibinfo{person}{Adedoyin~Tolulope Oyewole}, \bibinfo{person}{Bisola~Beatrice Oguejiofor}, \bibinfo{person}{Nkechi~Emmanuella Eneh}, \bibinfo{person}{Chidiogo~Uzoamaka Akpuokwe}, {and} \bibinfo{person}{Seun~Solomon Bakare}.} \bibinfo{year}{2024}\natexlab{}.
\newblock \showarticletitle{Data privacy laws and their impact on financial technology companies: a review}.
\newblock \bibinfo{journal}{\emph{Computer Science \& IT Research Journal}} \bibinfo{volume}{5}, \bibinfo{number}{3} (\bibinfo{year}{2024}), \bibinfo{pages}{628--650}.
\newblock


\bibitem[Patel et~al\mbox{.}(2022)]%
        {patel2022model}
\bibfield{author}{\bibinfo{person}{Neel Patel}, \bibinfo{person}{Reza Shokri}, {and} \bibinfo{person}{Yair Zick}.} \bibinfo{year}{2022}\natexlab{}.
\newblock \showarticletitle{Model explanations with differential privacy}. In \bibinfo{booktitle}{\emph{Proceedings of the 2022 ACM Conference on Fairness, Accountability, and Transparency}}. \bibinfo{pages}{1895--1904}.
\newblock


\bibitem[Qardaji et~al\mbox{.}(2013)]%
        {qardaji2013understanding}
\bibfield{author}{\bibinfo{person}{Wahbeh Qardaji}, \bibinfo{person}{Weining Yang}, {and} \bibinfo{person}{Ninghui Li}.} \bibinfo{year}{2013}\natexlab{}.
\newblock \showarticletitle{Understanding hierarchical methods for differentially private histograms}.
\newblock \bibinfo{journal}{\emph{Proceedings of the VLDB Endowment}} \bibinfo{volume}{6}, \bibinfo{number}{14} (\bibinfo{year}{2013}), \bibinfo{pages}{1954--1965}.
\newblock


\bibitem[Rawal and Lakkaraju(2020)]%
        {rawal2020beyond}
\bibfield{author}{\bibinfo{person}{Kaivalya Rawal} {and} \bibinfo{person}{Himabindu Lakkaraju}.} \bibinfo{year}{2020}\natexlab{}.
\newblock \showarticletitle{Beyond individualized recourse: Interpretable and interactive summaries of actionable recourses}.
\newblock \bibinfo{journal}{\emph{Advances in Neural Information Processing Systems}}  \bibinfo{volume}{33} (\bibinfo{year}{2020}), \bibinfo{pages}{12187--12198}.
\newblock


\bibitem[Sarawagi et~al\mbox{.}(1998)]%
        {sarawagi1998discovery}
\bibfield{author}{\bibinfo{person}{Sunita Sarawagi}, \bibinfo{person}{Rakesh Agrawal}, {and} \bibinfo{person}{Nimrod Megiddo}.} \bibinfo{year}{1998}\natexlab{}.
\newblock \showarticletitle{Discovery-driven exploration of OLAP data cubes}. In \bibinfo{booktitle}{\emph{EDBT}}.
\newblock


\bibitem[Stemmer and Kaplan(2018)]%
        {stemmer2018differentially}
\bibfield{author}{\bibinfo{person}{Uri Stemmer} {and} \bibinfo{person}{Haim Kaplan}.} \bibinfo{year}{2018}\natexlab{}.
\newblock \showarticletitle{Differentially private k-means with constant multiplicative error}.
\newblock \bibinfo{journal}{\emph{Advances in Neural Information Processing Systems}}  \bibinfo{volume}{31} (\bibinfo{year}{2018}).
\newblock


\bibitem[Strack et~al\mbox{.}(2014)]%
        {diabetes_paper}
\bibfield{author}{\bibinfo{person}{Beata Strack}, \bibinfo{person}{Jonathan~P DeShazo}, \bibinfo{person}{Chris Gennings}, \bibinfo{person}{Juan~L Olmo}, \bibinfo{person}{Sebastian Ventura}, \bibinfo{person}{Krzysztof~J Cios}, {and} \bibinfo{person}{John~N Clore}.} \bibinfo{year}{2014}\natexlab{}.
\newblock \showarticletitle{Impact of HbA1c measurement on hospital readmission rates: analysis of 70,000 clinical database patient records}.
\newblock \bibinfo{journal}{\emph{BioMed research international}} \bibinfo{volume}{2014}, \bibinfo{number}{1} (\bibinfo{year}{2014}), \bibinfo{pages}{781670}.
\newblock


\bibitem[Su et~al\mbox{.}(2016)]%
        {su2016differentially}
\bibfield{author}{\bibinfo{person}{Dong Su}, \bibinfo{person}{Jianneng Cao}, \bibinfo{person}{Ninghui Li}, \bibinfo{person}{Elisa Bertino}, {and} \bibinfo{person}{Hongxia Jin}.} \bibinfo{year}{2016}\natexlab{}.
\newblock \showarticletitle{Differentially private k-means clustering}. In \bibinfo{booktitle}{\emph{Proceedings of the sixth ACM conference on data and application security and privacy}}. \bibinfo{pages}{26--37}.
\newblock


\bibitem[Su et~al\mbox{.}(2017)]%
        {su2017differentially}
\bibfield{author}{\bibinfo{person}{Dong Su}, \bibinfo{person}{Jianneng Cao}, \bibinfo{person}{Ninghui Li}, \bibinfo{person}{Elisa Bertino}, \bibinfo{person}{Min Lyu}, {and} \bibinfo{person}{Hongxia Jin}.} \bibinfo{year}{2017}\natexlab{}.
\newblock \showarticletitle{Differentially private k-means clustering and a hybrid approach to private optimization}.
\newblock \bibinfo{journal}{\emph{ACM Transactions on Privacy and Security (TOPS)}} \bibinfo{volume}{20}, \bibinfo{number}{4} (\bibinfo{year}{2017}), \bibinfo{pages}{1--33}.
\newblock


\bibitem[Tang et~al\mbox{.}(2017a)]%
        {tang2017extracting}
\bibfield{author}{\bibinfo{person}{Bo Tang}, \bibinfo{person}{Shi Han}, \bibinfo{person}{Man~Lung Yiu}, \bibinfo{person}{Rui Ding}, {and} \bibinfo{person}{Dongmei Zhang}.} \bibinfo{year}{2017}\natexlab{a}.
\newblock \showarticletitle{Extracting top-k insights from multi-dimensional data}. In \bibinfo{booktitle}{\emph{Proceedings of the 2017 ACM International Conference on Management of Data}}. \bibinfo{pages}{1509--1524}.
\newblock


\bibitem[Tang et~al\mbox{.}(2017b)]%
        {tang2017privacy}
\bibfield{author}{\bibinfo{person}{Jun Tang}, \bibinfo{person}{Aleksandra Korolova}, \bibinfo{person}{Xiaolong Bai}, \bibinfo{person}{Xueqiang Wang}, {and} \bibinfo{person}{XiaoFeng Wang}.} \bibinfo{year}{2017}\natexlab{b}.
\newblock \showarticletitle{Privacy Loss in Apple's Implementation of Differential Privacy on MacOS 10.12}.
\newblock \bibinfo{journal}{\emph{CoRR}}  \bibinfo{volume}{abs/1709.02753} (\bibinfo{year}{2017}).
\newblock
\showeprint[arXiv]{1709.02753}
\urldef\tempurl%
\url{http://arxiv.org/abs/1709.02753}
\showURL{%
\tempurl}


\bibitem[Tao et~al\mbox{.}(2022)]%
        {tao2022dpxplain}
\bibfield{author}{\bibinfo{person}{Yuchao Tao}, \bibinfo{person}{Amir Gilad}, \bibinfo{person}{Ashwin Machanavajjhala}, {and} \bibinfo{person}{Sudeepa Roy}.} \bibinfo{year}{2022}\natexlab{}.
\newblock \showarticletitle{DPXPlain: privately explaining aggregate query answers}.
\newblock \bibinfo{journal}{\emph{Proceedings of the VLDB Endowment}} \bibinfo{volume}{16}, \bibinfo{number}{1} (\bibinfo{year}{2022}), \bibinfo{pages}{113--126}.
\newblock


\bibitem[Tutay and Somech(2023)]%
        {Cluster-Explorer}
\bibfield{author}{\bibinfo{person}{Sariel Tutay} {and} \bibinfo{person}{Amit Somech}.} \bibinfo{year}{2023}\natexlab{}.
\newblock \showarticletitle{Cluster-Explorer: An interactive Framework for Explaining Black-Box Clustering Results}. In \bibinfo{booktitle}{\emph{Proceedings of the 32nd ACM International Conference on Information and Knowledge Management}} (Birmingham, United Kingdom) \emph{(\bibinfo{series}{CIKM '23})}. \bibinfo{publisher}{Association for Computing Machinery}, \bibinfo{address}{New York, NY, USA}, \bibinfo{pages}{5106–5110}.
\newblock
\showISBNx{9798400701245}
\urldef\tempurl%
\url{https://doi.org/10.1145/3583780.3614734}
\showDOI{\tempurl}


\bibitem[Vieira et~al\mbox{.}(2011)]%
        {vieira2011query}
\bibfield{author}{\bibinfo{person}{Marcos~R Vieira}, \bibinfo{person}{Humberto~L Razente}, \bibinfo{person}{Maria~CN Barioni}, \bibinfo{person}{Marios Hadjieleftheriou}, \bibinfo{person}{Divesh Srivastava}, \bibinfo{person}{Caetano Traina}, {and} \bibinfo{person}{Vassilis~J Tsotras}.} \bibinfo{year}{2011}\natexlab{}.
\newblock \showarticletitle{On query result diversification}. In \bibinfo{booktitle}{\emph{2011 IEEE 27th International Conference on Data Engineering}}. IEEE, \bibinfo{pages}{1163--1174}.
\newblock


\bibitem[Willmott and Matsuura(2005)]%
        {MAE}
\bibfield{author}{\bibinfo{person}{Cort~J Willmott} {and} \bibinfo{person}{Kenji Matsuura}.} \bibinfo{year}{2005}\natexlab{}.
\newblock \showarticletitle{Advantages of the mean absolute error (MAE) over the root mean square error (RMSE) in assessing average model performance}.
\newblock \bibinfo{journal}{\emph{Climate research}} \bibinfo{volume}{30}, \bibinfo{number}{1} (\bibinfo{year}{2005}), \bibinfo{pages}{79--82}.
\newblock


\bibitem[Wongsuphasawat et~al\mbox{.}(2016)]%
        {wongsuphasawat2016voyager}
\bibfield{author}{\bibinfo{person}{Kanit Wongsuphasawat}, \bibinfo{person}{Dominik Moritz}, \bibinfo{person}{Anushka Anand}, \bibinfo{person}{Jock Mackinlay}, \bibinfo{person}{Bill Howe}, {and} \bibinfo{person}{Jeffrey Heer}.} \bibinfo{year}{2016}\natexlab{}.
\newblock \showarticletitle{Voyager: Exploratory analysis via faceted browsing of visualization recommendations}.
\newblock \bibinfo{journal}{\emph{TVCG}} (\bibinfo{year}{2016}).
\newblock


\bibitem[Xiao et~al\mbox{.}(2012)]%
        {xiao2012dpcube}
\bibfield{author}{\bibinfo{person}{Yonghui Xiao}, \bibinfo{person}{Li Xiong}, \bibinfo{person}{Liyue Fan}, {and} \bibinfo{person}{Slawomir Goryczka}.} \bibinfo{year}{2012}\natexlab{}.
\newblock \showarticletitle{DPCube: Differentially private histogram release through multidimensional partitioning}.
\newblock \bibinfo{journal}{\emph{arXiv preprint arXiv:1202.5358}} (\bibinfo{year}{2012}).
\newblock


\bibitem[Xu et~al\mbox{.}(2013)]%
        {xu2013differentially}
\bibfield{author}{\bibinfo{person}{Jia Xu}, \bibinfo{person}{Zhenjie Zhang}, \bibinfo{person}{Xiaokui Xiao}, \bibinfo{person}{Yin Yang}, \bibinfo{person}{Ge Yu}, {and} \bibinfo{person}{Marianne Winslett}.} \bibinfo{year}{2013}\natexlab{}.
\newblock \showarticletitle{Differentially private histogram publication}.
\newblock \bibinfo{journal}{\emph{The VLDB journal}}  \bibinfo{volume}{22} (\bibinfo{year}{2013}), \bibinfo{pages}{797--822}.
\newblock


\bibitem[Youngmann et~al\mbox{.}(2022)]%
        {youngmann2022guided}
\bibfield{author}{\bibinfo{person}{Brit Youngmann}, \bibinfo{person}{Sihem Amer-Yahia}, {and} \bibinfo{person}{Aurelien Personnaz}.} \bibinfo{year}{2022}\natexlab{}.
\newblock \showarticletitle{Guided exploration of data summaries}.
\newblock \bibinfo{journal}{\emph{Proceedings of the VLDB Endowment (PVLDB)}} \bibinfo{volume}{15}, \bibinfo{number}{9} (\bibinfo{year}{2022}), \bibinfo{pages}{1798--1807}.
\newblock


\bibitem[Zhang et~al\mbox{.}(2017)]%
        {zhang2017privbayes}
\bibfield{author}{\bibinfo{person}{Jun Zhang}, \bibinfo{person}{Graham Cormode}, \bibinfo{person}{Cecilia~M Procopiuc}, \bibinfo{person}{Divesh Srivastava}, {and} \bibinfo{person}{Xiaokui Xiao}.} \bibinfo{year}{2017}\natexlab{}.
\newblock \showarticletitle{Privbayes: Private data release via bayesian networks}.
\newblock \bibinfo{journal}{\emph{ACM Transactions on Database Systems (TODS)}} \bibinfo{volume}{42}, \bibinfo{number}{4} (\bibinfo{year}{2017}), \bibinfo{pages}{1--41}.
\newblock


\end{thebibliography}

\ifpaper

\else
% \newpage
\appendix
\newpage
\nobalance

\section{Theorems and Proofs}
\label{appendix:proofs}
Quality functions for histogram-based explanations used in prior work on are closely related to the $L_1$ distance between histograms, viewed as vectors. This connection will also be used in our analysis. Hence, we now formally define these concepts.
A histogram $h_A(D)$ can be viewed as a $\abs{\dom(A)}$-dimensional vector with value $\cnt_{A=a}(D)$ in its $a$'th entry for every $a\in \dom(A)$.
The $L_1$-norm  of a vector $v\in\R^d$, denoted $\norm{v}_1$, is defined as $\sum_{i=1}^d\abs{v_i}$. For histogram vectors, we always have $\norm{h_A(D)}_1 = \abs{D}$, as the $L_1$-norm is simply the sum of counts of all domain elements. The following is implied immediately from the definition:
\begin{corollary}\label{lemma:interest_vector}
    For a dataset $D$ and a clustering function $f:\dom(R)\to C$, a cluster label $c$, and an attribute $A$, the interestingness score (\Cref{def:low-sens-interest}) is 
$$
\interest( D, f, c, A) = \frac{1}{2}\norm{ \hist_A(D_c) - \frac{|D_c|}{\abs{D}} \hist_A(D)}_1
$$
\end{corollary}

\begin{corollary}\label{lemma:div_vector}
    Let $D$ be a dataset, $f$ a clustering function, $A$ an attribute, and $D_{c}, D_{c'}$ non-empty clusters such that $\abs{D_c} \le \abs{D_{c'}}$.
    The pairwise-diversity score (\Cref{def:pairwise_diversity}) is 
    \[
      d(D, f, c, c' , A, A) = \frac{1}{2}\norm{ \hist_A(D_c) - \frac{|D_c|}{|D_{c'}|} \hist_A (D_{c'})}_1 
    \]
\end{corollary}

Our global quality measures are defined as the average of low-sensitivity functions. Consequently, our analysis  frequently uses the following Lemma
\begin{lemma}\label{lemma:convex-comb}
    For $i=1,\dots, m$ let $\alpha_i\in \R$ be a scalar and $f_i:\cD\to\R$ be a function with sensitivity $\Delta_i$. Then, the function $f = \sum_{i=1}^m \alpha_i \cdot f_i$ has sensitivity bounded by $\sum_{i=1}^m \abs{\alpha_i}\Delta_i$.
\end{lemma}
\begin{proof}
Let $D\nbr D'$ be two neighboring datasets. By the triangle inequality,
\begin{align*}
    \abs{f(D) - f(D')} &= \abs{ \sum_{i=1}^m \alpha_i \cdot f_i(D) - \sum_{i=1}^m \alpha_i \cdot f_i(D') }\\
                        & \le \sum_{i=1}^m \abs{\alpha_i}\cdot \abs{ f_i(D) -f_i(D')  }\\
                        & \le \sum_{i=1}^m \abs{\alpha_i}\Delta_i.
\end{align*}
    
\end{proof}

\subsection{Interestingness} 
\label{appendix:interest}
We prove next the high sensitivity of the total variation distance interestingness measure \eqref{eq:tvd}, restated below for convenience.
\begin{align*}
    \tvd(\proj_A(D),\proj_A(D_c)) = \frac{1}{2}\sum_{\mathclap{a\in\dom(A)}} \abs{ \frac{\cnt_{A=a}(D)}{\abs{ D}} - \frac{\cnt_{A=a}(D_c)}{|D_c|}}
\end{align*}
\IntHighSens*
\begin{proof}
Since the range bound is standard (see, e.g., \cite{levin2017markov}), we focus on the sensitivity lower bound.
Let $ D$ be a dataset of size $n\ge1$, and $A$  an attribute. Suppose that for all tuples $t\in  D$ it holds that $t[A]=a$ for some $a\in \dom(A)$. That is, $\cnt_{A=a}(D)=n$. Let $D_c\subseteq  D$ be a cluster of size $1$. In this case $ \tvd(\proj_A(D),\proj_A(D_c)) = 0$, as $\proj_A(D)$ and $\proj_A(D_c)$ define the same distribution (the value $a$ has probability $1$). 
    
    Now, let $ D' =  D\cup\set{t'}$ and $D'_c = D_c\cup\set{t'}$ for a tuple $t'$ with $t'[A]=a'$ for $a'\neq a$. We now have
    \begin{align*}
    2\cdot \tvd(\proj_A(D'),\proj_A(D'_c)) &= \abs{ \frac{\cnt_{A=a}(D')}{\abs{ D'}} - \frac{\cnt_{A=a}(D'_c)}{\abs{D'_c}}}\\
                &\qquad+ \abs{ \frac{\cnt_{A=a'}(D')}{\abs{ D'}} - \frac{\cnt_{A=a'}(D'_c)}{\abs{D'_c}}} \\
                & = \abs{\frac{n}{n+1} - \frac{1}{2}} + \abs{\frac{1}{n+1} - \frac{1}{2}} \\
                & = {\frac{n}{n+1} - \frac{1}{2}} + \frac{1}{2}-\frac{1}{n+1} \\
                & = 1 -\frac{2}{n+1}
    \end{align*}
Therefore, 
\begin{align*}
    \tvd(\proj_A(D'),\proj_A(D'_c)) = \frac{1}{2} - \frac{1}{n+1}   
\end{align*}
Since sensitivity is defined as the supremum over all datasets and dataset sizes, the lemma follows.
\end{proof}

Since previous work has also considered the Jensen-Shannon distance  \cite{lin1991divergence} as an interestingness metric, we show that it is highly sensitive as well, making it unsuitable for the DP setting. 
\begin{definition}[Jensen-Shannon Divergence \cite{lin1991divergence}]
For two distributions $p$ and $q$ over the same domain, their Jensen–Shannon \emph{divergence}  is defined as 
\[
JSD(p,q) = H\left(\frac{p+q}{2}\right)-\frac{1}{2}H(q) - \frac{1}{2}H(q)
\]
where $H(\cdot)$ is the Shannon entropy  and $\frac{p+q}{2}$ is the mixture distribution of $p$ and $q$.
The Jensen–Shannon \emph{distance}, denoted $d_{JS}$, is the square root of the Jensen–Shannon divergence.
\end{definition}
Let $P_A(D)$ and $P_A(D_c)$ be the probability distributions of the values in the columns of $\proj_A(D)$ and $\proj_A(D_c)$, respectively, determined by the relative frequencies of each value.
Define $d_{JS}(P_A(D),P_A(D_c))$ as the Jensen-Shannon distance between these distributions. We prove the following:
\begin{proposition}
   The sensitivity of $d_{JS}$ is at least $\frac{1}{2}$ and its range is $[0,1]$.
\end{proposition}
\begin{proof}
The proof for the range bound can be found in \cite{lin1991divergence}. We proceed with the sensitivity analysis.
Let $D$  be a dataset of size $n$, and $A$ an attribute. Suppose that for all tuples $t\in D$ it holds that $t[A]=a$ for some $a\in \dom(A)$. Let $D_c = \set{t}$ be a cluster of size $1$. In this case, $ d_{JS}(P_A(D),P_A(D_c)) = 0$, as $P_A(D)$ and $P_A(D_c)$ are the same distribution (the value $a$ has probability $1$). 
Now, let $D' = D\cup\set{t'}$ and $D_c' = D_c\cup\set{t'}$ for a tuple $t'$ with $t'[A]=a'$ where $a'\neq a$. We now have
\[
P_A(D') = \begin{cases}
    a & \text{w.p $\frac{n}{n+1}$} \\
    a' & \text{w.p $\frac{1}{n+1}$} \\
\end{cases}
\]
and 
\[
P_A(D_c') = \begin{cases}
    a & \text{w.p $\frac{1}{2}$} \\
    a' & \text{w.p $\frac{1}{2}$}
\end{cases}
\]
therefore, the mixture distribution satisfies
\[
\frac{P_A(D') +P_A(D_c')}{2}  = \begin{cases}
    a & \text{w.p $\frac{n}{2(n+1)} + \frac{1}{4}$} \\
    a' & \text{w.p $\frac{1}{2(n+1)} + \frac{1}{4}$}
\end{cases}
\]
Since all three distributions are supported on $\set{a,a'}$, we have
\[
JSD(P_A(D'),P_A(D_c')) = H_b\left(\frac{1}{2(n+1)} + \frac{1}{4}\right)-\frac{H_b\left(\frac{1}{n+1} \right) + H_b\left(\frac{1}{2}\right)}{2} 
\]
where $H_b$ is the binary entropy function. Note that by continuity of $H_b$ we have
\begin{align*}
    \lim_{n\to\infty}JSD(P_A(D'),P_A(D_c')) &= H_b\left(\frac{1}{4}\right) - \frac{1}{2} \\
    & \approx 0.311
\end{align*}
In particular, for sufficiently large $n$, $JSD(P_A(D'),P_A(D_c')) > 0.3$, and hence $d_{JS}(P_A(D'),P_A(D_c')) > \frac{1}{2}$.
Therefore, we find that $d_{JS}$ has sensitivity greater than $\frac{1}{2}$. 
\end{proof}
Intuitively, since the range of $d_{JS}$ is $[0,1]$ \cite{lin1991divergence}, its sensitivity is relatively high.
We proceed with the analysis of our low sensitivity interestingness function.
\IntLowSens*

\begin{proof}
Let $f:\dom(R)\to C$ be a clustering function, and $c\in C$ be a cluster label.
Let  $A\in \cA$ and $ D\nbr  D'$ be neighboring datasets such that $ D' =  D\cup\set{t}$. 
Denote $a = t[A]$, and let $\hist_a = \hist_A(t)$ 
be the $\abs{\dom(A)}$-dimensional histogram with $1$ in its $a$'th entry and $0$ elsewhere.
Let $D_c$ (respectively, $D'_c$) denote the set of tuples in $D$ (respectively, $D'$) that are mapped to $c$ by the function $f$, and note that $D'_c$ is either $D_c$ or  $D_c\cup\set{t}$.

We first consider the case that $D_c = D'_c$. By \Cref{lemma:interest_vector} and the triangle inequality, we have
\begin{align*}
&\abs{\interest( D',f, c, A) - \interest( D,f, c, A) } \\
&= \frac{1}{2}\abs{\norm{ \hist_A(D_c) - \frac{|D_c|}{\abs{D'}} \hist_A(D')}_1- \norm{\hist_A (D_c) - \frac{|D_c|}{\abs{D}} \hist_A(D)}_1} \\
&\le \frac{1}{2}\norm{ \frac{|D_c|}{\abs{D}} \hist_A(D)- \frac{|D_c|}{\abs{D'}} \hist_A(D') }_1
\end{align*}
Substituting $\hist_A(D') = \hist_A(D) + \hist_a$ into the inequality above and applying the triangle inequality again, we obtain
\begin{align*}
&= \frac{1}{2}\norm{  \frac{|D_c|}{\abs{D}+1}\hist_a - \frac{|D_c|}{\abs{ D}(\abs{D}+1)} \hist_A(D)}_1 \\
&\le  \frac{1}{2}\left(\norm{ \hist_a}_1 + \frac{1}{\abs{ D}} \norm{\hist_A(D)}_1\right)
\end{align*}
% where in the equality we have used that $\abs{D'}=\abs{D}+1$, and therefore $|D_c|/\abs{D} - |D_c|/\abs{D'} = |D_c|/(\abs{D}(\abs{D}+1))$, and in the inequality we applied the triangle inequality, and used that $|D_c|\le |D|+1$.
where for the equality, we used the fact that $\abs{D'} = \abs{D} + 1$, which implies  
$
\frac{\abs{D_c}}{\abs{D}} - \frac{\abs{D_c}}{\abs{D'}} = \frac{\abs{D_c}}{\abs{D}(\abs{D} + 1)}.
$  
For the inequality,  we also  bound $\abs{D_c} \leq \abs{D} + 1$.
Next, we substitute 
$\norm{\hist_a}_1 =1$ and $\norm{\hist_A(D)}_1=|D|$ to obtain
\begin{align*}
= \frac{1}{2}\left(1 + \frac{\abs{D}}{\abs{D}}\right) = 1
\end{align*}
Now, consider the case that  $D'_c = D_c\cup\set{t}$. Applying the triangle inequality, we have
\begin{align*}
&\abs{\interest( D',f, c, A)  - \interest( D,f, c, A)} \\
&\le \frac{1}{2}\norm{ \hist_A(D'_c) - \frac{\abs{D_c'}}{\abs{D'}} \hist_A(D')- \hist_A(D_c)+ \frac{|D_c|}{\abs{D}} \hist_A(D)}.
\end{align*}
Substituting  $\hist_A(D') = \hist_A(D) + \hist_a$ and  $\hist_A(D'_c) = \hist_A(D_c) + \hist_a$, and rearranging, we obtain
\begin{align*}
&= \frac{1}{2}\norm{  \hist_a - \frac{\abs{D'_c}}{\abs{D'}} \hist_A(D) -\frac{\abs{D'_c}}{\abs{D'}} \hist_a + \frac{|D_c|}{\abs{D}} \hist_A(D)} \\
& = \frac{1}{2}\norm{  \left( 1 -  \frac{\abs{D'_c}}{\abs{D'}}\right)\hist_a - \left(\frac{\abs{D'_c}}{\abs{D'}} - \frac{|D_c|}{\abs{D}}\right) \hist_A(D)} \\
\end{align*}
Recall that $\abs{D'_c}=|D_c|+1$ and $\abs{D'}=\abs{D}+1$. Therefore, $\abs{D'_c} / \abs{D'} - |D_c| / \abs{D} =(\abs{D}-|D_c|) / (\abs{D}(\abs{D}+1))$, which is non-negative, as $|D_c|\le \abs{D}$. Since $\abs{D'_c}\le \abs{D'}$ holds as well, we have
\begin{align*}
& \le  \frac{1}{2}\left( 1 -  \frac{\abs{D'_c}}{\abs{D'}}\right)\norm{ \hist_a}_1 + \frac{1}{2}\left(\frac{\abs{D}-|D_c|}{\abs{D}(\abs{D}+1)}\right)\norm{\hist_A(D)}\\
&\le  \frac{1}{2}\left( 1 -  \frac{\abs{D'_c}}{\abs{D'}}\right)\norm{ \hist_a}_1 + \frac{1}{2}\left(\frac{1}{\abs{D}+1}\right)\norm{\hist_A(D)}.\\
\end{align*}
Since $\norm{\hist_a}_1 =1$ and $\norm{\hist_A(D)}_1=|D|$, we conclude
\begin{align*}
 &   = \frac{1}{2}\left( 1 -  \frac{\abs{D'_c}}{\abs{D'}}\right) + \ \frac{1}{2}\left(\frac{1}{\abs{D}+1}\right) |D| \\
    & \le  \frac{1}{2} + \frac{1}{2} = 1.
\end{align*}
For the range upper bound, we use \Cref{lemma:interest_vector} to obtain
\begin{align*}
   \interest( D,f, c, A)  &= \frac{1}{2} \norm{\hist_A(D_c)- \frac{|D_c|}{\abs{D}} \hist_A(D) }_1 \\
   & \le \frac{1}{2} \left(\norm{\hist_A(D_c)}_1 + \frac{|D_c|}{\abs{D}} \norm{\hist_A(D)}_1 \right) \\
    &= \abs{D_c}
\end{align*}

\end{proof}

\subsection{ Sufficiency}
\label{appendix:suf}

We revisit the notion of sufficiency from prior work and provide its sensitivity analysis.
For a tuple $t$, an attribute $A$, and a cluster label $c$, Let
$r(t,A) = \cnt_{A=t[A]}(D_{f(t)})/ \cnt_{A=t[A]}(D)$. Observe that $r(t, A)$ equals the probability that a uniformly random tuple sampled from $D$, conditioned on having the same value in attribute $A$ as $t$, belongs the same  cluster as $t$. The quantity $r(t, A)$ is used in \cite{TabEE} to quantify the extent to which an HBE that employs attribute $A$ to explain a cluster labeled $f(t)$ applies to the tuple $t$. The local sufficiency of an attribute combination $\AC$ at a tuple $t$ is thus defined as 
\begin{align}\label{eq:suff-old-local}
     m^s_{\AC}(t) =  \frac{\displaystyle \sum_{t'\in D}  \one_{\set{f(t')=f(t)}}\cdot r(t',\AC({f(t)})), } {\displaystyle \sum_{t'\in D} r(t', \AC({f(t)}))}.
\end{align}
Let $c = f(t)$ be the cluster of a tuple $t$, and $A_c = \AC(c)$ the attribute used to explain this cluster by attribute combination $\AC$. Observe that $m^s_{\mathcal{AC}}(t)$ attains its maximum value of $1$ when the value $t[A_c]$ appears only in the cluster $D_c$, capturing the notion of maximal sufficiency.

To measure the \emph{global} sufficiency of $\AC$, \cite{TabEE} averages the local sufficiency across all tuples:
\begin{align} \label{eq:suff-old-global}
    \sufTabEE(D, f, \AC) = \frac{1}{\abs{D}} \sum_{t\in D}  m^s_{\AC}(t).
\end{align}
In the unbounded-DP variant, where the dataset size is not fixed and which we adopt in this work, this function exhibits high sensitivity relative to its range. We remark that this issue does not arise in the bounded-DP variant, as can be seen in the proof of \Cref{prop:suff-sensitivity}.  However, there is still motivation to modify this function due to the other reasons outlined in \Cref{subsec:suff}.
\SufHighSens*

\begin{proof}
To see that the range is $[0,1]$, note that by definition in~\eqref{eq:suff-old-local}, $m^s_{\AC}(t) \in [0,1]$ for every tuple~$t$, and~\eqref{eq:suff-old-global} averages these values across all tuples. We proceed with the sensitivity analysis.
Consider a dataset $D=\set{t_1}$ and two clusters $D_1=\set{t_1}$ and an empty cluster $D_2=\emptyset$. Let $\AC$ be an attribute combination, and suppose denote $\AC(i)=A$ for $i=1,2$.
Let us denote $a=t_1[A]$.
In this case $ \suf( D, f, 1,  A)=1$, as the value $a$ appears only inside the cluster $D_1$. 
Hence, using the the equality from item \textit{(1)} of \Cref{prop:suff-sensitivity}, 
\begin{align*}
\sufTabEE(D, \AC, f)=  \frac{1}{ \abs{D}} \sum_{c\in C}  \suf(D, f, c, \AC(c)) = 1.
\end{align*}
Now, suppose a tuple $t_2$ is added to the cluster $D_2$. That is, $D'=\set{t_1,t_2}$, $D'_1=\set{t_1}$ and $D'_2=\set{t_2}$. Suppose further that $t_2[A]=a$.
In this case, for $i\in\set{1,2}$
\begin{align*}
\suf(D', f, i, A) &= \sum_{b\in\dom_{D'_i}(A)}\frac{(\cnt_{A=b}(D'_i))^{2}}{\cnt_{A=b}(D')}  =  \frac{1}{2}.
\end{align*}
% Moreover,
% \begin{align*}
% \suf(D', f, 2, A_2) &= \sum_{a\in\dom_{D'_2}(A_2)}\frac{(\cnt_{A_2=a}(D'_2))^{2}}{\cnt_{A_2=a}(D')} \\
% &=  \frac{1}{1} + \frac{1}{2} =  \frac{3}{2}.
% \end{align*}
Hence, we find that
\begin{align*}
\sufTabEE(D', \AC, f) &=  \frac{1}{ \abs{D'}} \sum_{c\in \set{1,2}}  \suf(D', f, c, \AC(c)) = \frac{1}{2}\cdot (\frac{1}{2}+\frac{1}{2})\\
& = \frac{1}{2}.
\end{align*}

\end{proof}

\SufLowSens*
% Before analyzing the sensitivity $\SufTabEE$, we now prove \Cref{eq:equiv_form_suff}, which establishes the relation of our low-sensitivity sufficiency function $\Suf$ (\Cref{def:low_sens_suff}) to  $\SufTabEE$. This lemma will be used for the sensitivity analysis
\begin{proof}[Proof of \textit{(1)}]
% Let $\AC$ be an attribute combination and $t\in D$. Denote $c=f(t)$, and $A_c=\AC(c)$.
Let $\AC$ be an attribute combination. Denote $A_c=\AC(c)$.
Observe that the right-hand-side of \Cref{eq:suff-old-local} depends only on the cluster assignment $f(t)$, therefore it is equal across all tuples $t$ of the same cluster. Thus,
plugging in the definition of $r(t',A)$,
\begin{align}\label{eq:equiv_form_suff}
   \sum_{t\in D} m^s_E(t) =  \sum_{c\in C} \abs{D_c}\cdot \frac{\sum_{t'\in D_c}\frac{\cnt_{A_c=t'[A_c]}(D_c)}{\cnt_{A_c=t['A_c]}(D)}}{\sum_{t'\in D}\frac{\cnt_{A_c=t'[A_c]}(D_c)}{\cnt_{A_c=t['A_c]}(D)}}.
\end{align}
Let us consider each cluster separately, and
% Using the definition for $\Pr[D_c|A=t[A]]$, we have
% \begin{align*}
% \frac{\sum_{t\in D_c}\Pr[D_c|A_c=t[A_c]]}{\sum_{t\in D}\Pr[D_c|A_c=t[A_c]]} =\frac{\sum_{t'\in D_c}\frac{\cnt_{A_c=t[A_c]}(D_c)}{\cnt_{A_c=t[A_c]}(D)}}{\sum_{t\in D}\frac{\cnt_{A_c=t[A_c]}(D_c)}{\cnt_{A_c=t[A_c]}(D)}}.
% \end{align*}
% Now,
observe that each value $\cnt_{A_c=a}(D_c) / \cnt_{A_c=a}(D)$ appears in the numerator exactly $\cnt_{A_c=a}(D_c)$ times, and $\cnt_{A_c=a}(D)$ times in the denominator. Hence, by changing the order of summation
\begin{align*}
& \frac{\sum\limits_{t\in D_c}\frac{\cnt_{A_c=t'[A_c]}(D_c)}{\cnt_{A_c=t['A_c]}(D)}}{\sum\limits_{t'\in D}\frac{\cnt_{A_c=t'[A_c]}(D_c)}{\cnt_{A_c=t['A_c]}(D)}}\\
 & =\frac{\sum\limits_{a\in\dom_{D_c}(A)}\cnt_{A_c=a}(D_c)\cdot\frac{\cnt_{A_c=a}(D_c)}{\cnt_{A_c=a}(D)}}{\sum\limits_{a\in\dom_D(A_c)}\cnt_{A_c=a}(D)\cdot\frac{\cnt_{A_c=a}(D_c)}{\cnt_{A_c=a}(D)}}\\
  & =\frac{\sum\limits_{a\in\dom_{D_c}(A_c)}\frac{(\cnt_{A_c=a}(D_c))^2}{\cnt_{A_c=a}(D)}}{\sum\limits_{a\in\dom_D(A_c)}\cnt_{A_c=a}(D_c)}
\end{align*}
Since the denominator equals $|D_c|$, the following equality holds:
\begin{align*}
 & =\frac{1}{|D_c|}\sum_{a\in\dom_{D_c}(A_c)}\frac{(\cnt_{A_c=a}(D_c))^{2}}{\cnt_{A_c=a}(D)} \\
  & = \frac{1}{|D_c|}  \suf( D, f, c,  A).
 \end{align*}
The proposition follows by substituting the equality into \Cref{eq:equiv_form_suff}.
\end{proof}

% \ron{TODO: Add motiv
To prove item \textit{(2)} of \Cref{prop:suff-sensitivity}, we introduce the following to lemmas.
\begin{lemma}\label{claim:suff_properties}
The range of $\suf( D, f, c,  A) $ is  $ [0, |D_c|]$.
\end{lemma}

\begin{proof}
We have
\begin{align*}
     \suf( D, f, c,  A) &=  \sum_{a\in\dom_{D_c}(A)} \frac{\cnt_{A=a}(D_c)^2}{\cnt_{A=a}(D)} \\
                    & \le \sum_{a\in\dom_{D_c}(A)} \cnt_{A=a}(D_c)\\
                    & = |D_c|
\end{align*}
    where the inequality holds since $\cnt_{A=a}(D_c) \le \cnt_{A=a}(D)$.
\end{proof}

\begin{lemma}\label{lem:suff}
    Let $a,b\in\R$ such that $b>0$ and $0\le a\le b$. Then
    \begin{enumerate}[label=(\roman*),font=\itshape]
        \item  $
        \abs{ \frac{a^2}{b} - \frac{(a+1)^2}{b+1}}\le 1
    $, 
    \item $
        \abs{ \frac{a^2}{b} - \frac{a^2}{b+1}}\le 1
    $
    \end{enumerate}

\end{lemma}

\begin{proof}
For item \textit{(i)},  observe that our assumption $a\le b$ implies that $ \frac{(a+1)^2}{b+1} \ge \frac{a^2}{b}$. 
Indeed,
\begin{align*}
\frac{a^2}{b} &= \frac{a^2(b+1)}{b(b+1)} \\
              &=   \frac{a^2b+a^2}{b(b+1)} 
              \le   \frac{a^2b+2ab+b}{b(b+1)} 
              =   \frac{b(a^2+2a+1)}{b(b+1)}             
              = \frac{(a+1)^2}{b+1} .    
\end{align*}
where the in the inequality we used $a^2 \le ab \le 2ab+b$.
Now, we have 
\begin{align*}
\frac{(a+1)^2}{b+1} - \frac{a^2}{b} &= \frac{b(a+1)^2-a^2(b+1)}{b(b+1)}\\
            &= \frac{2ab+b-a^2}{b(b+1)}\le \frac{b^2+b}{b(b+1)} = 1
\end{align*}
where the inequality follows from $2ab-a^2\le b^2$ which holds for any $a,b\in\R$.

For \textit{(ii)}, note that
   $$ \frac{a^2}{b} - \frac{a^2}{b+1} = \frac{a^2}{b(b+1)} < 1$$
   as $a\le b$.
\end{proof}

\begin{proof}[Proof of \Cref{prop:suff-sensitivity} item \textit{(2)}]

% Consider the function $f_c( D,A)= |D_c|\suf_c( D,A)$. That is, we drop the normalizing factor. We prove that $f_p$ as global sensitivity $1$.
Let $f:\dom(R)\to C $ be a clustering function, and $c\in C$ be a cluster label. Fix an attribute $A$.  Let $ D\nbr  D'$ be two neighboring datasets such that $ D'= D \cup \set{t}$. Let $D_c$ (respectively, $D'_c$) denote the set of tuples in $D$ (respectively, $D'$) that are mapped to $c$ by the function $f$, and note that $D'_c$ is either $D_c$ or  $D_c\cup\set{t}$.
First, note that if $D_c$ is empty, 
\[
\abs{\suf( D, f, c,  A)-\suf( D', f, c,  A)} \le 1.
\]
Indeed, in this case $\abs{D'_c}\le 1$, and 
by \Cref{claim:suff_properties} we have $\suf( D, f, c,  A) = 0$ and $\suf( D', D'_c, A) \le 1$.

Second, if $\cnt_{A=t[A]}(D)=0$, then clearly  $\cnt_{A=t[A]}(D_c)=0$ and $\cnt_{A=t[A]}(D'_c)\le 1$.
In this case
\begin{align*}
    &\abs{\suf( D, f, c,  A)-\suf( D', f, c,  A)} \\
        &= \abs{ \sum_{a\in\dom_{D_c}(A)} \frac{\cnt_{A=a}(D_c)^2}{\cnt_{A=a}(D)} -  \sum_{a\in\dom_{D'_c}(A)} \frac{\cnt_{A=a}(D'_c)^2}{\cnt_{A=a}(D')}} \\
        &= \frac{\cnt_{A=t[A]}(D'_c)^2}{\cnt_{A=t[A]}(D')}\\
        & \le 1.    
\end{align*}
where the second equality holds because all summands are identical in both sums, except for the term corresponding to $a = t[A]$, which appears only in the second sum. The inequality holds since the numerator is at most $1$.

Hence, we assume that $D_c$ is not empty and that $\cnt_{A=t[A]}(D)>0$.
We consider two cases according to whether or not $t$ is added to $D_c$.

First assume that $D'_c = D_c$.  We have
\begin{align*}
&\abs{\suf( D, f, c,  A)-\suf( D', f, c,  A)}\\
&=  \abs{\frac{\cnt_{A=t[A]}(D_c)^2}{\cnt_{A=t[A]}(D)} - \frac{\cnt_{A=t[A]}(D_c)^2}{\cnt_{A=t[A]}(D)+1}} \le 1
\end{align*}
where in the equality we have used that $\cnt_{A=a}(D_c)=\cnt_{A=a}(D'_c)$ for all $a$ since $D'_c=D_c$. We also used that the only summand in the definition of $\suf$ that changes is the one corresponding to $t[A]$. The inequality uses  item \textit{(i)} of \Cref{lem:suff}.

If $D'_c = D_c\cup\set{t}$, we use item \textit{(ii)} of \Cref{lem:suff} to obtain
\begin{align*}
&\abs{\suf( D, f, c,  A)-\suf( D', f, c,  A)}\\
&=  \abs{\frac{\cnt_{A=t[A]}(D_c)^2}{\cnt_{A=t[A]}(D)} - \frac{(\cnt_{A=t[A]}(D_c)+1)^2}{\cnt_{A=t[A]}(D)+1}} \le 1.
\end{align*}

\end{proof}

\subsection{Diversity}
\label{appendix:div}
We begin with a sensitivity analysis of the diversity function from prior work, and proceed with the analysis of our low sensitivity diversity measure (\Cref{def:glob_diversity}).

The diversity from \cite{TabEE} is defined as follows.
Let $\AC:C\to\cA$ be an attribute combination.
For an attribute $A$, let Let $ExpBy(\AC, A) = \AC^{-1}(\set{A})$. That is, the set of cluster labels assigned to $A$ by $\AC$ (which can also by empty).

For a finite set $S$, let $Perm(S)$ denote its set of permutations, containing all bijections  $p:\set{1,\dots,|S|}\to S$.
For a permutation $p\in Perm(ExpBy(\AC, A))$, its diversity is defined as \cite{TabEE}:
\[
    PermDiv_A(p) = \sum_{i=1}^{\abs{ExpBy(\AC, A)}}\min_{j<i} \tvd(\proj_A(D_{p(i)}, \proj_A(D_{p(j)}))
\]
and if $\abs{ExpBy(\AC, A)}=1$, $PermDiv_A(p)$ is set as $1$.

Now, the diversity measure is defined as\footnote{to obtain a value between $0$ and $1$, this function can be normalized by the number of clusters $\abs{C}$.}
\[
    \diversityTabEE(D,f,\AC) = \sum_{A\in \cA} \sum_{p\in Perm(ExpBy(\AC, A))} \frac{PermDiv_A(p)}{\abs{ExpBy(\AC, A)}!}
\]
\begin{proposition}
    The sensitivity of $\diversityTabEE(D,f,\AC) $ is at least $\frac{1}{2}$ and its range is $[0,\abs{C}]$.
\end{proposition}
Note that since the number of clusters $\abs{C}$ is typically a small constant, the sensitivity is relatively high compared to the range.
\begin{proof}
Let $D$ be a dataset of size $n$ and $f:\dom(R)\to C$ be a clustering function such that $D_1 = \set{t_1}$ is a cluster of size 1. 
Let $A$ be an attribute and suppose that there exists $a\in \dom(A)$ such that $t[A]=a$ for all tuples $t\in D$. 
Let $\AC$ be an attribute combination mapping all cluster labels to $A$. 

In this case, we have $ExpBy(\AC, A) = C$.  Moreover, for any $c,c'\in C$ we have that $\tvd(\proj_A(D_{c}), \proj_A(D_{c'})) = 0$, as the distributions of values of the column $A$ is identical among all clusters. 
Therefore, for any permutation $p\in Perm(C)$, we find that $PermDiv_A(p) = 0$. Hence, $ \diversityTabEE(D,f,\AC) = 0$.

Now, suppose a new tuple is added to cluster 1. Denote $D' = D\cup\set{t_2}$ and $D_1' = \set{t_1, t_2}$. Suppose further that $t_2[A]=a'$ for $a'\neq a$.  Now, for every $c\neq 1$ we have  
\begin{align}\label{eq:dist}
    \tvd(\proj_A(D_{1}), \proj_A(D_{c}))=\frac{1}{2}
\end{align}
and for every $c,c'\in C\setminus\set{1}$ we have 
\begin{align}\label{eq:dist2}
    \tvd(\proj_A(D_{c}), \proj_A(D_{c'}))=0
\end{align}
as for other clusters the distribution of values is unchanged. Hence, for any permutation $p\in Perm(C)$, we find that
\[
    PermDiv_A(p) = \sum_{i=1}^{\abs{C}}\min_{j<i} \tvd(\proj_A(D_{p(i)}, \proj_A(D_{p(j)})) = \frac{1}{2},
\]
where we used that exactly one of the summands has the form (\ref{eq:dist}) and the rest have the form (\ref{eq:dist2}). Therefore, we conclude that $\diversityTabEE(D,f,\AC) =1/2$.

For the range bound, observe that $\diversityTabEE(D, f, \AC)$ is maximized when all pairwise distances equal $1$. This can occur, for instance, when $\AC$ maps each cluster to a distinct attribute. Alternatively, when any two clusters assigned with the same attribute $A$ satisfy $\tvd(\proj_A(D_{1}), \proj_A(D_{c})) = 1$. Assuming this is the case, we have
\[
    PermDiv_A(p) = \sum_{i=1}^{\abs{ExpBy(\AC, A)}} 1 =\abs{ExpBy(\AC, A)}.
\]
Thus,
\begin{align*}
      \diversityTabEE(D,f,\AC) &= \sum_{A\in \cA} \sum_{p\in Perm(ExpBy(\AC, A))} \frac{PermDiv_A(p)}{\abs{ExpBy(\AC, A)}!}\\
        &= \sum_{A\in \cA} \sum_{p\in Perm(ExpBy(\AC, A))} \frac{\abs{ExpBy(\AC, A)}}{\abs{ExpBy(\AC, A)}!}\\
        &= \sum_{A\in \cA} \abs{ExpBy(\AC, A)} \\
        &= \abs{C}.
\end{align*}
\end{proof}
We now proceed with the sensitivity analysis of our diversity measure (\Cref{def:glob_diversity}).
\begin{lemma}\label{claim:divers-upperbound}
 For two clusters $D_c,D_{c'}\subseteq D$ and  attributes $A_c, A_{c'}$, it holds that $d(D, f, c, c', A_c, A_{c'}) \in [0, \min\set{|D_c|, |D_{c'}|}] $
\end{lemma}
\begin{proof}
If one of the two clusters is empty, or if $A_c \neq A_{c'}$, then the claim trivially holds. Hence, we assume that both are not empty and that $A_c= A_{c'} = A$. In this case
\begin{align*}   
            &\frac{1}{2}\sum_{a\in \dom(A)} \abs{ \frac{\cnt_{A=a}(D_c)}{\max\set{\abs{ D_c},1}} - \frac{\cnt_{A=a}(D_{c'})}{\max\set{|D_{c'}|,1}}} \\
            & \le \frac{1}{2}\sum_{a\in \dom(A)} \abs{ \frac{\cnt_{A=a}(D_c)}{\abs{ D_c}}} + \frac{1}{2}\sum_{a\in \dom(A)} \abs{ \frac{\cnt_{A=a}(D_{c'})}{\abs{ D_{c'}}}} \\
             & = \frac{1}{2} + \frac{1}{2} = 1
\end{align*}
and therefore $d(D, f, c, c', A_c, A_{c'})\le \min\set{|D_c|, |D_{c'}|}$.
\end{proof}

\begin{proposition}\label{lem:pairwise-diverse-sensitivity}
$d$ has sensitivity  $1$.
\end{proposition}

The proof is similar to that of \Cref{prop:interest-sensitivity}, with the main difference being that we now consider the distance between histograms of disjoint subsets of the data, instead of one subset and the entire dataset.
\begin{proof}
Let $f:\dom(R)\to C$ be a clustering function, and $c,c'\in C$ be two cluster labels.
Fix attributes $A_c, A_{c'}$, and let  $ D\nbr  D'$ be neighboring datasets such that $ D' =  D\cup\set{t}$. 

Let $D_c$ and $D_{c'}$ (respectively, $D'_c$ and $D'_c$) denote the sets of tuples in $D$ (respectively, $D'$) that are mapped to $c$ and $c'$ by the function $f$.

First, observe that when $A_c \neq A_{c'}$, we have $d(D, f, c, c', A_c, A_{c'}) = \min\set{|D_c|, |D_{c'}|}$, which changes by at most $1$ when we add or remove one tuple from $D$. Hence, we proceed with the assumption that $A_c=A_{c'} = A$. To simplify notation, we let $d(D, c, c', A) =d(D, f, c, c', A, A)$.
If $t$ has not been added to either cluster, i.e, $D_c=D'_c$ and $D_{c'}=D'_{c'}$, then clearly 
$$\abs{d(D, c, c', A)- d(D', c, c', A)} = 0.$$
Otherwise, consider the case that $t$ has been added to one of the clusters. Without loss of generality, suppose that $D'_{c'} = D_{c'}\cup\set{t}$ and  $D_c' = D_c$. 
If either of $D_c$ or $D_{c'}$ is empty, then by \Cref{claim:divers-upperbound}, we have $d(D_c,D_{c'}, A) = 0$, and $d(D_c,D'_{c'}, A) \le 1$, and so
\[
\abs{d(D, c, c', A)- d(D', c, c', A)} \le 1.
\]
Hence, we proceed with the assumption that both are not empty.
Denote $a = t[A]$, and let $\hist_a = \hist_A(t)$ be the $\abs{\dom(A)}$-dimensional histogram with $1$ in its $a$'th entry and $0$ elsewhere.
First, consider the case that $|D_c| \le |D_{c'}|$. Recalling that $|D'_{c'}|= |D_{c'}|+1$, we have
\begin{align*}
&\abs{d(D, c, c', A)- d(D', c, c', A)} \\
&= \frac{1}{2}\abs{\norm{ \hist_A(D_c) - \frac{|D_c|}{|D_{c'}|} \hist_A (D_{c'})}_1 - \norm{ \hist_A (D_c) -\frac{|D_c|}{|D_{c'}|+1} \hist_A(D'_{c'})}_1}
% &\le \frac{1}{2}\norm{ \hist_A(D') - \frac{|D_c|}{\abs{D'}} \hist_A(D_c)- \hist_A(D)+ \frac{|D_c|}{\abs{D}} \hist_A(D_c)}
\end{align*}
By the triangle inequality,
\begin{align*}
&\le \frac{1}{2}\norm{ \hist_A(D_c) - \frac{|D_c|}{|D_{c'}|} \hist_A (D_{c'}) - \hist_A (D_c) + \frac{|D_c|}{|D_{c'}|+1} \hist_A(D'_{c'})}
\end{align*}
Substituting $ \hist_A(D'_{c'})=  \hist_A(D_{c'}) + \hist_a$ into the inequality above, we obtain
\begin{align*}
&= \frac{1}{2}\norm{ \frac{|D_c|}{|D_{c'}|+1}\hist_a - \left(\frac{|D_c|}{|D_{c'}|} - \frac{|D_c|}{|D_{c'}|+1}\right) \hist_A (D_{c'})}_1 \\
% &\le  \frac{1}{2}\left(\norm{ \hist_a}_1 + \frac{|D_c|}{\abs{ D}(\abs{D}+1)} \norm{\hist_A(D_c)}_1\right)
\end{align*}
Applying the triangle inequality and noting that $\frac{|D_c|}{|D_{c'}|} - \frac{|D_c|}{|D_{c'}|+1} = \frac{|D_c|}{|D_{c'}|(|D_{c'}|+1)}$, we have
\begin{align*}
&\le  \frac{1}{2}\left(\norm{ \hist_a}_1 +  \frac{|D_c|}{|D_{c'}|(|D_{c'}|+1)} \norm{\hist (D_{c'})}_1\right)
\end{align*}
substituting 
$\norm{\hist_a}_1 =1$ and $\norm{\hist (D_{c'})}_1=|D_{c'}|$,
\begin{align*}
= \frac{1}{2}\left(1 + \frac{|D_c|\cdot |D_{c'}|}{|D_{c'}|(|D_{c'}|+1)}\right) \le 1
\end{align*}
where we have also used our assumption that $|D_c|\le |D_{c'}|$.

Now, consider the case that $|D_c| > |D_{c'}|$, and therefore also $|D_c| \ge |D'_{c'}| = |D_{c'}|+1$. We have
\begin{align*}
&\abs{d(D, c, c', A)- d(D', c, c', A)} \\
&= \frac{1}{2}\abs{\norm{  \frac{|D_{c'}|}{|D_c|} \hist_A(D_c) - \hist_A(D_{c'})}_1 - \norm{ \frac{|D_{c'}|+1}{|D_c|}\hist_A (D_c) - \hist_A(D'_{c'})}_1}
\end{align*}
By the triangle inequality,
\begin{align*}
&\le \frac{1}{2}\norm{  \frac{|D_{c'}|}{|D_c|} \hist_A(D_c) - \hist_A (D_{c'}) - \frac{|D_{c'}|+1}{|D_c|}\hist_A (D_c) + \hist_A(D'_{c'})}_1
\end{align*}
Substituting $ \hist_A(D'_{c'})=  \hist_A(D_{c'}) + \hist_a$ into the inequality above and rearranging, we obtain
\begin{align*}
&\le \frac{1}{2} \norm{  \frac{-1} {|D_c|} \hist_A(D_c) + \hist_a}_1
\end{align*}
Next, we once again use 
$\norm{\hist_a}_1 =1$ and $\norm{\hist (D_c)}_1=|D_c|$
\begin{align*}
&\le \frac{1}{2}\left(\frac{1}{|D_c|}\norm{\hist_A(D_c)}_1 + \norm{\hist_a}_1 \right)\\
& = 1
\end{align*}
\end{proof}

\DivLowSens*
\begin{proof}
    The sensitivity bound follows from \Cref{lem:pairwise-diverse-sensitivity} using \Cref{lemma:convex-comb}, as $\diversity$ is a convex combination of sensitivity-1 functions. For the range bound, note that $d(D, f, c, c', A_c, A_{c'})$ attains its upper-bound (\Cref{claim:divers-upperbound}), in particular, when $A_c\neq A_{c'}$. Hence, Let $\AC$ be an attribute combination assigning a different attribute for each cluster. It follows that $\diversity$ is maximized, attaining value
    \begin{align}\label{diversitysum}
         \frac{1}{\binom{\numclust}{2}}\sum_{\set{c,c'}\subseteq C} \min\set{|D_c|, |D_{c'}|}.
    \end{align}
    Now, let $|D_{c_1}|\le \dots \le |D_{c_{|C|}}|$ be an ordering of the clusters by increasing size.
    For each $i=1,\dots, |C|$, the value $|D_{c_i}|$ appears as a summand in (\ref{diversitysum})
    $|C|-i$ times, each for a coupling of $c_i$ with $c_j$ for $j < i$. By changing the order of summation, we obtain that (\ref{diversitysum}) equals
    \begin{align*}
         \frac{1}{\binom{\numclust}{2}} \sum_{i=1}^{\numclust} (\numclust-i)\abs{D_{c_{i}}}
    \end{align*}
    
\end{proof}

\subsection{Combining All Quality Functions}
\label{appendix:all}
\SingLowSens*
\begin{proof}
The proposition follows directly from \Cref{lemma:convex-comb} and the sensitivity bounds of the sufficiency and interestingness functions (\Cref{prop:interest-sensitivity}, \Cref{prop:suff-sensitivity}), as $\score_\gamma$ is a convex combination of the two.
\end{proof}

\GlobLowSens*
\begin{proof}
The proposition follows directly from \Cref{lemma:convex-comb} and the sensitivity bounds of the sufficiency, interestingness and diversity functions (\Cref{prop:interest-sensitivity}, \Cref{prop:suff-sensitivity}, \Cref{prop:sens-div}), as $\globscore_\lambda$ is a convex combination of the three. For the range bounds, note that $   \frac{1}{\abs{C}}\sum_{c\in C}   \suf( D, f, c,  A)$ and $ \frac{1}{\abs{C}}\sum_{c\in C}   \suf( D, f, c,  A)$ both have range $[0,  \frac{1}{\abs{C}}\sum_{c\in C}  \abs{D_c}]$. The range upper bound for $\diversity$, $R_{\diversity}$, is given by $\Cref{prop:sens-div}$.
\end{proof}

\subsection{Proof of \Cref{prop:topk-single privacy}}
\label{appendix:single}
\CandPrivUtility*

\begin{proof}
    \textbf{(1) Privacy. } For each cluster, the distribution of the selected top-$k$ is equivalent to iteratively applying $k$ exponential mechanisms \cite{durfee2019practical}, where each satisfies $\eps = \eps_{CandSet}/(\numclust\cdot k)$. Overall, we have $\numclust$ applications of Top-$k$, hence the output distribution is equivalent to $\numclust\cdot k$ exponential mechanisms. by sequential composition (\autoref{prop:composition}), \Cref{alg:topk-single} satisfies overall $\eps_{CandSet}$-DP.

    \textbf{(2) Utility Bound. } Let $c\in C$, and $\cA_{\ell}$ be the set of remaining attributes after the $(\ell-1)$-th selection for this cluster. The distribution over the selected sequence $A^{(1)}_c,\dots,A^{(k)}_c$ is equal to applying $k$ exponential mechanisms \cite{durfee2019practical} where  $A_c^{(\ell)}$ is selected from $\cA_{\ell}$. Note that we always have $\max_{A\in \cA_{\ell}} \score(c,A) \ge  \OPT_c^{(\ell)}$. Hence, for $\eps = \eps_{CandSet}/(\numclust\cdot k)$, we apply the utility theorem of exponential mechanism (Theorem 3.11 in \cite{dwork2014algorithmic}) to obtain
    \begin{align*}
           &\Pr\left[ \score(c,A_c^{(\ell)}) \le  \OPT_c^{(\ell)} -  \frac{2}{\eps}(\ln{\abs{\cA}}+t)\right]  \\
              &\le \Pr\left[ \score(c,A_c^{(\ell)}) \le \max_{A\in \cA_{\ell}} \score(c,A) - \frac{2}{\eps}(\ln{\abs{\cA_{\ell}}}+t)\right] \le e^{-t}
    \end{align*}
    where we have used that $\Delta_{\score} \le 1$.
\end{proof}

\section{Generating Multiple Explanations per Cluster}
\label{appendix:multexp}
In this section we show how \ourframework\ can be extended to output multiple histograms per cluster in a global explanation.
To this end, we extend the definition of an \emph{attribute combination} to a mapping $\AC:C\to\set{S \subseteq \cA \mid \abs{S}=\ell}$, assigning to each cluster label a set of $\ell$ attributes. Our goal is thus to find a high-quality attribute combination such that the histograms of the corresponding attributes form a high-quality HBE, where the global explanation contains $\ell$ histograms per cluster. This requires extending the global score function \Cref{def:global_score} to attribute combinations with larger outputs.

We measure the overall diversity as follows. Let \[Cand(\AC) = \set{ (c,A) \mid c\in C, A\in \AC(c)}\]
and define
\begin{align*}
   \mathrm{Div}_{\ell}(D, f, \AC) = \frac{1}{\binom{\abs{Cand(\AC) }}{2}} \sum_{\substack{(c,A),(c',A')}} d(D, f,c,c',A_c,A_{c'})
\end{align*}
where the sum is over all different pairs  $\set{(c,A),(c',A')} \subseteq Cand(\AC) $. Note that the definition coincides with \Cref{def:glob_diversity} when $\ell = 1$.

The sufficiency and interestingness in the global score function \Cref{def:global_score} were averages across all candidates. With the new definition, they remain averages, but are now taken over a larger set of candidates. Overall, we have:
\begin{align*}
    \globscore_{\lambda}(D,f, \AC) &=   \lambda_{\intTabEE}\cdot \mathrm{Int}_{\ell}(D, f, \AC)+\lambda_{\sufTabEE}\cdot \mathrm{Suf}_{\ell}(D, f, \AC)  \\
     &\qquad \qquad +  \lambda_{\diversityTabEE}\cdot \mathrm{Div}_{\ell}(D, f, \AC)
\end{align*}
where we extend \[\mathrm{Int}_{\ell}(D, f, \AC) = \frac{1}{\abs{Cand(\AC)}} \sum_{(c,A)\in Cand(\AC)} \interest( D, f, c, A))\] and \[\mathrm{Suf}_{\ell}(D, f, \AC) = \frac{1}{\abs{Cand(\AC)}} \sum_{(c,A)\in Cand(\AC)} \suf( D, f, c, A). \] 
 Note that the definition coincides with \Cref{def:global_score} when $\ell = 1$. The sensitivity of this function is bounded by $1$, as it remains a convex combination of sensitivity-$1$ functions, with an analysis analogous to that of \Cref{prop:global_score-sensitivity}.

Stage-1 of \ourframework\ is unchanged, as its purpose is to narrow the search space to high-quality candidates for each cluster, which then form the pool for a global explanation. For Stage-2, we may use the exponential mechanism to select a high-scoring attribute combination $\AC:C\to\set{S \subseteq \cA \mid \abs{S}=\ell}$ with respect to the extended low sensitivity global score.
Finally, As in the case of $\ell=1$, noisy histograms are generated only for the $\abs{C}\times \ell $ selected attributes.

However, the drawback of this approach is that the set of all possible attribute combinations, after the filtering performed in Stage-1, now has a size of ${k \choose \ell}^{\abs{C}}$. This size can be large, and computing the global score for all attribute combinations may require significant computational time.
While the utility guarantee of \Cref{alg:topk-single} remains the same, the candidate set provided to the exponential mechanism in \Cref{alg:gen_global_explanation} now has a size of ${k \choose \ell}^{\abs{C}}$ instead of $k^{\abs{C}}$, resulting in a larger additive error term of the EM utility bound. Indeed,  letting $\OPT$ be the highest score of an attribute combination $\AC:C\to\set{S \subseteq \cA \mid \abs{S}=\ell}$ and $\AC$ be the one selected by the EM in \Cref{alg:gen_global_explanation}, Theorem 3.11 in \cite{dwork2014algorithmic} implies that 
 \begin{align*}
           &\Pr\left[  \globscore_{\lambda}(\AC) \le \OPT -  \frac{2\abs{C}}{\eps}\left(\ln{{k \choose \ell}}+t\right)\right]  \le e^{-t}
    \end{align*}

\section{Supplementary Experiments}
\label{appendix-exp}

\paratitle{Pre-processing of the Diabetes  dataset}
To ensure the interpretability of the generated histograms with a bounded number of bins, we apply several preprocessing steps to the Diabetes dataset \cite{diabetes}. We remove the unique identifiers \texttt{`patient\_nbr`} and \texttt{`encounter} \texttt{\_id`}. Numerical attributes including \texttt{`num\_lab\_procedures`} and \texttt{`num\_medications`} are binned. The attribute \texttt{`medical\_specialty`} is mapped to broader categories such as ``General Practice" and ``Surgery", following the categorization in \cite{diabetes_paper} that introduced the dataset. Additionally, each ICD code in the attributes \texttt{`diag\_1`}, \texttt{`diag\_2`}, and \texttt{`diag\_3`} is replaced with its corresponding diagnostic category (e.g., values in the range 390–459 are mapped to "Circulatory") according to the mapping defined in \cite{diabetes_paper}. The preprocessing code is publicly available in \cite{dpclustex_github}.

\paratitle{Pre-processing of the Stack Overflow dataset}
We chose the 2018 Survey due to its larger sample size compared to more recent years. Both numerical and categorical attributes were considered, while attributes containing textual values or multiple-choice combinations were excluded. One potential approach was to expand the multiple-choice answers into binary attributes. However, this led to a significant increase in dimensionality, causing the DP-k-means algorithm to fail in clustering the data within a reasonable privacy budget, and other methods to fail due to scalability limitations. Additionally, we discard attributes with more than $60\%$ of missing values. 
The numerical attribute  \texttt{`ConvertedSalary`} is binned. The preprocessing code is available in \cite{dpclustex_github}.

\subsection{Selected Attributes Quality score and Error}
We present additional results for the Diabetes dataset with $3$ and $7$ clusters.
\Cref{fig:diabetes-3-7-quality} shows the trend of $Quality$ values of the selected attribute combination as the total privacy budget $\eps$ varies. \Cref{fig:diabetes-3-7-mae} shows the trend of $ \mathrm{MAE}$ values of the selected attribute combination as the total privacy budget $\eps$ varies.

\begin{figure*}
    \includegraphics[width=\linewidth]{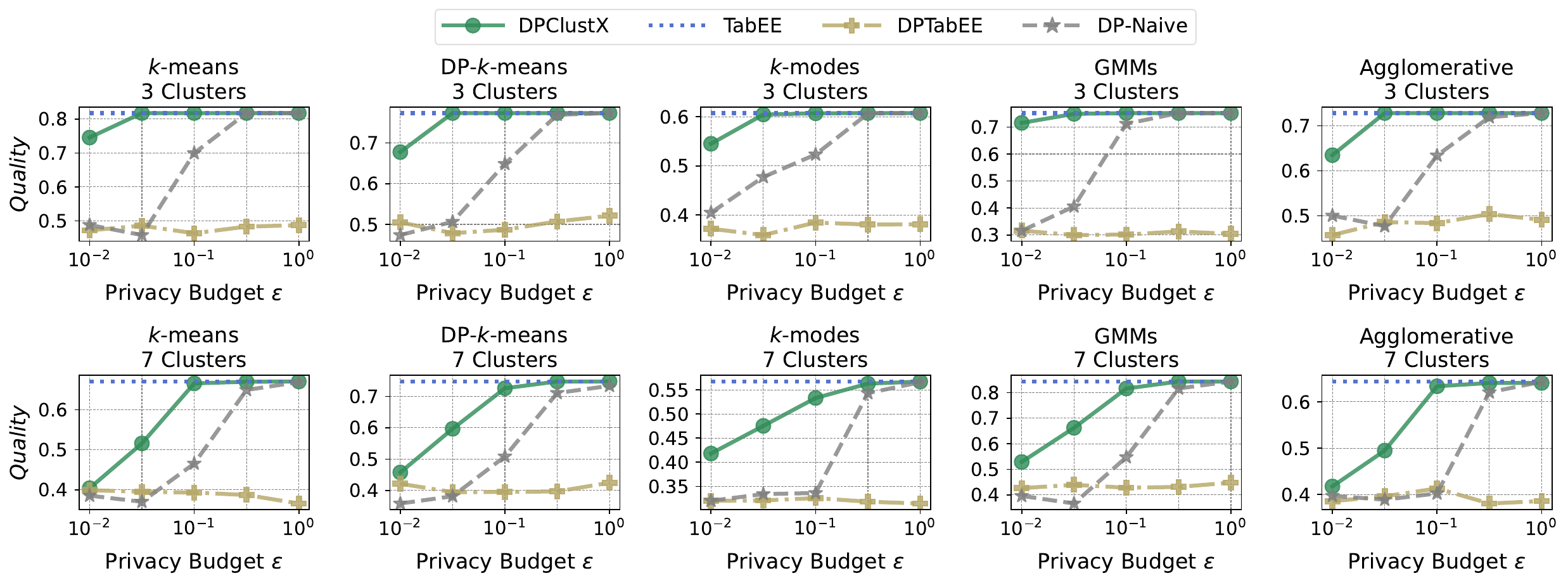}
    \caption{$Quality$ values of the selected attribute combination for the Diabetes dataset,  as the total privacy budget $\eps$ varies.}
    \label{fig:diabetes-3-7-quality}
\end{figure*}

\begin{figure*}
    \includegraphics[width=\linewidth]{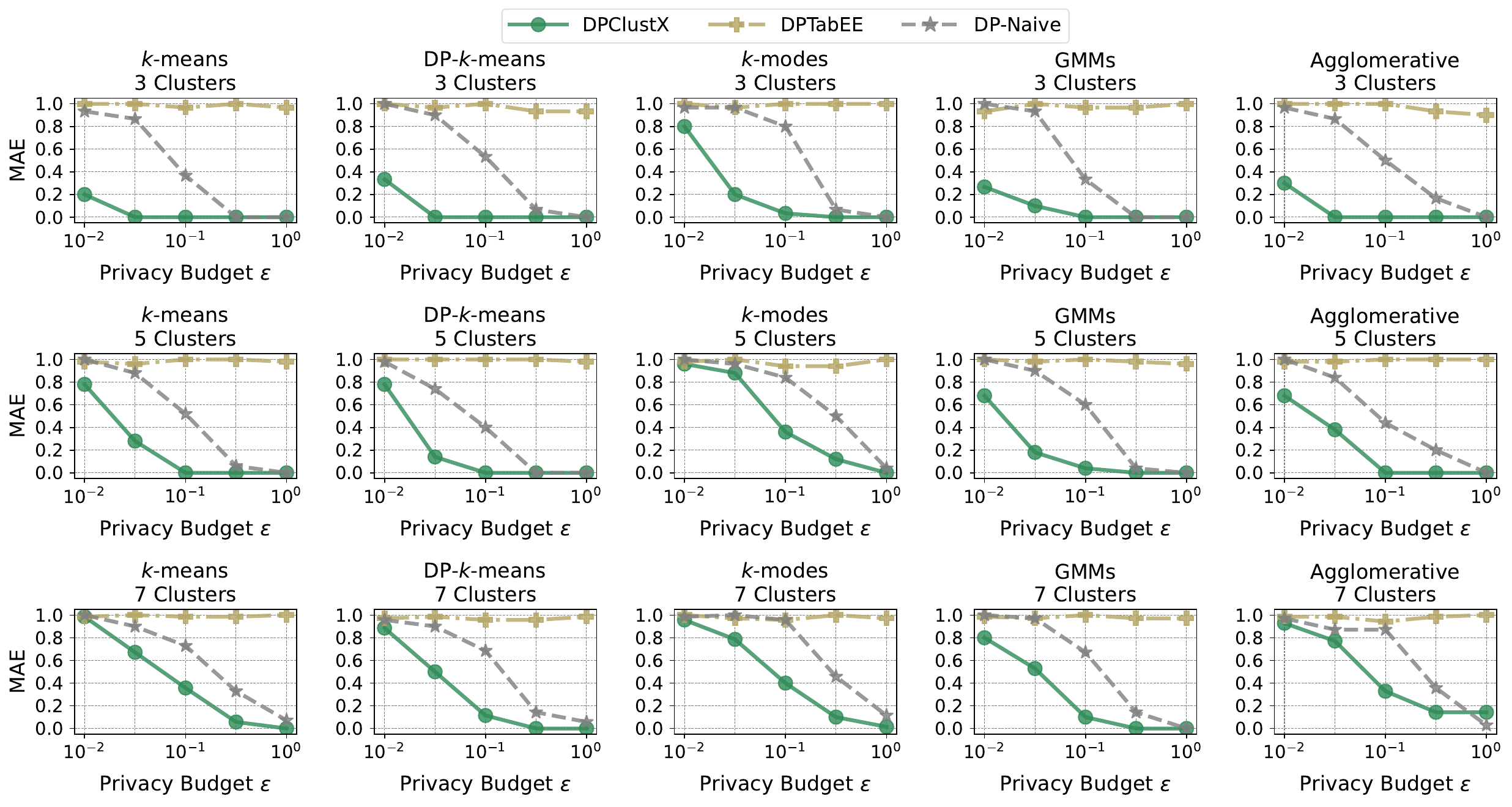}
    \caption{$\mathrm{MAE}$ values of the selected attribute combination for the Diabetes dataset,  as the total privacy budget $\eps$ varies.}
    \label{fig:diabetes-3-7-mae}
\end{figure*}

\subsection{Quality for different choices of weights}
We present the missing empirical results on the Diabetes and the Census datasets in \Cref{table:weight_exp}.

%%%%%%%%%%%%%%%%%%%%%%%%%%%%%%%%%%%%%% 3 Clusters %%%%%%%%%%%%%%%%%%%%%%%%%%%%%%%%%%%%%%%%%
{\fontsize{6.5}{9.5}\selectfont
\begin{table*}
\centering
\begin{minipage}[t]{0.48\textwidth}
\centering
\begin{tabular}[t]{c|c|c|c|c|c|c}
\hline
\textbf{\# Clusters} &\textbf{Clustering}  & \textbf{Explainer}   & Equal & \textbf{$\lambda_{\intTabEE} = 0$} & \textbf{$\lambda_{\sufTabEE} = 0$} & \textbf{$\lambda_{\diversityTabEE} = 0$} \\ \hline
3 & $k$-means            & \ourframework\        & 0.8176  & 0.9208   & 0.7882   & 0.7439   \\
&                       & TabEE                 & 0.8176  & 0.9208   & 0.7882   & 0.7439   \\ \cline{2-7}
& $k$-modes            & \ourframework\        & 0.6076  & 0.7493   & 0.6618   & 0.4114   \\
&                       & TabEE                 & 0.6076  & 0.7493   & 0.6621   & 0.4114   \\ \cline{2-7}
& Agglomerative        & \ourframework\        & 0.7281  & 0.8291   & 0.7362   & 0.6547   \\
&                       & TabEE                 & 0.7281  & 0.8291   & 0.7362   & 0.6547   \\ \cline{2-7}
& DP-$k$-means         & \ourframework\        & 0.7735  & 0.8701   & 0.7494   & 0.7010   \\
&                       & TabEE                 & 0.7735  & 0.8701   & 0.7662   & 0.7010   \\ \cline{2-7}
& GMMs  & \ourframework\        & 0.7515  & 0.8709   & 0.7563   & 0.6294   \\
&                       & TabEE                 & 0.7515  & 0.8709   & 0.7563   & 0.6294   \\\cline{1-7}
5& $k$-means         & \ourframework\ & 0.6874 & 0.7498 & 0.7805 & 0.5319 \\
&                   & TabEE          & 0.6874 & 0.7498 & 0.7805 & 0.5319 \\ \cline{2-7}
& $k$-modes         & \ourframework\ & 0.5592 & 0.6615 & 0.6821 & 0.3500 \\
&                   & TabEE          & 0.5639 & 0.6641 & 0.6828 & 0.3503 \\ \cline{2-7}
& Agglomerative     & \ourframework\ & 0.7255 & 0.7804 & 0.7899 & 0.6061 \\
&                   & TabEE          & 0.7255 & 0.7836 & 0.7899 & 0.6061 \\ \cline{2-7}
& DP-$k$-means      & \ourframework\ & 0.7735 & 0.8263 & 0.8310 & 0.6633 \\
&                   & TabEE          & 0.7735 & 0.8263 & 0.8310 & 0.6633 \\ \cline{2-7}
& GMMs              & \ourframework\ & 0.8164 & 0.8708 & 0.8518 & 0.7267 \\
&                   & TabEE          & 0.8164 & 0.8708 & 0.8518 & 0.7267 \\\cline{1-7}
7 & $k$-means            & \ourframework\        & 0.6664  & 0.7040   & 0.7893   & 0.5304   \\
&                       & TabEE                 & 0.6706  & 0.7063   & 0.7893   & 0.5307   \\ \cline{2-7}
& $k$-modes            & \ourframework\        & 0.5613  & 0.6380   & 0.6994   & 0.3527   \\
&                       & TabEE                 & 0.5673  & 0.6461   & 0.7049   & 0.3533   \\ \cline{2-7}
& Agglomerative        & \ourframework\        & 0.6396  & 0.6957   & 0.7481   & 0.5068   \\
&                       & TabEE                 & 0.6438  & 0.6926   & 0.7630   & 0.5068   \\ \cline{2-7}
& DP-$k$-means         & \ourframework\        & 0.7442  & 0.7696   & 0.8340   & 0.6429   \\
&                       & TabEE                 & 0.7474  & 0.7798   & 0.8340   & 0.6429   \\ \cline{2-7}
& GMMs  & \ourframework\        & 0.8440  & 0.8694   & 0.8967   & 0.7660   \\
&                       & TabEE                 & 0.8440  & 0.8694   & 0.8967   & 0.7660   \\ \cline{1-7}

\end{tabular}

\vspace{2mm}\subcaption{Diabetes dataset.}
\label{table:weight_exp_diabetes_3}
\end{minipage}%
\hspace{0.02\textwidth} % Adjust spacing between the two tables
\begin{minipage}[t]{0.48\textwidth}
\centering
\begin{tabular}[t]{c|c|c|c|c|c|c}
\hline
\textbf{\# Clusters} &\textbf{Clustering}  & \textbf{Explainer}   & Equal & \textbf{$\lambda_{\intTabEE} = 0$} & \textbf{$\lambda_{\sufTabEE} = 0$} & \textbf{$\lambda_{\diversityTabEE} = 0$} \\ \hline
3& $k$-means            & \ourframework\        & 0.8785  & 0.9888   & 0.8289   & 0.8187   \\
&                      & TabEE                 & 0.8785  & 0.9888   & 0.8289   & 0.8187   \\ \cline{2-7}
&$k$-modes            & \ourframework\        & 0.8749  & 0.9859   & 0.8265   & 0.8131   \\
&                      & TabEE                 & 0.8749  & 0.9859   & 0.8265   & 0.8131   \\ \cline{2-7}
& DP-$k$-means         & \ourframework\        & 0.8889  & 1.0000   & 0.8333   & 0.8333   \\
&                      & TabEE                 & 0.8889  & 1.0000   & 0.8333   & 0.8333   \\ \cline{2-7}
& GMMs  & \ourframework\        & 0.5438  & 0.7122   & 0.6066   & 0.3157   \\
&                      & TabEE                 & 0.5438  & 0.7122   & 0.6066   & 0.3157   \\ \cline{1-7}
5& $k$-means     & \ourframework\ & 0.8637 & 0.9195 & 0.8768 & 0.7987 \\
&               & TabEE          & 0.8643 & 0.9197 & 0.8768 & 0.7987 \\ \cline{2-7}
& $k$-modes     & \ourframework\ & 0.8247 & 0.8981 & 0.8390 & 0.7449 \\
&               & TabEE          & 0.8248 & 0.8981 & 0.8390 & 0.7449 \\ \cline{2-7}
& DP-$k$-means  & \ourframework\ & 0.8827 & 0.9552 & 0.8541 & 0.8451 \\
&               & TabEE          & 0.8828 & 0.9658 & 0.8619 & 0.8451 \\ \cline{2-7}
& GMMs          & \ourframework\ & 0.4822 & 0.6296 & 0.6009 & 0.2258 \\
&               & TabEE          & 0.4820 & 0.6296 & 0.6009 & 0.2258 \\\cline{1-7}
7& $k$-means     & \ourframework\ & 0.8521 & 0.8825 & 0.8995 & 0.7922 \\
&               & TabEE          & 0.8538 & 0.8865 & 0.9016 & 0.7922 \\ \cline{2-7}
& $k$-modes     & \ourframework\ & 0.7798 & 0.8506 & 0.8293 & 0.6719 \\
&               & TabEE          & 0.7806 & 0.8513 & 0.8293 & 0.6726 \\ \cline{2-7}
& DP-$k$-means  & \ourframework\ & 0.8981 & 0.9376 & 0.9053 & 0.8515 \\
&               & TabEE          & 0.8985 & 0.9376 & 0.9064 & 0.8515 \\ \cline{2-7}
& GMMs          & \ourframework\ & 0.4944 & 0.6174 & 0.6271 & 0.2430 \\
&               & TabEE          & 0.4944 & 0.6174 & 0.6271 & 0.2430 \\ \cline{1-7}

\end{tabular}

\vspace{2mm}\subcaption{Census dataset.}
\label{table:weight_exp_census_3}
\end{minipage}
\caption{$Quality$ values for with different choices of parameters $\lambda_{\intTabEE}$, $\lambda_{\sufTabEE}$, and $\lambda_{\diversityTabEE}$.}
\label{table:weight_exp}
\end{table*}
}

\fi

\end{document}